\documentclass[a4paper,11pt]{article}
\pdfoutput=1 
\usepackage{jheppub1} 
\usepackage[T1]{fontenc} 

\usepackage{subfigure}
\usepackage{braket,comment}

\renewcommand{\Im}{\operatorname{Im}}

\newcommand{\bcc}{\operatorname{b.c.c.}}
\newcommand{\arctanh}{\operatorname{arctanh}}
\newcommand{\sn}{\operatorname{sn}}
\newcommand{\cn}{\operatorname{cn}}
\newcommand{\dn}{\operatorname{dn}}
\newcommand{\arcsinh}{\operatorname{arcsinh}}

\newcommand{\calK}{\mathcal{K}}
\newcommand{\calT}{\mathcal{T}}
\newcommand{\boldg}{\boldsymbol{g}}
\newcommand{\boldI}{\boldsymbol{I}}

\title{\boldmath Subsystem Complexity and Measurements in Holography}

\author[a]{Shao-Kai Jian}
\author[b,c]{and Yuzhen Zhang}

\affiliation[a]{Department of Physics and Engineering Physics, Tulane University, \\New Orleans, Louisiana, 70118, USA}

\affiliation[b]{Institute for Advanced Study, Tsinghua University, \\Beijing 100084, China}

\affiliation[c]{Zhili College, Tsinghua University, \\Beijing 100084, China}

\emailAdd{sjian@tulane.edu}
\emailAdd{zhang-yz20@mails.tsinghua.edu.cn}

\abstract{We investigate the impact of measuring one subsystem on the holographic complexity of another. 
While a naive expectation might suggest a reduction in complexity due to the collapse of the state to a trivial product state during quantum measurements, our findings reveal a counterintuitive result: in numerous scenarios, measurements on one subsystem can amplify the complexity of another.
We first present a counting argument elucidating this complexity transition in random states.
Then, employing the subregion ``complexity=volume'' (CV) proposal, we identify a complexity phase transition induced by projection measurements in various holographic CFT setups, including CFT vacuum states, thermofield double states, and the joint system of a black hole coupled to a bath. 
According to the AdS/BCFT correspondence, the post-measurement dual geometry involves an end-of-the-world brane created by the projection measurement.  
The complexity phase transition corresponds to the transition of the entanglement wedge to the one connected to the brane. 
In the context of the thermofield double setup, complete projection on one side can transform the other side into a boundary state black hole with higher complexity or a pure AdS with lower complexity.
In the joint system of a black hole coupled to a nongraviting bath, where (a part of) the radiation is measured, the BCFT features two boundaries: one for the black hole and the other for the measurement. 
We construct the bulk dual involving intersecting or non-intersecting branes, and investigate the complexity transition induced by the projection measurement. 
Notably, for a subsystem that contains the black hole brane, its RT surface may undergo a transition, giving rise to a complexity jump.

}

\begin{document}

\maketitle
\flushbottom

\newpage

\section{Introduction}
Quantum complexity is the minimal number of simple gates required to prepare the target state from a simple product state. 
In a generic random quantum circuit, complexity increases with time~\cite{Brown2017second,Brandao:2019sgy,Haferkamp:2021uxo,Jian:2022pvj}. 
Excessive measurements in the quantum circuit can lead to the collapse of the time-evolved state into trivial product states, resulting in a reduction in global state complexity.
Hence, measurements can reduce complexity in the global state. 
If one continuously varies the measurement rate, there could be a transition~\cite{Suzuki:2023wxw}. 
However, a more unconventional phenomenon arises, wherein measuring a subsystem has the unexpected consequence of increasing the complexity of its complement.
This was revealed by recent research in condensed matter and quantum information~\cite{Ho:2021dmh,Choi:2021npc,Ippoliti:2022bsj,Cotler:2021pbc,Claeys:2022hts,aguado2008creation,Bolt:2016nww,piroli2021quantum,Tantivasadakarn:2021vel,Verresen:2021wdv,Bravyi:2022zcw,Lu:2022jax,Tantivasadakarn:2022ceu,Tantivasadakarn:2022hgp,Lee:2022xss,Zhu:2022bpk,Iqbal:2023shx,Foss-Feig:2023uew,Lu:2023mpl}. 
In this study, we aim to systematically explore and understand these phenomena within the framework of holography.

Consider a generic (Haar random) state on $N$ qubits, with $N$ large. 
For a subsystem $A$ that is smaller than the half of the total system size, it has almost vanishing complexity, because the density matrix on $A$ is approximately maximally mixed. 
As shown by Page~\cite{Page:1993df},
\begin{equation}
	\left\lVert\rho_A-\frac{I_A}{d_A}\right\lVert_1\le\sqrt{\frac{d_A}{d_B}}
\end{equation}
where $d_{A,B}$ are the dimension of the Hilbert space on $A$ and $B$, respectively. 
The maximally mixed state contains almost no information, and is easy to prepare---we simply introduce one ancillary qubit for every qubit in our system and form EPR pairs between them. It is a simple state.

On the other hand, measuring the complement system $B$ will teleport information into $A$, making it more complex. 
More rigorously, given a global pure state $\ket{\psi}$, one can measure $B$ in different computational basis to form an ensemble of pure states in $A$. 
When $\frac{d_B}{d_A}$ gets bigger, the ensemble will form an approximate $k$-design with high probability~\cite{Cotler:2021pbc}. 
The mean complexity of states in this ensemble is lower bounded from below by a linear dependence on $k$~\cite{Roberts:2016hpo,Brandao:2019sgy}. 
If we further assume that the initial state $\ket{\psi}$ is Haar random, then the post-measurement ensemble on $A$ becomes Haar random with exponential complexity. 
Hence, measurements create complexity. 

\begin{figure}
    \centering
    \subfigure[]{\includegraphics[width=0.63\textwidth]{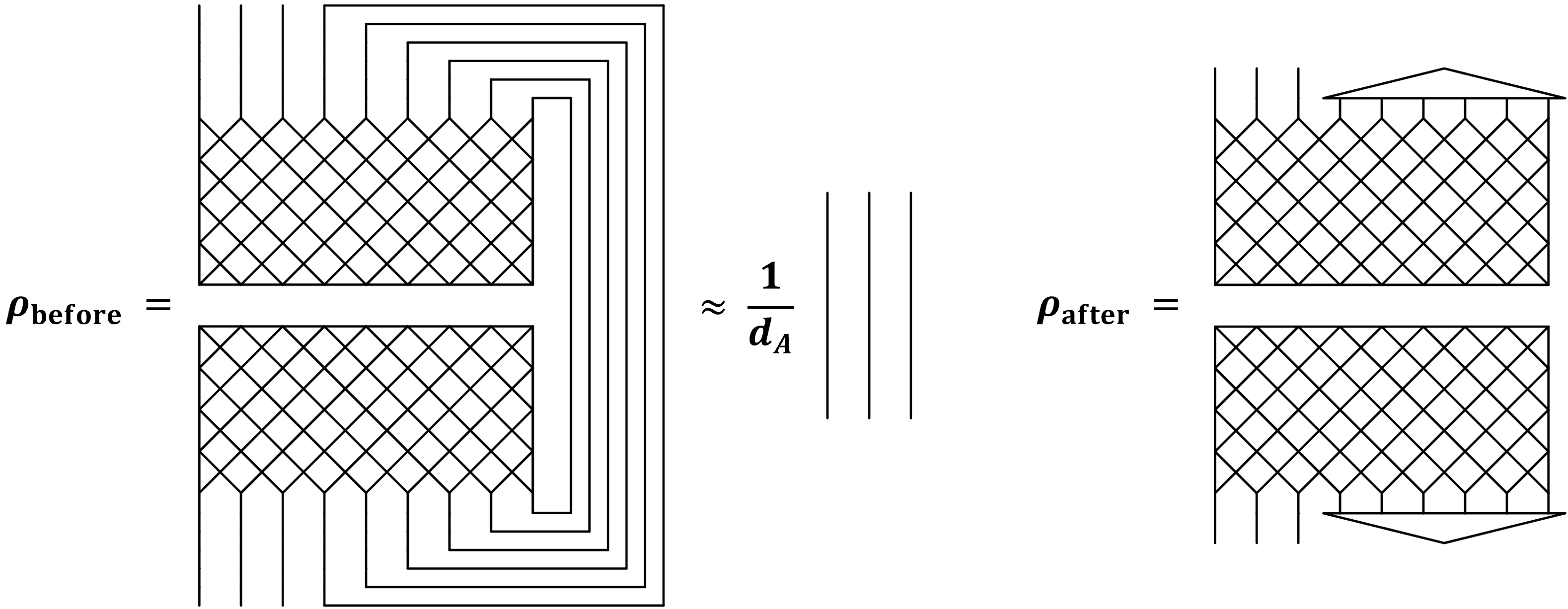}\label{fig:cartoon_tensor_network}} \qquad 
    \subfigure[]{\includegraphics[width=0.3\textwidth]{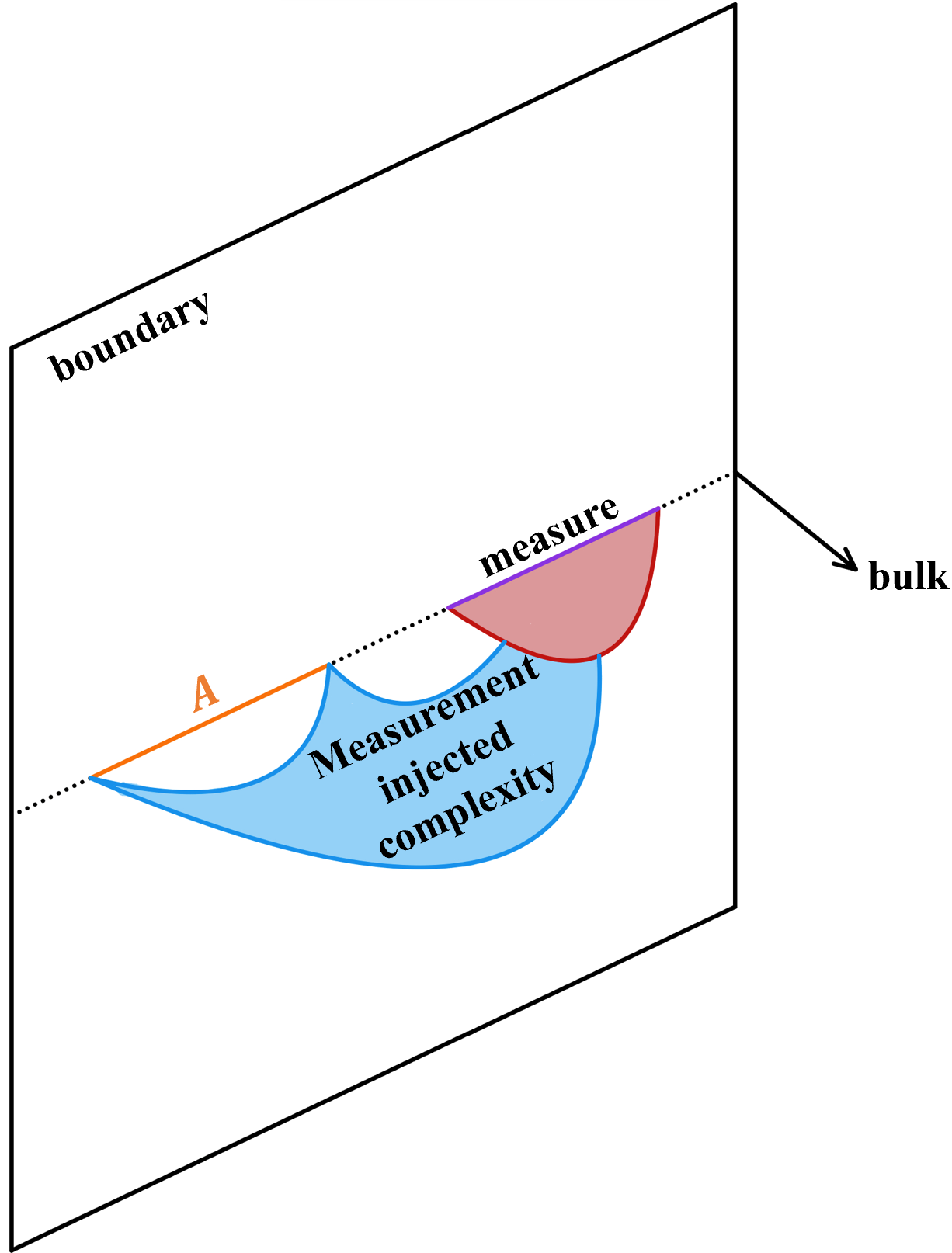}\label{fig:cartoon_holographic_measurement_injected_complexity}} 
    \caption{(a) The density matrix in generic states before and after measurements---a tensor network cartoon. (b) Measurement injected complexity in holography.}
\end{figure}

We can use tensor networks to gain some intuition. 
All states can be represented (approximately) by some tensor network. 
If the initial state is a random state, then it takes a very complex tensor network to describe it. 
However, if we only look at a small subsystem $A$ and trace out its complement $B$, then we get an approximately maximally mixed state that has a very simple tensor network representation---a few lines that represent the identity, or, EPRs between the bra and ket. 
See the cartoon in figure \ref{fig:cartoon_tensor_network}. 
Taking the trace amounts to connecting qubits in $B$ in the bra and ket. 
These connections miraculously allow the tensor network to simplify into the EPR lines. 
But if we measure $B$ instead of tracing it out, we cut off these connections between the bra and ket. 
Now the TN cannot be represented as EPR lines. 
Perhaps the TN can be somewhat simplified, but we are sure that it is of maximal complexity for subsystem $A$,  because the pure states we get on $A$ are Haar random. 

In the realm of quantum information and condensed matter, there has been a surge of interest in using measurements to simplify state preparation. 
For instance, projection on a subsystem $B$ can be used to prepare approximate $k$-designs on the subsystem $A$~\cite{Ho:2021dmh,Choi:2021npc,Ippoliti:2022bsj,Cotler:2021pbc,Claeys:2022hts}. 
Moreover, measuring some ancillary qubits can help prepare long-range entangled states~\cite{aguado2008creation,Bolt:2016nww,piroli2021quantum,Tantivasadakarn:2021vel,Verresen:2021wdv,Bravyi:2022zcw,Lu:2022jax,Tantivasadakarn:2022ceu,Tantivasadakarn:2022hgp,Lee:2022xss,Zhu:2022bpk,Iqbal:2023shx,Foss-Feig:2023uew,Lu:2023mpl}. 
Our study sheds light on understanding the success of these state preparation protocols from a complexity perspective in the context of holography. 

In holographic CFTs, we can invoke AdS/CFT duality to geometrize information theoretic quantities in the field theory. 
For pure states, the geometrization of complexity is achieved by the Complexity-Volume (CV) proposal~\cite{Stanford:2014jda}. 
For mixed states, complexity is a somewhat more ambiguous object, and there are several different definitions of mixed state complexity~\cite{Agon:2018zso}. 
Here we will adopt the subregion CV proposal proposed in~\cite{Alishahiha:2015rta} and further studied in~\cite{Carmi:2016wjl,Ben-Ami:2016qex,Mazhari:2016yng,Abt:2017pmf,Abt:2018ywl,Chapman:2018bqj,Caceres:2019pgf,Hernandez:2020nem}.
Notice that holographic complexity has been extensively studied in various situations~\cite{Susskind:2014jwa,Brown:2015bva,Aaronson:2016vto,Couch:2016exn,Chapman:2016hwi,susskind2016entanglement,Carmi:2017jqz,Susskind:2018pmk,Chapman:2018hou,Hernandez:2020nem,Belin:2021bga}.\footnote{This list is by no means complete. 
Interested readers may also look into the references in therein.}
The proposal states that the complexity of a boundary interval $A$ is given by the maximal volume bounded by $A$ and its RT surface:
\begin{equation}
    C=\max\frac{V}{G_NL},
\end{equation}
where $L$ is the curvature radius of AdS, and $G_N$ is the Newton's constant. 
This is very intuitive because in tensor network models of holography~\cite{Swingle:2009bg,Pastawski:2015qua,Hayden:2016cfa}, the complexity is the size of the most efficient tensor network that represents the subsystem~\cite{Caputa:2017yrh,Abt:2017pmf}. 
The holographic dual of a special type of subregion projection is studied in~\cite{Miyaji:2014mca, Numasawa:2016emc}, see also~\cite{Antonini:2022lmg,Antonini:2022sfm,Antonini:2023aza,Sun:2023hlu}. 
The bulk dual after the projection contains an end-of-the-world brane, on which the RT surface can end. 
In some scenarios, the entanglement wedge can undergo a transition from a disconnected one to the one connected to the brane. 
Complexity then jumps to a larger value due to the extra volume. 
See the cartoon in figure \ref{fig:cartoon_holographic_measurement_injected_complexity}. 
It is also possible for the measurements to make complexity smaller, where the brane has negative tension and ``bends'' toward the region $A$ to eat up some volume. 

Based on the subregion CV correspondence, we study complexity transitions induced by subregion projection measurements in various holographic setups.
The paper is organized as follows. 
In section~\ref{sec:Measurement induced complexity transition in random states}, we give a more quantitative estimate of the complexity transition based on counting arguments for random states. 
In particular, if the size of the subsystem is less than half of the total system, the complexity of this subsystem will go through a jump from near zero to an exponentially big number when its complement is gradually projected out.

In section~\ref{sec:Measurements on the vacuum}, we study the subsystem complexity with projection measurements in the holographic CFT vacuum. 
The projection measurement is modeled by a slit, which upon conformal transformation is mapped onto an upper half plane.  
Then, according to AdS/BCFT, we arrive at an AdS$_3$ bulk with an end-of-the-world brane terminated at the boundary (corresponding to the measurement region). 
The subregion complexity, given by the volume enclosed by the subsystem at the boundary and its RT surface, is evaluated using the Gauss-Bonnet theorem. 
We find that when the measurement region is increased, the subsystem complexity can feature a jump to a higher value, which originates from the exchange of RT surfaces. 

In section \ref{sec:Thermal field double state, measure left system}, we extend our investigation to thermofield double state. 
One of the goals is to explore the effect of entanglement on the subsystem complexity upon measurement.
We investigate subsystem complexity with projection measurements in an infinite TFD state.
In certain limits, the subsystem complexity becomes linear in the subsystem size and in the temperature, in contrast to the case of measurements in the CFT vacuum state.
We also investigate measuring entirely one side of the thermofield double state and study how the complexity of the other side is affected. 
The measurement can either transform the other side into a boundary state black hole with higher complexity or pure AdS with lower complexity.

In section \ref{sec:Black hole coupled to zero temperature bath}, we couple a quantum dot to a semi-infinite bath CFT, and study measurements on the bath. 
This is a toy model for ``a black hole coupled to a bath'' setup, where the radiation (bath) is measured. 
We focus on the zero temperature case, and comment on the finite temperature case. 
The joint system is modeled by a BCFT with two boundaries---one for the system, the other for the bath. 
We construct the bulk dual with intersecting or non-intersecting branes associated with the two boundaries. 
For a subsystem that contains the system brane, its RT surface may undergo a transition, giving rise to a complexity jump.

In section \ref{sec:conclusion}, we conclude the paper with a few future research directions.

\section{Measurement induced complexity transition in random states}
\label{sec:Measurement induced complexity transition in random states}

\subsection{Measuring the entire complement gives maximal complexity}

We consider $N$ qubits divided into subsystem $A$ with $N_A$ qubits and subsystem $B$ with $N_B$ qubits. 
Suppose the entire system is in a random pure state $\ket{\psi}$. 
Before doing any measurements, the state in the subsystem $A$ is close to the maximally mixed state when $\frac{d_A}{d_B}$ gets large: $\left\lVert\rho_A-\frac{I_A}{d_A}\right\lVert_1\le\sqrt{\frac{d_A}{d_B}}$. 
This state is very simple and can be prepared by introducing $N_A$ ancillary qubits and forming Bell pairs with $A$. 

Now we measure subsystem $B$. 
When the entire complement of $A$ is measured, we can get pure states, whose complexity is somewhat less ambiguous than the complexity of mixed states. 
The pure state ensemble is labelled by two parameters: the measurement outcomes $z$ and the random initial states $\ket{\psi}$. This ensemble is actually the ensemble of Haar random pure states on $A$. 
Here is an explanation. 
First, we show that for a fixed measurement result $z$, the ensemble of pure states $\ket{\phi_{A}(z)}$ that we get on $A$ is Haar distributed. 
We expand the global state in some basis $\ket{\psi}=\sum_ic_i\ket{i}$.
\begin{equation}
	\ket{\psi}=\begin{pmatrix}
		c_1 & \cdots & c_{d_{A}} & c_{d_{A}+1} & \cdots & c_{d_{AB}}.
	\end{pmatrix}^T
\end{equation}
Since $\ket{\psi}$ is a random state on $AB$, the entries are (independently) Gaussian distributed before normalization. 
Suppose the first $d_{A}$ entries are coefficients before the basis $\ket{i_{A}}\ket{z}$ for some specific $z$. 
Projecting on $\ket{z}$ gives
\begin{equation}
	\ket{\phi_{A}(z)}\propto\langle z|\psi\rangle=\begin{pmatrix}
		c_1 & \cdots & c_{d_{A}}
	\end{pmatrix}^T
\end{equation}
Now $c_1,\cdots, c_{d_{A}}$ are still Gaussian variables, so they define random states on $A$ after normalization. 
We have thus shown that for a fixed outcome, the post-measurement state is Haar random. 
Hence, the entire ensemble is Haar random. 
As we have explained, since we obtained Haar random states on $A$, the complexity of those states are exponential in $N_A$---it is the maximal complexity for a pure state. 
A natural question to ask is what happens when we do not measure the entire complement of $A$. 
We can increase the number of measured qubits one by one and ask if complexity undergoes a transition. 
This time we have a mixed state on $A$. We explore the complexity of these mixed states in the next subsection.

\subsection{Counting argument for a complexity jump}\label{sec:Counting argument for a complexity jump}

We divide the whole system into $A$, $B$ and $C$, with $N_A$, $N_B$, $N_C$ qubits. 
Let $d_A$, $d_B$ and $d_C$ denote the corresponding Hilbert space dimensions. 
We measure the qubits in $B$ and study the complexity of $A$. 
With $A$ fixed, we tune the number of measured qubits $N_B$ and ask how the complexity of $A$ changes with it.

One way to give an estimate of complexity is to count the number of distinct states in the ensemble~\cite{Roberts:2016hpo,Brandao:2019sgy}. 
Let's quickly review the method to estimate the complexity of random pure states in a $d$-dimensional Hilbert space. 
The states live in the space of $CP(d-1)$. Suppose we coarse-grain the space and count every small ball of radius $\epsilon$ as a distinct state. 
Then there are $\exp(d)$ distinct states in total~\cite{Susskind:2018pmk}. 
This is a very big number compared with the number of low complexity states. 
Suppose in every application of a gate we have $M$ options. 
Then the number of states that have complexity $C\le r$ is upper bounded by $M^r\sim\exp(r)$. 
When $\exp(d)$ is much bigger than $\exp(r)$, we can say that most of the states have $C\ge r$. 
Hence, the logarithm of the number of distinct states is roughly a lower bound for complexity. 
Now we have an ensemble of mixed states, and we expect that the argument still works here. 
If you are uncomfortable about the fact that they are mixed, think about the Choi isomorphism that maps them to pure states. 

In our setting, we get Haar random states on $AC$ after measuring $B$. The ensemble is invariant under arbitrary unitaries on $A$, so it can be cast in the following form~\cite{Nechita_2007}
\begin{equation}
	\rho=U_A\Lambda U_A^\dagger,\qquad \Lambda=\operatorname{diag}\{\lambda_1,\cdots,\lambda_{d_A}\},
\end{equation}
where $U_A$ are Haar random matrices on acting on $A$. When we take $N\rightarrow\infty$ with $N_A/N_C$ fixed, the eigenvalue distribution converges to the Marchcenko--Pastur distribution~\cite{Nechita_2007}
\begin{equation}
	D(\lambda)=\max\left\{0,1-\frac{d_C}{d_A}\right\}\delta(\lambda)+\frac{d_C}{2\pi\lambda}\sqrt{\left[\lambda-\left(d_A^{-\frac{1}{2}}-d_C^{-\frac{1}{2}}\right)^2\right]\left[\left(d_A^{-\frac{1}{2}}+d_C^{-\frac{1}{2}}\right)^2-\lambda\right]}.
\end{equation}
When $d_A/d_C$ is far greater or far smaller than one, the distribution is approximated by delta functions
\begin{equation}
	D(\lambda)\approx\begin{cases}
		\delta\left(\lambda-\frac{1}{d_A}\right),& d_A\ll d_C \\
		\left(1-\frac{d_C}{d_A}\right)\delta(\lambda)+	\frac{d_C}{d_A}\delta\left(\lambda-\frac{1}{d_C}\right), & d_A\gg d_C
	\end{cases}
\end{equation}
this happens when the size of $A$ and $C$ differ by a few qubits. 
Therefore, we can approximate the density matrix with
\begin{equation}
	\rho_A\approx\begin{cases}
		\frac{1}{d_A}\boldI_{d_A\times d_A},& d_A\ll d_C \\
		U_A\begin{pmatrix}
			\frac{1}{d_C}\boldI_{d_C\times d_C} & 0 \\ 0 & 0
		\end{pmatrix}U_A^\dagger, &d_A\gg d_C
	\end{cases}
     \label{eq:haar_density_matrix}
\end{equation}
In the first phase, all states look approximately like the maximally mixed state, because the identity is invariant under any unitary. 
In the second phase, the density matrix looks like (rescaled) random projection operators. 
Note that unitaries that are related by
\begin{equation}
	U_A=U_A'\begin{pmatrix}
		U_p & 0 \\ 0 & U_n
	\end{pmatrix}
\end{equation}
give the same density matrix, where $U_p$ is a $d_C\times d_C$ unitary and $U_n$ is a $(d_A-d_C)\times(d_A-d_C)$ unitary. 
To count the number of distinct states, we should count the number of $U_A$ and quotient over $U_p$ and $U_n$.
\begin{equation}
	\mbox{\#\ of\ states}\sim\frac{(\mbox{\#\ of\ $U_A$})}{(\mbox{\#\ of\ $U_p$})(\mbox{\#\ of\ $U_n$})}\sim\exp(2(d_A-d_C)d_C)\approx\exp(2d_Ad_C)\sim\exp\exp(N_A+N_C),
\end{equation}
where we used the fact that the number of distinct $d\times d$ unitaries is $\exp(d^2)$~\cite{Susskind:2018pmk}. 
The number $d_Ad_C$ appeared because we have $d_C$ eigenvectors of the density matrix. 
These vectors are almost independent when $d_A\gg d_C$, and each of them has $d_A$ degrees of freedom. 
In conclusion, we have
\begin{equation}
	\mbox{\#\ of\ states}\sim\begin{cases}
		1,& d_A\ll d_C \\
		\exp\exp(N_A+N_C), &d_A\gg d_C
	\end{cases}
\end{equation}
Hence our complexity estimate is
\begin{equation}
	C\gtrsim\begin{cases}
		0,& N_A<N_C \\
		?,& N_A\approx N_C \\
		\exp(N_A+N_C), &N_A>N_C
	\end{cases}
\end{equation}
Here $N_A>N_C$ means that they differ by some qubits that is enough to make $d_A\gg d_C$. 
One should not be bothered by the zero in the $N_A<N_C$ case, because in the purification definition of mixed state complexity, it is $\Omega(N_A)$ because one only needs to introduce one ancilla for every qubit and form EPR pairs between them. 
It is exponentially smaller than the complexity in the $N_A>N_C$ phase. 
If $N_A<\frac{N}{2}$ before doing any measurements, the state on $A$ will go through a transition from $C\gtrsim 0$ to $C\gtrsim\exp(N_A+N_C)$ when we continuously increase the number of measured qubits.

Let's comment on some previously defined notions of mixed state complexity~\cite{Agon:2018zso}. 
The \emph{purifcation complexity} is defined as the minimal number of gates required to prepare the purification of the desired state, where the initial state is a tensor product of $\ket{0}$'s in our system and the ancillas. 
To get the purification of \eqref{eq:haar_density_matrix}, we can first prepare $\min\{N_A,N_C\}$ EPR pairs between $A$ and ancilla qubits. 
This can be done with $\Omega(N_A)$ gates. 
In the $N_A<N_C$ phase, this is enough. In the $N_A>N_C$ phase, we still need to apply $U_A$. Therefore, the purification complexity is essentially the complexity of $U_A$, which is $\Omega(\exp N_A)$. The \emph{spectrum approach} decomposes the complexity into two parts. The first part is the \emph{spectrum complexity} that counts the minimal number of gates acting on our system and the ancilla to prepare a density matrix $\rho_{\mbox{spec}}$ with the same spectrum. 
The second part is the \emph{basis complexity} the counts the minimal number of gates on our system to convert $\rho_{\mbox{spec}}$ to our desired state. 
In the case of \eqref{eq:haar_density_matrix}, the spectrum complexity is $\Omega(N_A)$ and the basis complexity is exactly the complexity of $U_A$.

\section{Measurements on vacuum}\label{sec:Measurements on the vacuum}

In this section, we study measurements on holographic CFTs. 
We compute the complexity with the subsystem CV proposal~\cite{Alishahiha:2015rta,Carmi:2016wjl,Ben-Ami:2016qex,Mazhari:2016yng,Abt:2017pmf,Agon:2018zso,Abt:2018ywl,Chapman:2018bqj,Caceres:2019pgf,Hernandez:2020nem}.
In particular, we focus on static geometries and the complexity of a subsystem $A$ is dual to the maximal codimension-1 volume that is enclosed by its RT surface and the cut-off surface located at the asymptotic boundary,
\begin{equation}
	C_{A}=\max\frac{V}{G_NL}.
\end{equation}
In tensor network models of holography, the volume measures the size of the tensor network that is needed to describe the subsystem on the boundary.

\subsection{Infinite size system}

\begin{figure}
    \centering
    \includegraphics[width=6cm]{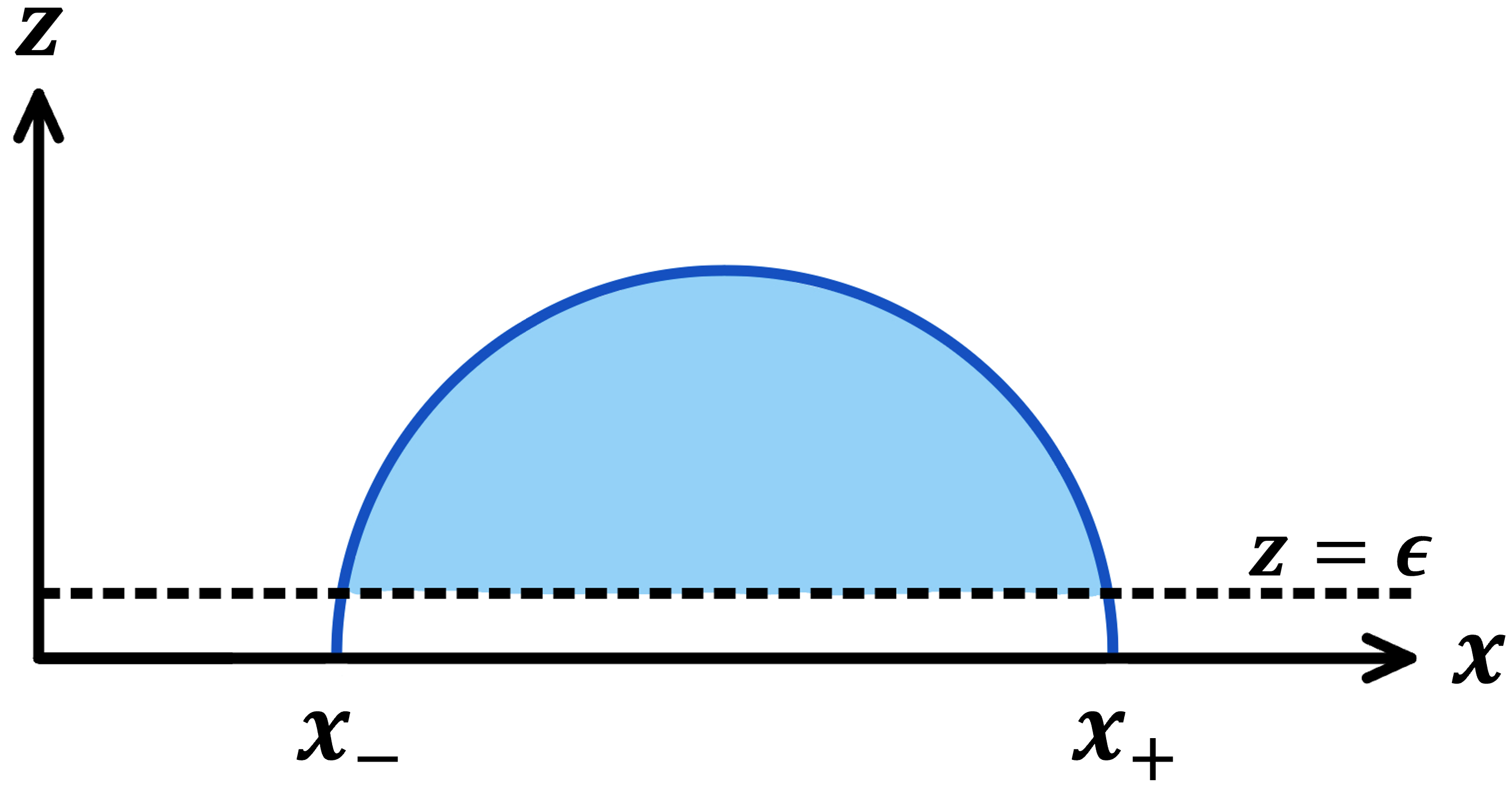}
    \caption{The volume in an asymptotic AdS slice without measurements.}
    \label{fig:vacuum_infinite_line_no_measurement}
\end{figure}

Consider a $2d$ CFT vacuum that is dual to Poincare AdS$_3$ with the metric
\begin{equation}
	ds^2=\frac{1}{z^2}(-dt^2+dx^2+dz^2).
\end{equation} 
Let region $A$ be a boundary subregion with a length $l$. 
Its RT surface is a semi-circle $z_{RT}=\sqrt{l^2/4-x^2}$ that lies on the $t=0$ slice. 
The cut-off surface is located at $z=\epsilon$. 
The volume of the maximal surface enclosed by the RT surface and the cut-off surface, see figure \ref{fig:vacuum_infinite_line_no_measurement}, is given by
\begin{equation}
   V_0=\int_{x^-}^{x^+}dx\int_\epsilon^{z_{RT}(x)}dz\frac{L^2}{z^2}=L^2\int_{x^-}^{x^+}dx\left(\frac{1}{\epsilon}-\frac{1}{\sqrt{l^2/4-x^2}}\right)=L^2\left(\frac{l}{\epsilon}-\pi\right),
\end{equation}
where $x_\pm$ denote the coordinates of the intersects of the RT surface and the cut-off surface.
Using the Brown-Henneaux relation $c=\frac{3L}{2G_N}$~\cite{cmp/1104114999}, we obtain the subsystem complexity
\begin{equation}
	C_0=\frac{2c}{3}\left(\frac{l}{\epsilon}-\pi\right).
\end{equation}

\begin{figure}
    \centering
    \includegraphics[width=\textwidth]{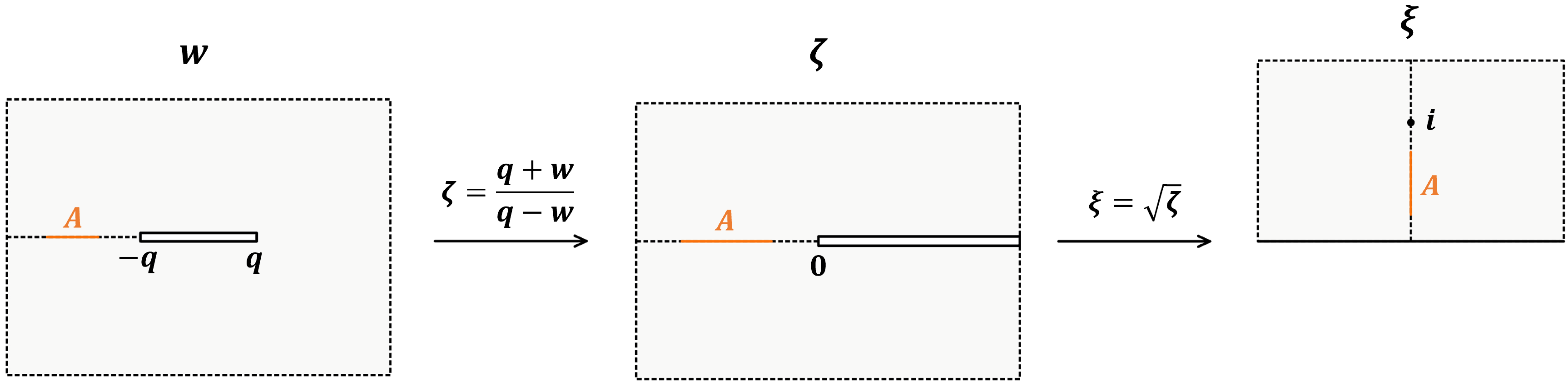}
    \caption{Conformal map to the upper half plane. The projection measurement in $(-q,q) $ is modeled by a slit. $A$ is the subsystem of which the complexity will be calculated.}
    \label{fig:vacuum_infinite_line_map}
\end{figure}

Now we measure another region $B: -q<x<q$ by projecting onto a Cardy state $\ket{\psi_B}$. 
Region $A:-q-l_2<x<-q-l_1$ does not overlap with $B$. 
The Euclidean path integral that computes
\begin{equation}
	\langle 0|\psi_B\rangle\langle \psi_B|0\rangle
\end{equation}
is realized by inserting a slit along the measured region. 
To construct the bulk dual of this boundary manifold, we follow the approach in~\cite{Numasawa:2016emc}, see also~\cite{Antonini:2022sfm,Antonini:2023aza}. 
The idea is to map the boundary manifold with a slit to an upper half plane, whose bulk dual is well studied~\cite{Takayanagi:2011zk,Fujita:2011fp}. 
To this end, we first map the $w=x+iy$ with a slit to a semi-infinite slit through the map
\begin{equation}
	\zeta=\frac{w+q}{q-w}.
\end{equation} 
Then, it is mapped to the upper half plane with coordinates $\xi=\xi_1+i\xi_2$ through the square root,
\begin{equation}
    \xi=\sqrt{\zeta}=\sqrt{\frac{w+q}{q-w}}\equiv f(w) \label{eq:vacuum_infinite_line_map}
\end{equation}
The comformal map is depicted in Fig \ref{fig:vacuum_infinite_line_map}.

Let $\ket{B}$ denote the state defined by the boundary data on the real axis. 
We choose $\ket{B}$ to be a boundary state~\cite{Cardy:2004hm}---it preserves the conformal symmetry of the upper half plane. 
By specifying $\ket{B}$ and the conformal map \eqref{eq:vacuum_infinite_line_map}, we have also specified the state $\ket{\psi_B}$. 

\begin{figure}
    \centering
    \includegraphics[width=0.5\textwidth]{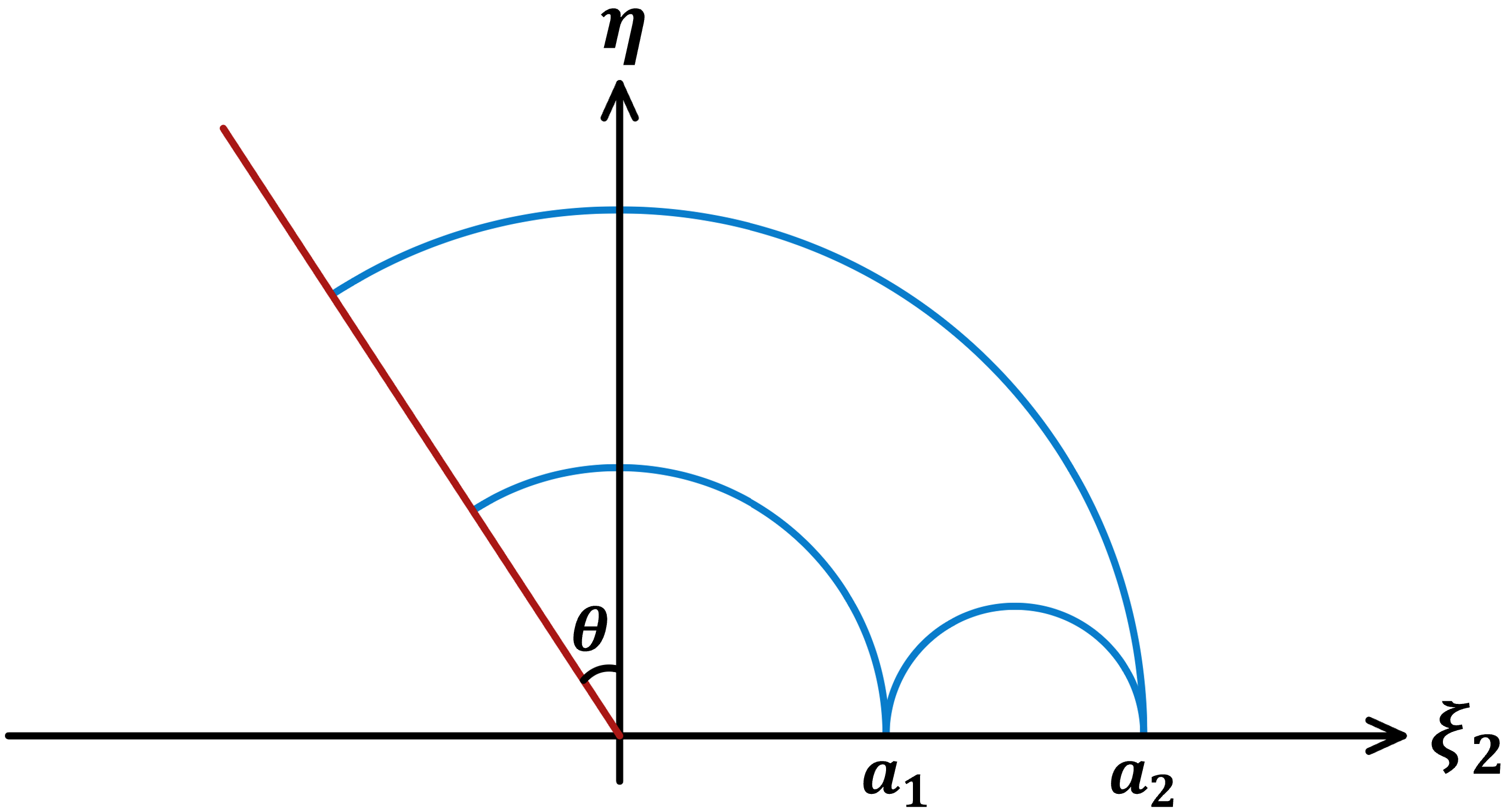}
    \caption{The bulk in $(\xi,\bar{\xi},\eta)$ coordinates. The red line is the end-of-the-world brane. The blue lines are the two candidates for the minimal surface.}
    \label{fig:vacuum_infinite_line_bulk}
\end{figure}

The bulk dual of the upper half plane with a conformal boundary condition has been studied in the AdS/BCFT proposal~\cite{Fujita:2011fp,Takayanagi:2011zk}. 
The bulk dual is a Poincare AdS
\begin{equation}
	ds^2=\frac{L^2}{\eta^2}(d\eta^2+d\xi d\bar{\xi})
\end{equation}
with an end-of-the-world brane that shoots out radially from the $\xi_2=0$ axis~(figure~\ref{fig:vacuum_infinite_line_bulk}),
\begin{equation}
	 \eta_B(\xi)=-\cot\theta\cdot\xi_2,\qquad\sin\theta=LT\equiv\calT. \label{eq:vacuum_infinite_line_brane_location}
\end{equation}
The tension $\calT\in(-1,1)$ is controlled by the boundary entropy of the state $\ket{B}$~\cite{Fujita:2011fp,Takayanagi:2011zk}.

The bulk metric in the original $(w,\bar{w},z)$ coordinates can be obtained by the coordinate transformation~\cite{Roberts:2012aq}
\begin{equation}
	\begin{aligned}
		\xi&=f-\frac{2z^2f'^2\bar{f}''}{4f'\bar{f}'+z^2f''\bar{f}''} \\
		\bar{\xi}&=\bar{f}-\frac{2z^2\bar{f}'^2f''}{4f'\bar{f}'+z^2f''\bar{f}''}\\
		\eta&=\frac{4z(f'\bar{f}')^{\frac{3}{2}}}{4f'\bar{f}'+z^2f''\bar{f}''} \label{bulktrans}
	\end{aligned}
\end{equation}
The metric becomes
\begin{equation}
    ds^2=L^2\left[L(w)dw^2+\bar{L}(\bar{w})d\bar{w}^2+\left(\frac{1}{z^2}+z^2L(w)\bar{L}(\bar{w})\right)dwd\bar{w}+\frac{dz^2}{z^2}\right]
\end{equation}
\begin{equation}
    L(w)=-\frac{1}{2}\{f(w),w\},\qquad\{f(w),w\}=\frac{f'''}{f'}-\frac{3}{2}\left(\frac{f''}{f'}\right)^2.
\end{equation}
However, we will not refer to the metric in the original coordinate, because the new coordinates $(\xi,\bar{\xi},\eta)$ is more convenient.

Under the conformal transformation, the region $A:-q-l_2<x<-q-l_1$ is mapped to a segment, $a_1=\sqrt{\frac{l_1}{2q+l_1}}<\xi_2<\sqrt{\frac{l_2}{2q+l_2}}=a_2$, on the imaginary axis of the $\xi$ plane (see figure~\ref{fig:vacuum_infinite_line_bulk}). 
The cut-off at $z=\epsilon$ 
is mapped to
\begin{equation}
	\eta_\epsilon=\left|\frac{d\xi}{dw}\right|\epsilon=\left|\frac{(1+\xi^2)^2}{4q\xi}\right|\epsilon
\end{equation}
in the $(\xi,\bar{\xi},\eta)$ coordinates. 
On the $\xi_1=0$ slice, the cut-off surface is at $\eta_\epsilon(\xi_2)=\frac{(1-\xi_2^2)^2}{4q|\xi_2|}\epsilon$.

After spelling out the full detail of the conformal map, let us now compute complexity in the $(\xi,\bar{\xi},\eta)$ coordinates. 
There are two possible minimal surfaces. 
The first candidate is the usual semi-circle surface that starts and ends on the boundary. 
Note that it ceases to exist when the brane tension is so negative that it bends toward the semi-circle and has to intersect with it. 
This happens when 
\begin{equation}
	\frac{a_2-a_1}{2}>\frac{a_2+a_1}{2}|\cos\theta|\quad\Leftrightarrow \quad q>\frac{l_1l_2(\cot^4\frac{\theta}{2}-1)}{2(l_2-l_1\cot^4\frac{\theta}{2})}.
\end{equation}
The complexity given by the first surface is
\begin{equation}
\begin{aligned}
    C_1&=\frac{2c}{3}\int_{a_1}^{a_2}d\xi_2\left(\frac{1}{\eta_\epsilon}-\frac{1}{\eta_{RT}}\right) 
    =\frac{2c}{3}\left(\frac{l}{\epsilon}-\pi\right).	
\end{aligned}
\end{equation}
Exploiting the fact that the constant $\xi_2$ slice has fixed Gaussian curvature, the volume can be converted into topological quantities. 
We review the computation in appendix \ref{app:Computation of volume using Gauss-Bonnet}.
Thus, measurements will not change this volume because the topology of this surface stays the same~\cite{Abt:2017pmf}. 

The second candidate lands on the brane. 
The complexity is given by (see appendix \ref{app:Computation of volume using Gauss-Bonnet})
\begin{equation}
	C_2=\frac{2c}{3}\left(\frac{l}{\epsilon}+\frac{\calT}{\sqrt{1-\calT^2}}\cdot\log\sqrt{\frac{2q/l_1+1}{2q/l_2+1}}\right)
\end{equation}
For positive (negative) tension, $C_2$ is a monotonically increasing (decreasing) function of $q$. 
It diverges if we naively take $l_1\rightarrow 0$.\footnote{This divergence is due to the UV nature of the projection---it injects infinite energy into the system.} 
$C_2$ is always greater than $C_1$, as long as the first RT surface exists, i.e., the semicircle and the brane do not intersect. 
In $q\rightarrow 0$ and $q\rightarrow\infty$ limit
\begin{equation}
	C_2(q\rightarrow 0)=\frac{2c}{3}\frac{l}{\epsilon},\qquad 	C_2(q\rightarrow \infty)=\frac{2c}{3}\left(\frac{l}{\epsilon}+\frac{\calT}{\sqrt{1-\calT^2}}\cdot\log\sqrt{\frac{l_2}{l_1}}\right).
\end{equation}

\begin{figure}
    \centering
    \subfigure[]{\includegraphics[width=0.47\textwidth]{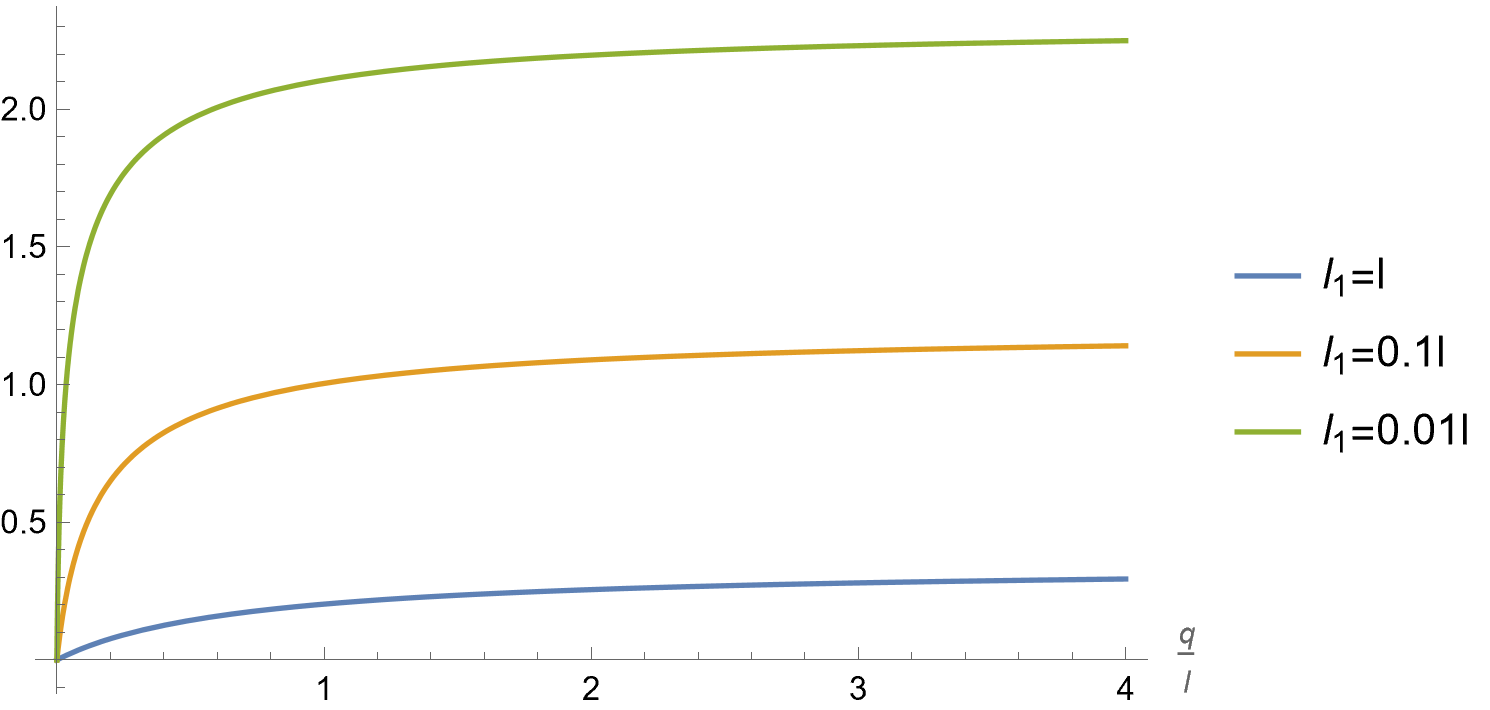}} \qquad 
    \subfigure[]{\includegraphics[width=0.47\textwidth]{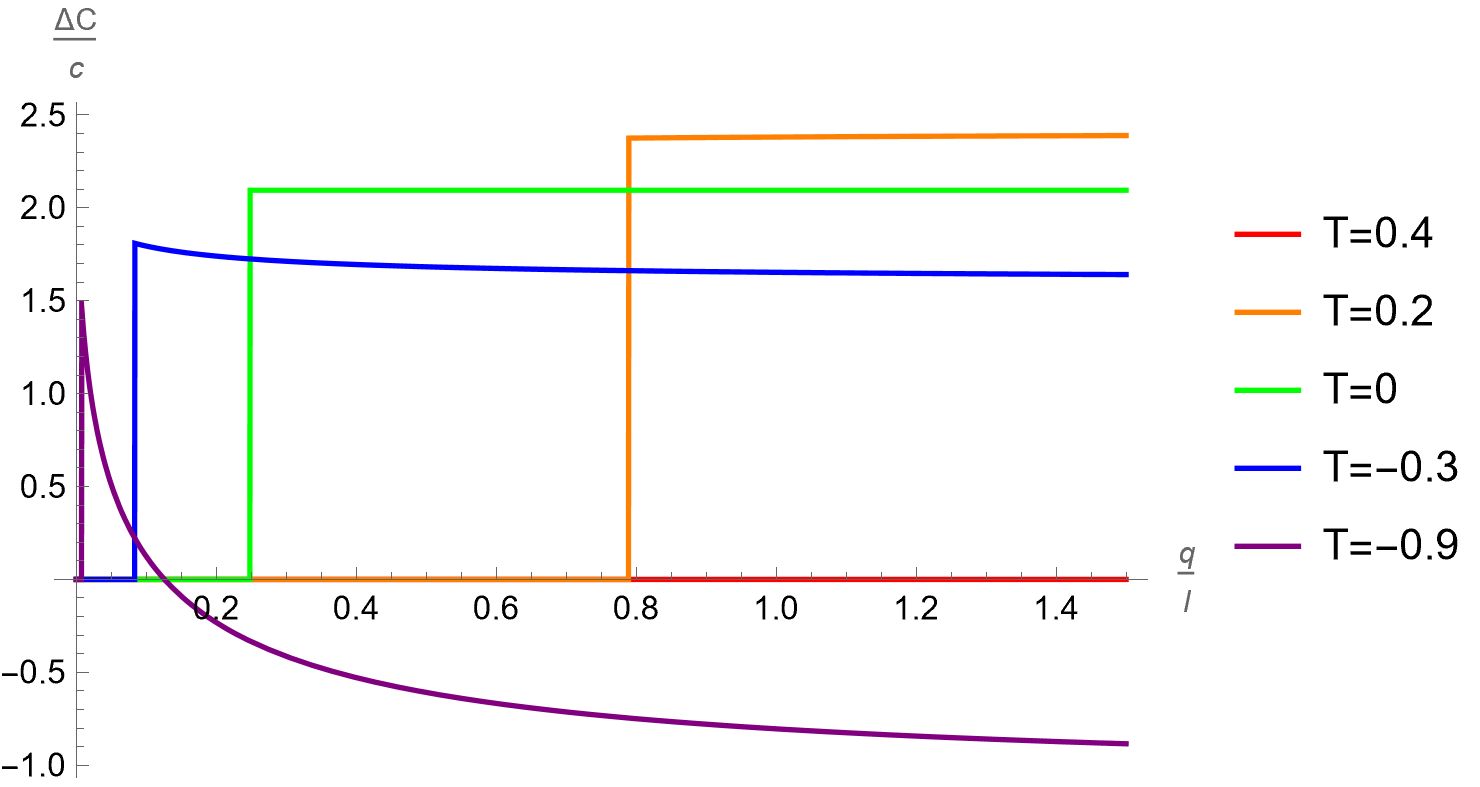}} 
    \caption{(a) The log term for different values of $l_1/l$ and $q$. 
    (b) The complexity changes with respect to the measurement length. $l_1=0.01l$ is used.}
    \label{fig:vacuum_infinite_line_complexity_mathematicaplot}
\end{figure}

Now we analyze which candidate is the minimal surface in detail. 
The entropy (proportional to the geodesic length) given by the two surfaces reads
\begin{equation}
	\begin{aligned}
		&S_1=\frac{c}{6}\log\frac{(a_2-a_1)^2}{\eta_\epsilon(a_1)\eta_\epsilon(a_2)},\qquad 
		S_2=\frac{c}{6}\log\frac{2a_1}{\eta_\epsilon(a_1)}+\frac{c}{6}\log\frac{2a_2}{\eta_\epsilon(a_2)}+\frac{c}{3}\log\sqrt\frac{1+\calT}{1-\calT}, \\
		&S_1-S_2=\frac{c}{3}\left[\log\frac{1}{2}\left(\sqrt{\frac{a_2}{a_1}}-\sqrt{\frac{a_1}{a_2}}\right)-\log\sqrt{\frac{1+\calT}{1-\calT}}\right] . \label{eq:vacuum_infinite_line_entropy}
	\end{aligned}
\end{equation}
$S_1-S_2$ is a monotonically increasing function of $q$. 
In $q\rightarrow 0$ and $q\rightarrow\infty$ limit,
\begin{equation}
\begin{aligned}
	S_1-S_2&\xrightarrow{q\rightarrow 0}
	\frac{c}{3}\left[\log\frac{q}{2}\left(\frac{1}{l_1}-\frac{1}{l_2}\right)-\log\sqrt{\frac{1+\calT}{1-\calT}}\right]
	\rightarrow -\infty,  \\
	S_1-S_2&\xrightarrow{q\rightarrow \infty}
	\frac{c}{3}\left[\log\frac{1}{2}\left(\left(\frac{l_2}{l_1}\right)^{1/4}-\left(\frac{l_1}{l_2}\right)^{1/4}\right)-\log\sqrt{\frac{1+\calT}{1-\calT}}\right].
\end{aligned}
\end{equation}
Even for a very small $q$, the candidate surfaces that land on the brane still exists. 
But this surface has to  ``reach'' very far to be able to land on the brane, which means that its geodesic length is longer than that of the semi-circle surface. 
Therefore, for a very small measurement length, the first surface is dominant, and the complexity does not change. 
At $q\rightarrow\infty$, $S_1-S_2$ can either be positive or negative. 
This value monotonically decreases with $l_2/l_1$. 
For small $l_2/l_1$, it is negative and $S_1$ dominates for all $q$. 
But for $l_2/l_1$ that satisfies
\begin{equation}
	\left(\frac{l_2}{l_1}\right)^{1/4}-\left(\frac{l_1}{l_2}\right)^{1/4}>2\sqrt{\frac{1+\calT}{1-\calT}},
\end{equation}
$S_1-S_2$ is positive and $S_2$ becomes dominant for a large enough $q$.

\begin{figure}
	\centering
	\includegraphics[width=\textwidth]{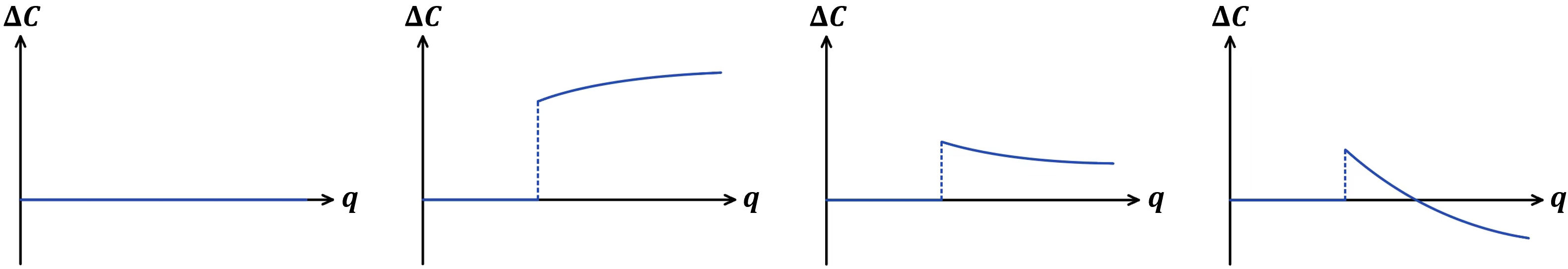}
	\caption{Possible complexity-$q$ curves with other parameters held fixed. (first panel) Phase 1: complexity does not change under measurements. 
    (second panel) Phase 2: complexity jumps to $C_2$, then increases and saturates. 
    (third panel) Phase 3: complexity jumps, then decreases but remains higher than the initial value. 
    (fourth panel) Phase 4: complexity jumps and then decreases to be smaller than the initial value.}
    \label{fig:vacuum_infinite_line_complexity_q_four_phases}
\end{figure}

Now we can give a complete description of the effect of measurement when we fix $l_1$, $l_2$, $T$ and change $q$. 
The possible $\Delta C$-$q$ curves are plotted in figure~\ref{fig:vacuum_infinite_line_complexity_q_four_phases}. 
For sufficiently small $q$, $S_1$ is dominant because the second candidate surface requires a greater length to reach the brane. 
In this case, the complexity is unchanged. 
If $\left(\frac{l_2}{l_1}\right)^{1/4}-\left(\frac{l_1}{l_2}\right)^{1/4}>2\sqrt{\frac{1+\calT}{1-\calT}}$, $S_2$ becomes dominant when $q$ surpasses a critical value and complexity jumps to a larger value. 
For $T>0$, when $q$ further increases, the complexity increases and approaches a fixed value at $q\rightarrow\infty$. 
If $\left(\frac{l_2}{l_1}\right)^{1/4}-\left(\frac{l_1}{l_2}\right)^{1/4}<2\sqrt{\frac{1+\calT}{1-\calT}}$, then $S_1$ is always dominant and measurement does not affect complexity.

\begin{figure}
	\centering
	\includegraphics[width=10cm]{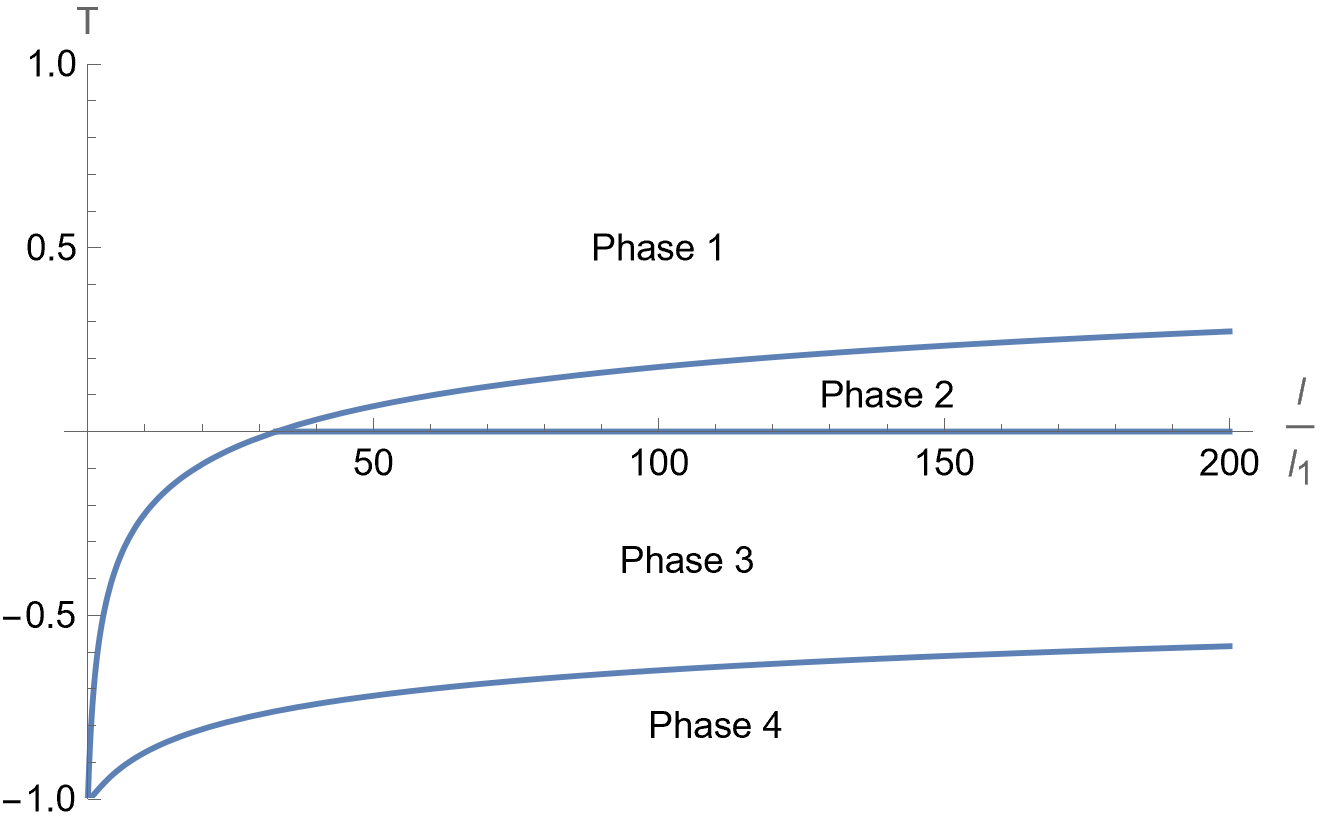}
	\caption{The phase diagram of the four phases in figure~\ref{fig:vacuum_infinite_line_complexity_q_four_phases}.}
	\label{fig:vacuum_infinite_line_phase_diagram}
\end{figure}

For $T<0$, 
if $\left(\frac{l_2}{l_1}\right)^{1/4}-\left(\frac{l_1}{l_2}\right)^{1/4}>2\sqrt{\frac{1+\calT}{1-\calT}}$, there is a phase transition as we increase $q$. This gives a complexity jump to a higher value. As we further increase $q$, the increment in complexity decreases with $q$, until it approaches a fixed value. 
If $\left(\frac{l_2}{l_1}\right)^{1/4}-\left(\frac{l_1}{l_2}\right)^{1/4}<2\sqrt{\frac{1+\calT}{1-\calT}}$, then $S_1$ is dominant for all $q$ and complexity stays the same. 
When $\frac{l_2}{l_1}>\cot^4\frac{\theta}{2}$, the first candidate surface ceases to exist for a large enough $q$. 
But we shouldn't worry about it because this happens when $S_2$ is already dominant, guaranteed by the requirement that entropy should be continuous. 
The $q\rightarrow\infty$ value of $C_2$ can be smaller than $C_1$ when $\frac{\calT}{\sqrt{1-\calT^2}}\log\sqrt{\frac{l_2}{l_1}}<-\pi$. This is the only case where measurement makes complexity smaller. 
The phase diagram is summarized in figure~\ref{fig:vacuum_infinite_line_phase_diagram}.


\subsection{Finite size system}

\begin{figure}
    \centering
    \includegraphics[width=\textwidth]{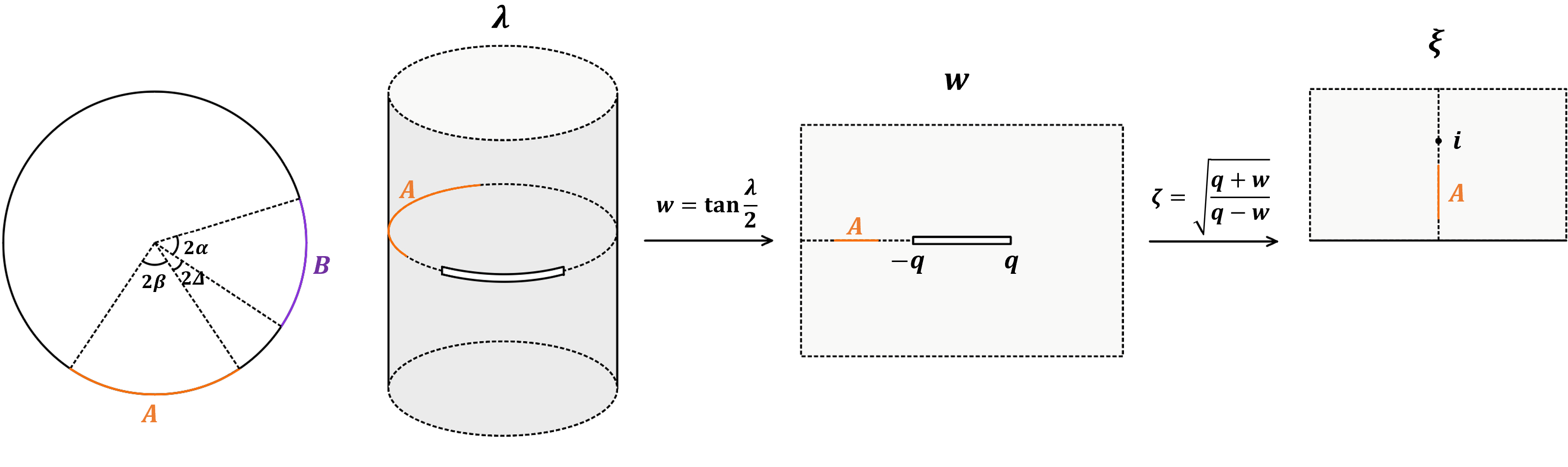}
    \caption{Path integral manifold and the conformal map to the upper half plane. $B$ is the measurement region and $A$ is the region whose complexity is calculated. 
    In the figure, we show the case for $0<\alpha+\beta_2<\pi$.}
    \label{fig:vacuum_finite_map}
\end{figure}

Consider a CFT on a circle with circumference $2\pi$. 
Let $\lambda=\lambda_1+i\lambda_2$ denote the boundary coordinates~\footnote{We set $\lambda_1\in[-\pi,\pi)$.} and $\mu$ denote the bulk direction. 
We project on the interval $B:-\alpha <\lambda_1<\alpha$. The Euclidean path integral that computes $\langle 0|\psi_B\rangle\langle \psi_B|0\rangle$ is given by a cylinder with a slit inserted at $\lambda_2=0$ along $B$. 
We will study the complexity of the region $A:-\alpha-\beta_2<\lambda_1<-\alpha-\beta_1 $. 
This geometry can be mapped to the upper half plane in the following way (figure \ref{fig:vacuum_finite_map}). 
First, we map the cylinder with a slit to a complex plane with a slit via~\cite{Stephan:2014nda,Antonini:2022sfm}
\begin{equation}
	w=\tan\frac{\lambda}{2}.
\end{equation}
The end points of $B$ are mapped to $\pm q=\pm\tan\frac{\alpha}{2}$. 
Then, we map it to the upper half plane with
\begin{equation}
	\xi=\sqrt{\frac{q+w}{q-w}}=\sqrt{\frac{\tan\frac{\alpha}{2}+\tan\frac{\lambda}{2}}{\tan\frac{\alpha}{2}-\tan\frac{\lambda}{2}}}.
\end{equation}
The region $A$ is mapped to
$a_1\equiv\sqrt{\frac{\sin\frac{\beta_1}{2}}{\sin\left(\alpha+\frac{\beta_1}{2}\right)}}<\xi_2<\sqrt{\frac{\sin\frac{\beta_2}{2}}{\sin\left(\alpha+\frac{\beta_2}{2}\right)}}\equiv a_2$. 
This expression still holds for $\alpha+\beta_i>\pi$.
Without the measurement, the complexity is~\cite{Abt:2017pmf} 
\begin{equation}
	C_1=\frac{2c}{3}\left(\frac{\beta_2-\beta_1}{\epsilon}-\pi\right).
\end{equation}
After the measurement, the complexity given by the semi-circle surface is still $C_1$, as topological quantities of the surface remain the same. 
The complexity given by the second surface is
\begin{equation}
\begin{aligned}
	C_2&=\frac{2c}{3}\left(\frac{\beta_2-\beta_1}{\epsilon}+\tan\theta\cdot\log\frac{a_2}{a_1}\right)
	=\frac{2c}{3}\left(\frac{\beta_2-\beta_1}{\epsilon}+\frac{\calT}{\sqrt{1-\calT^2}}\frac{1}{2}\log\frac{\sin(\Delta+\beta)\sin(\alpha+\Delta)}{\sin\Delta\sin(\alpha+\Delta+\beta)}\right),
\end{aligned}
\end{equation} 
where we defined $2\Delta=\beta_1$ that acts as the regulator and $2\beta=\beta_2-\beta_1$.
The log term goes to zero when $\alpha\rightarrow 0$, as expected.
\begin{equation}
	C_2(\alpha\rightarrow 0)=\frac{2c}{3}\frac{\beta_2-\beta_1}{\epsilon}.
\end{equation}

~\\
\noindent\textbf{Fix system size, symmetric measurement.}

We choose the measurement region to be symmetric with respect to $\lambda_1 = 0$
and the region $A$ is fixed to be $\pi - \beta <\lambda_1 < \pi + \beta$.
We start by measuring the furthest point from $A$, i.e., $\lambda_1=0$, and increase the measurement region in a symmetric way. 
In other words, we fix $\beta$ and the relation $\alpha+\beta+2\Delta=\pi$, while change $\alpha$ from $0$ to $\pi-\beta$.

The equation for entanglement in the two phases \eqref{eq:vacuum_infinite_line_entropy} still holds. 
The parameter that determines the phase is
\begin{equation}
	\frac{a_2}{a_1}=\frac{\sin(\beta+\Delta)}{\sin\Delta}=\frac{\cos\frac{\beta-\alpha}{2}}{\cos\frac{\beta+\alpha}{2}}=\sin\beta\cot\Delta+\cos\beta\in(1,+\infty)
\end{equation}
When $\alpha$ varies from $0$ to $\pi-\beta$ (or $\Delta$ varies from $\frac{\pi}{2}-\frac{\beta}{2}$ to $0$), $a_2/a_1$ increases monotonically from $1$ to $+\infty$. The transition from $C_1$ to $C_2$ happens at 
\begin{equation}
	\sqrt{\frac{a_2}{a_1}}-\sqrt{\frac{a_1}{a_2}}=2\sqrt{\frac{1+\calT}{1-\calT}}.
\end{equation}

\begin{figure}
    \centering
    \subfigure[]{\includegraphics[width=0.47\textwidth]{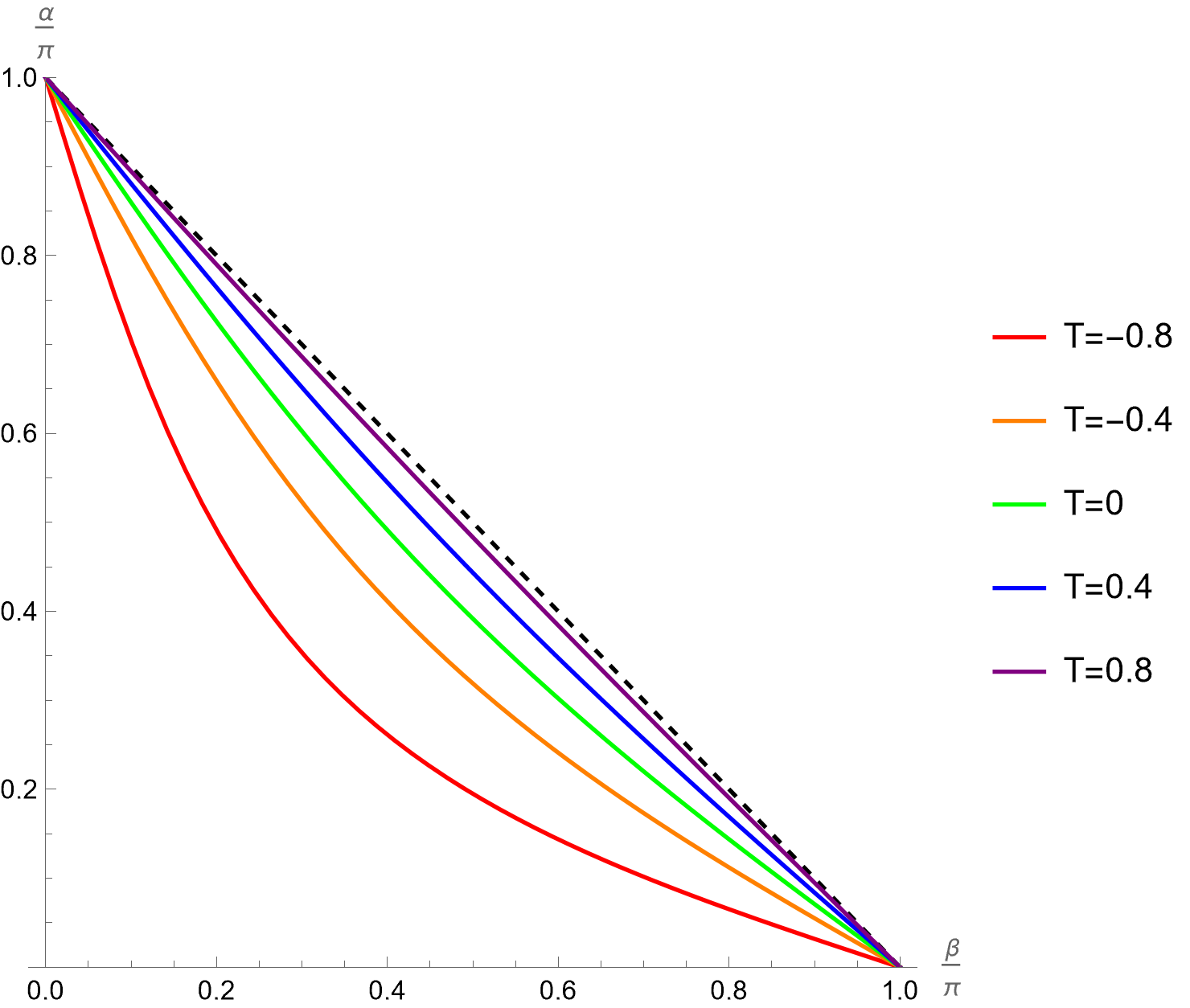}} \qquad 
    \subfigure[]{\includegraphics[width=0.47\textwidth]{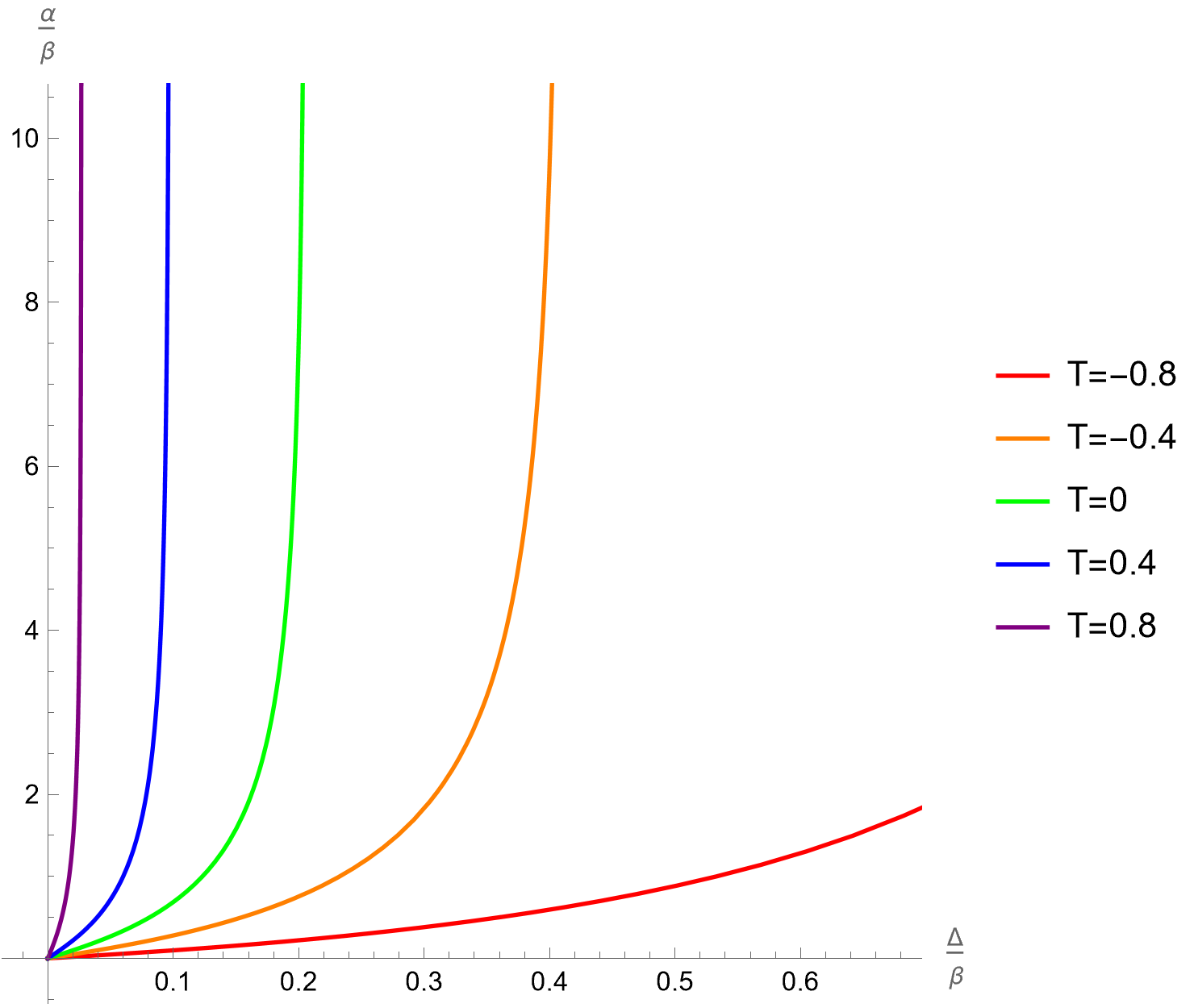}} 
    \caption{Phase transition in the asymmetric measurement case for different $T$. 
    Above (below) the lines are the $C_2$ ($C_1$) phases. (a) $\frac{\alpha}{\pi}$ against $\frac{\beta}{\pi}$. The dashed line is when $\Delta=0$. (b) $\frac{\alpha}{\beta}$ against $\frac{\Delta}{\beta}$.}
    \label{fig:vacuum_inite_line_phases}
\end{figure}
In this case, the complexity $C_2$ is simplified to be
\begin{equation}
	C_2(\alpha+\beta+2\Delta=\pi)
	=\frac{2c}{3}\left(\frac{2\beta}{\epsilon}+\frac{\calT}{\sqrt{1-\calT^2}}\log\frac{\sin(\Delta+\beta)}{\sin\Delta}\right).
\end{equation}
In order to measure the entire complement of region $A$, we should send $\Delta$ to zero.
But if we naively take set $\Delta\rightarrow0$, the expression diverges.
So, we take $\Delta=\epsilon$, which is the smallest parameter in our system. 
In this limit, the second term becomes a logarithmic divergence,
\begin{equation}
	C_2(\alpha+\beta=\pi-2\epsilon)=\frac{2c}{3}\left[\frac{2\beta}{\epsilon}+\frac{\calT}{\sqrt{1-\calT^2}}\log\frac{\sin\alpha}{\epsilon}\right].
\end{equation}

~\\
\noindent\textbf{Fix system size, fix $\beta_{1,2}$, vary $\alpha$ (asymmetric measurement).}

Another measurement scheme is to start by measuring the region that is close to $A$.
Then we fix $\Delta$ and vary $\alpha$ from $0$ to $\pi-\beta-2\Delta$. 
In this case, if $\alpha$ is able to reach a value above the lines in figure \ref{fig:vacuum_inite_line_phases} then there is a phase transition.

\begin{figure}
    \centering
    \includegraphics[width=8cm]{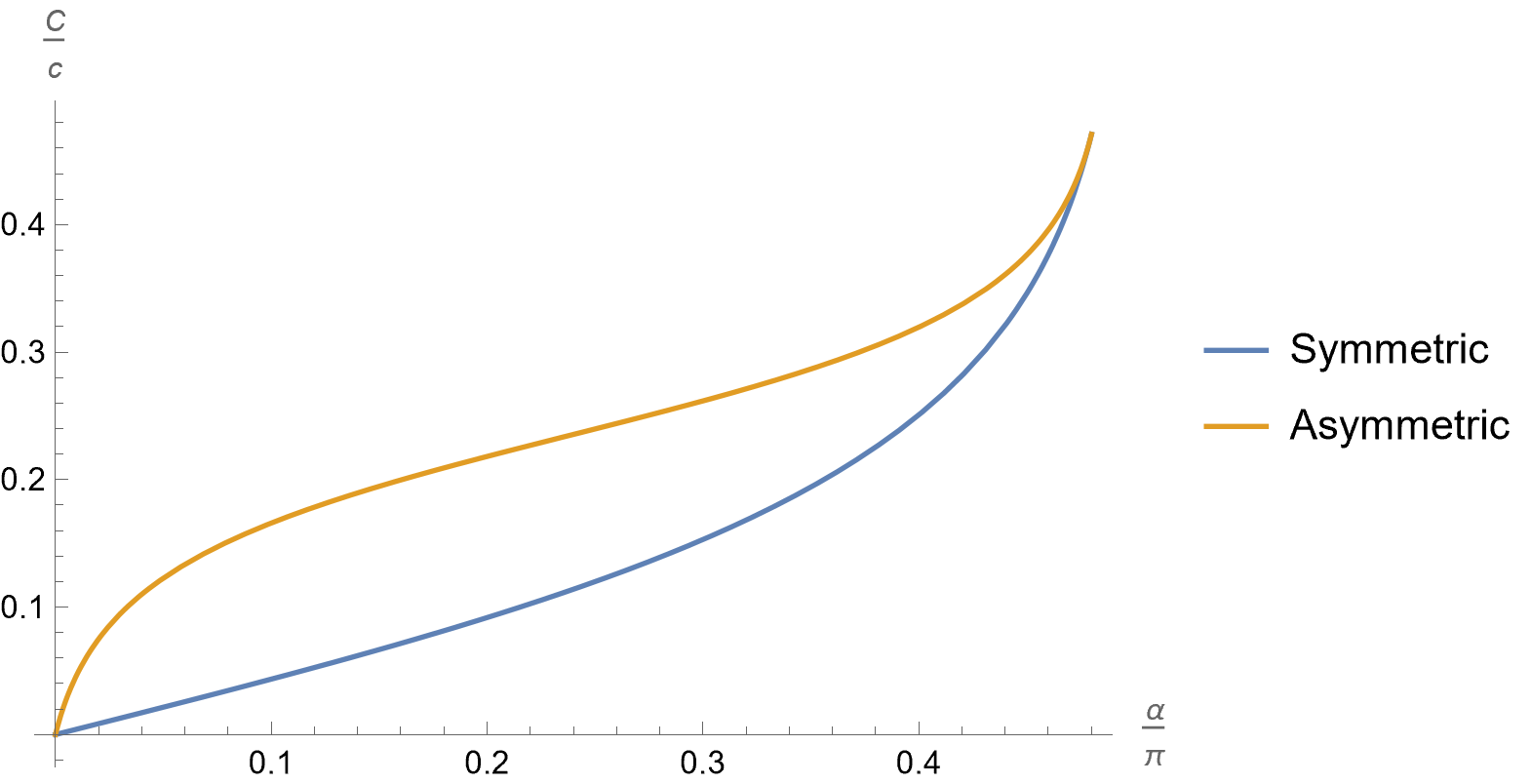}
    \caption{Log term for the symmetric measurements and the asymmetric measurements. $\beta=\frac{\pi}{2}$ and $\Delta=0.01\pi$.}
    \label{fig:vacuum_finite_log_term_beta=0.5pi_Delta=0.01pi}
\end{figure}

\begin{figure}
    \centering
    \subfigure[]{\includegraphics[width=0.32\textwidth]{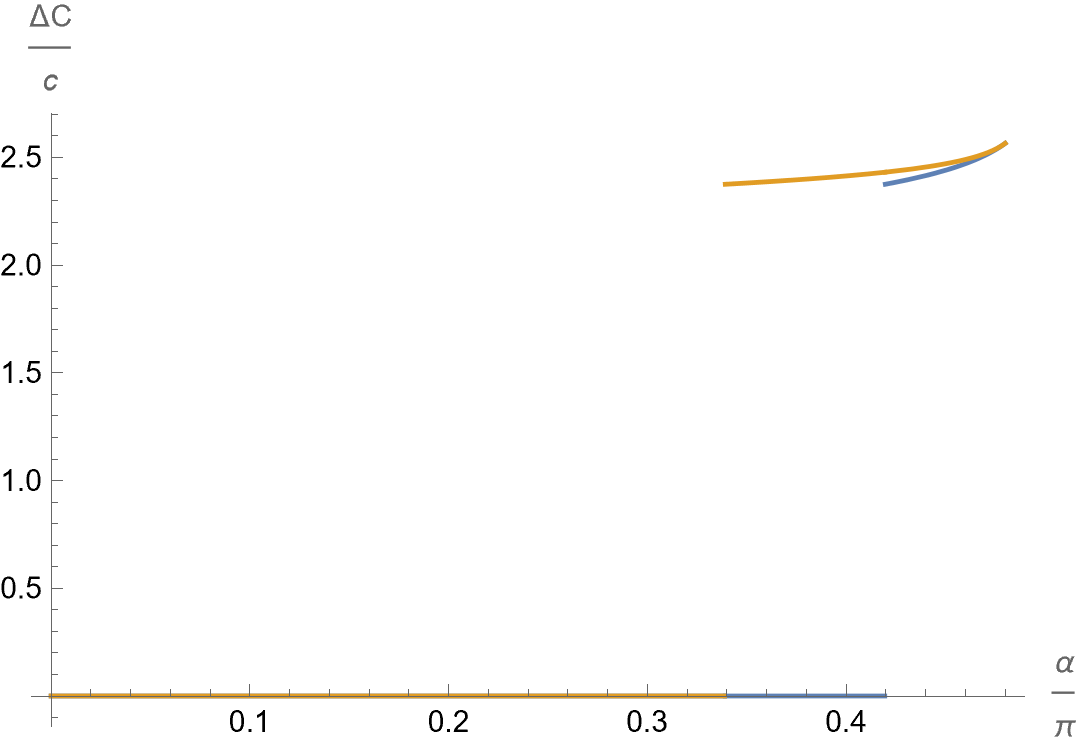}} 
    \subfigure[]{\includegraphics[width=0.32\textwidth]{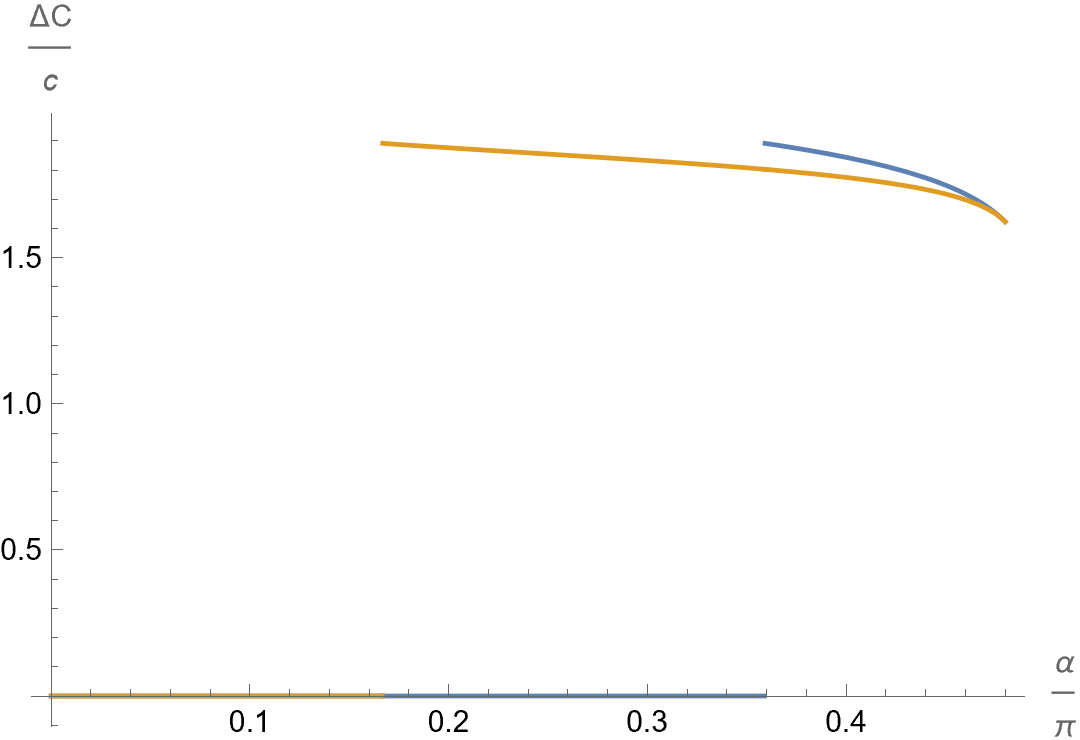}}
    \subfigure[]{\includegraphics[width=0.32\textwidth]{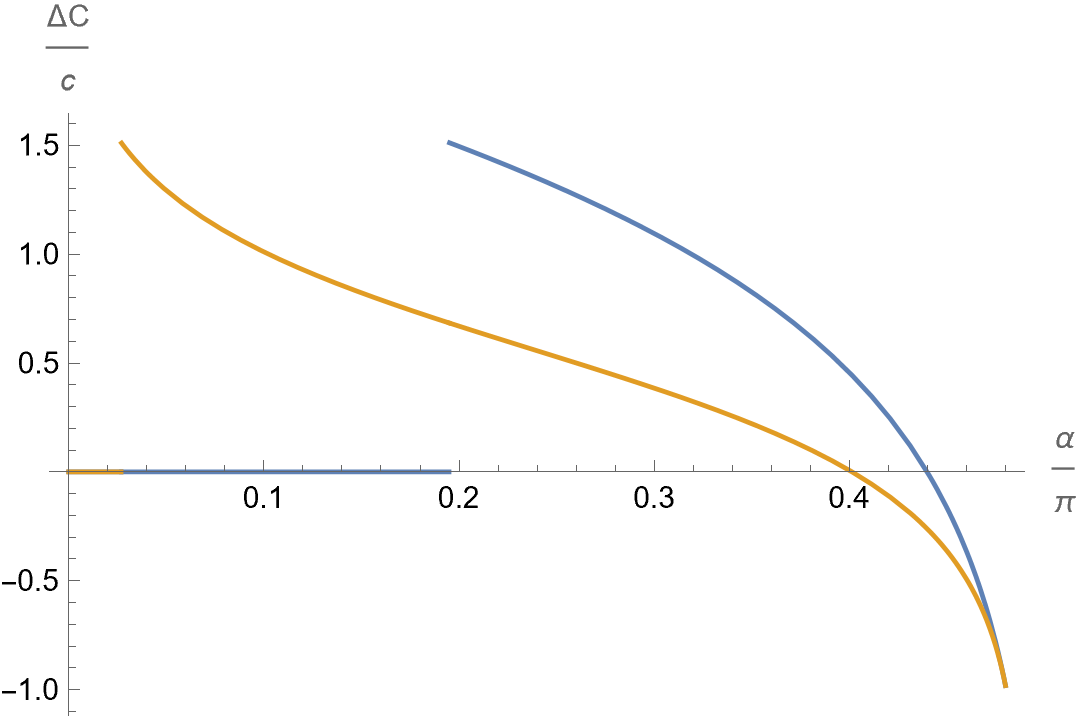}}
    \caption{The complexity change with respect to measurement length. The blue line is for the symmetric measurements; the orange line is for the asymmetric measurements we just described. $\beta=\frac{\pi}{2}$, $\Delta=0.01\pi$. 
 (a) $T=0.2$.  
 (b) $T=-0.2$. 
 (c) $T=-0.8$.  }
    \label{fig:vacuum_finite_complexity_comparing _sym&asymm}
\end{figure}

The log term in $C_2$ is plotted in figure \ref{fig:vacuum_finite_log_term_beta=0.5pi_Delta=0.01pi}. The complexity phase transition is plotted in \ref{fig:vacuum_finite_complexity_comparing _sym&asymm}, also with a comparison with the symmetric measurement scheme. Similar to the measurement in the infinite system, when the tension is positive (negative), the complexity change increases (decreases) as the measurement length increases.


~\\
\noindent\textbf{Fix size of $A$ and its distance from $B$, change size of $B$.}

\begin{figure}
    \centering
    \subfigure[]{\includegraphics[width=0.37\textwidth]{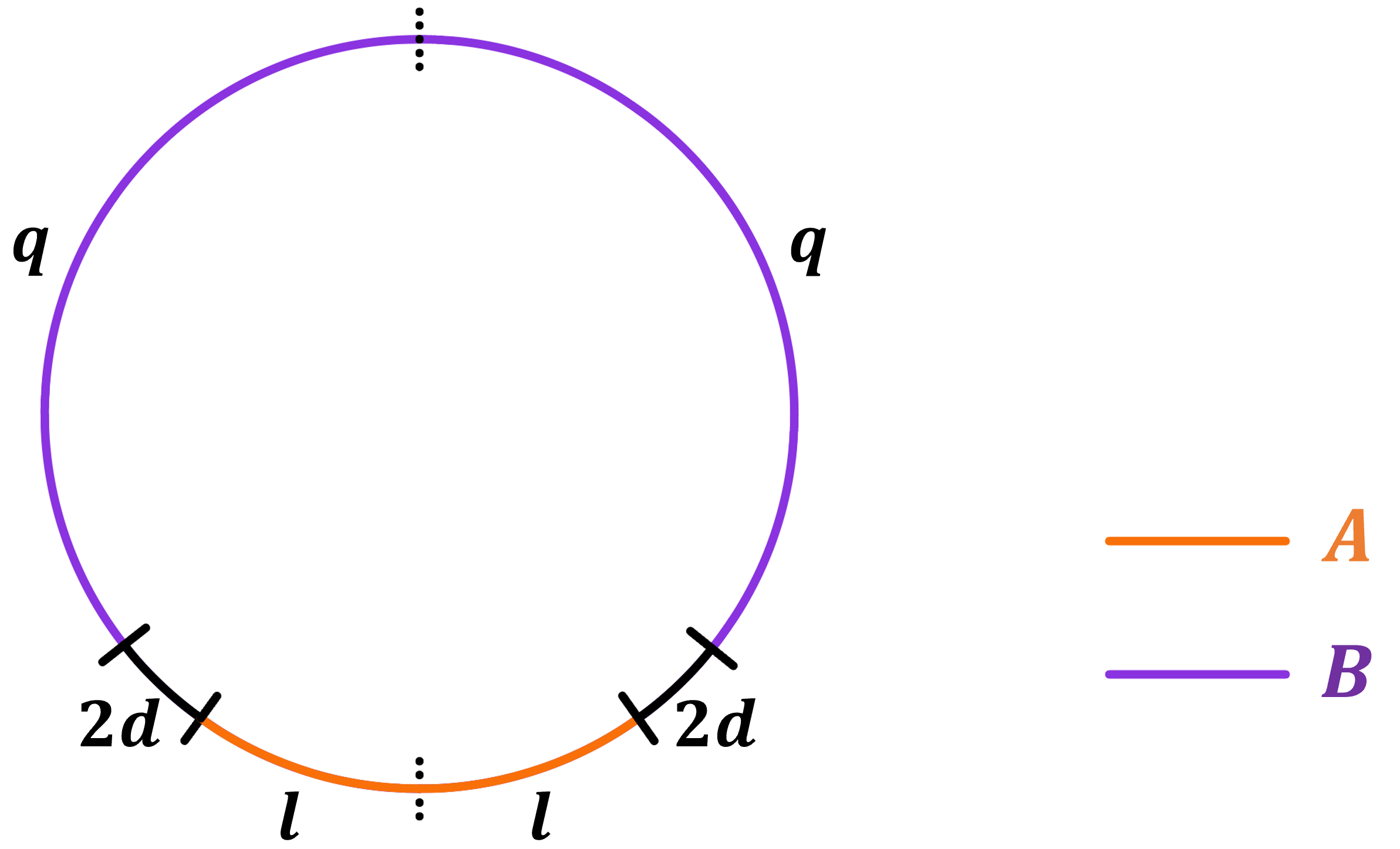}} 
    \qquad
    \subfigure[]{\includegraphics[width=0.37\textwidth]{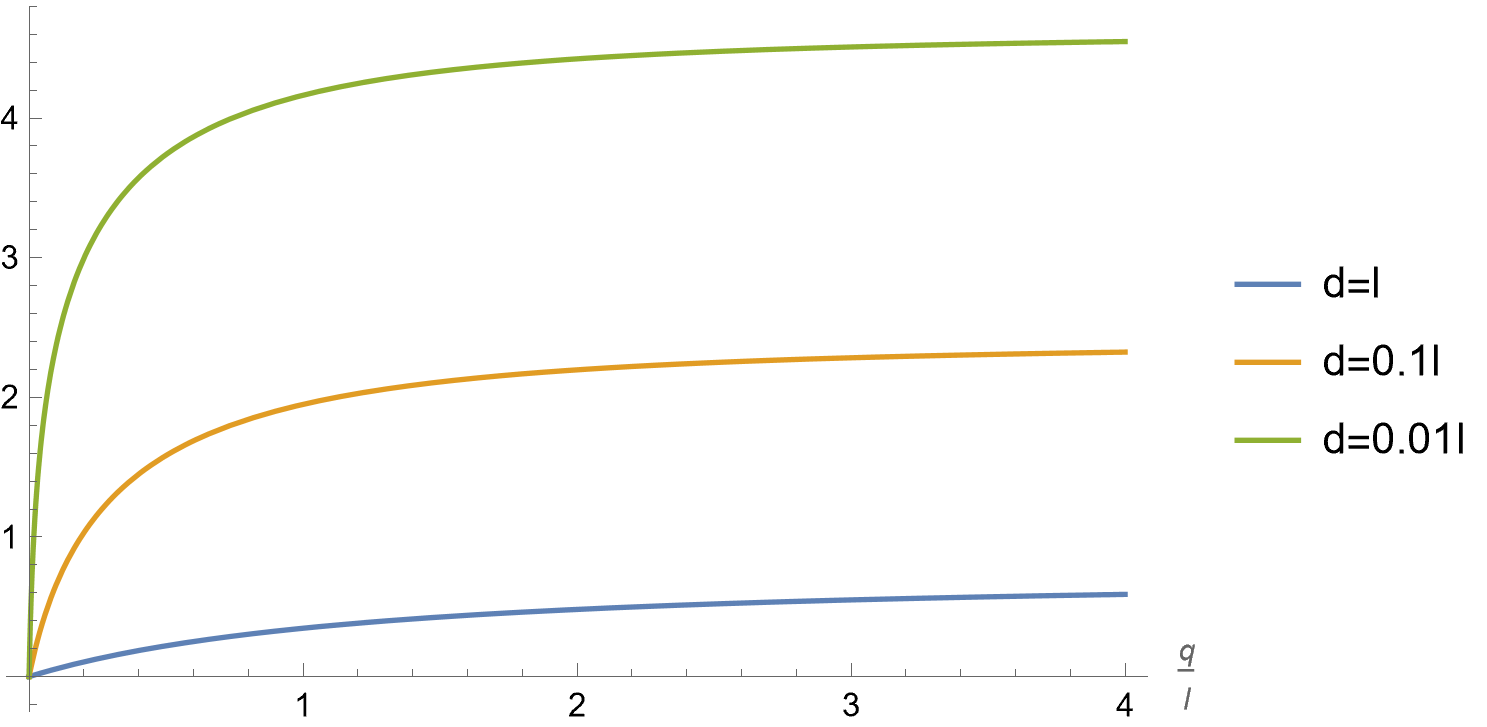}}
    \caption{(a) The setup. $d$ and $l$ are fixed, and the measurement region $q$ is increased.  
    (b) The log term with respect to $q$.}
    \label{fig:vacuum_finite_change_size}
\end{figure}

Another thing we can do is to fix the length of $A$ and its distance to the projected region, and tune the length of the projected region (in other words, we change the size of the entire system). This is like changing $N_A/N_B$ in many-body systems. 
To study this, we scale our boundary to a cylinder of circumference $R$. 
In the bulk, this amounts to re-scaling the cut-off parameter:
\begin{equation}
	\epsilon\rightarrow\frac{2\pi}{R}\epsilon,\qquad\beta\rightarrow\frac{2\pi l}{R},\qquad\Delta\rightarrow\frac{2\pi d}{R},\qquad\alpha\rightarrow\frac{2\pi q}{R}.
\end{equation}
The complexities are
\begin{equation}
	C_1=\frac{2c}{3}\left(\frac{l}{\epsilon}-\pi\right),\qquad
	C_2(q=R/2-l-2d)
	=\frac{2c}{3}\left(\frac{l}{\epsilon}+\frac{\calT}{\sqrt{1-\calT^2}}\log\frac{\sin\frac{\pi(l+d)}{q+l+2d}}{\sin\frac{\pi d}{q+l+2d}}\right).
\end{equation}
When we increase $q$, the log term increases. 
See figure \ref{fig:vacuum_finite_change_size} for an illustration. 
When $q\rightarrow\infty$, complexity saturates to 
\begin{equation}
	C_2(q=R/2-l-2d\rightarrow\infty)=\frac{2c}{3}\left[\frac{l}{\epsilon}+\frac{\calT}{\sqrt{1-\calT^2}}\log\left(\frac{l}{d}+1\right)\right].
\end{equation}
The phase diagram is given by figure \ref{fig:vacuum_infinite_line_phase_diagram}, but with the substitution $\frac{\alpha}{\beta}\rightarrow\frac{q}{l}$ and $\frac{\Delta}{\beta}\rightarrow\frac{d}{l}$. If we further take $d=\epsilon$, then
\begin{equation}
	C_2=\frac{2c}{3}\left(\frac{l}{\epsilon}+\frac{\calT}{\sqrt{1-\calT^2}}\log\frac{l}{\epsilon}\right).
\end{equation}

\section{Measurements on thermofield double state}\label{sec:Thermal field double state, measure left system}
\subsection{Infinite size system}

We would like to investigate the role of entanglement in the measurement induced complexity change. 
One way to tune entanglement is by introducing the thermofield double state between a left system and a right system. 
The entanglement between the left and the right systems varies with temperature. 

\begin{figure}
	\centering
	\includegraphics[width=\textwidth]{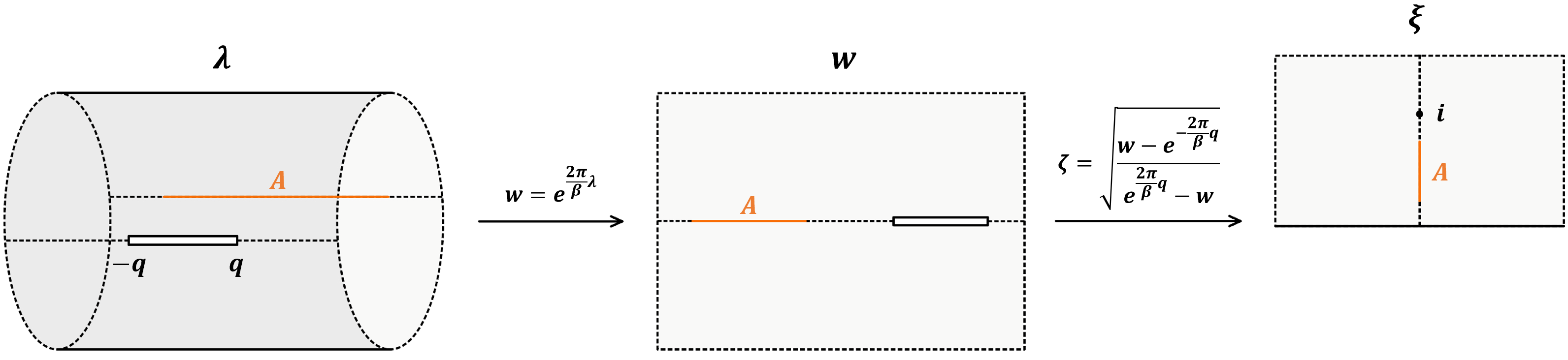}
    \caption{Conformal mapping to the upper half plane. The projection measurement in $(-q,q)$ is
    modeled by a slit. A is the subsystem of which the complexity will be calculated. In the first step, we map it to the complex plane with a slit by an exponential mapping.}
    \label{fig:TFD_infinite_map}
\end{figure}

We take the left and right system to have an infinite length. 
We measure the left system in region $B:-q<\lambda_1<q$ and compute the complexity of region $A:-l<\lambda_1<l$ in the right system. 
Some other measurement schemes are considered in~\cite{Antonini:2023aza}. 
The path integral manifold in this case is a horizontal cylinder with a slit. 
We use coordinates $\lambda=\lambda_1+i\lambda_2$ on the cylinder, with $\lambda_2$ having a periodicity of $\beta$. 
Let $\mu$ denote the bulk direction.
The left system corresponds to the $\lambda_2=0$ line, while the right system corresponds to the $\lambda_2=\frac{\beta}{2}$ line. 
We can map this manifold to the familiar infinite plane with a slit by an exponential function. 
Then we map it to the upper half plane in the usual way. 
The conformal map is given by
\begin{equation}
	\xi=\sqrt{\frac{e^{\frac{2\pi}{\beta}\lambda}-e^{-\frac{2\pi}{\beta}q}}{e^{\frac{2\pi}{\beta}q}-e^{\frac{2\pi}{\beta}\lambda}}}\equiv f(\lambda).
\end{equation}
Region $A$ is mapped to $a_1\equiv\sqrt{\frac{e^{-\frac{2\pi}{\beta}l}+e^{-\frac{2\pi}{\beta}q}}{e^{\frac{2\pi}{\beta}q}+e^{-\frac{2\pi}{\beta}l}}}<\xi_2<\sqrt{\frac{e^{\frac{2\pi}{\beta}l}+e^{-\frac{2\pi}{\beta}q}}{e^{\frac{2\pi}{\beta}q}+e^{\frac{2\pi}{\beta}l}}}\equiv a_2$ as illustrated in figure~\ref{fig:TFD_infinite_map}.

As usual, the two candidates for the minimal surface give two complexities
\begin{equation}
\begin{aligned}
	C_1&=\frac{2c}{3}\left(\frac{2l}{\epsilon}-\pi\right), \\
	C_2&=\frac{2c}{3}\left(\frac{2l}{\epsilon}+\tan\theta\cdot\log\frac{a_2}{a_1}\right)
	=\frac{2c}{3}\left(\frac{2l}{\epsilon}+\frac{\calT}{\sqrt{1-\calT^2}}\frac{1}{2}\log\frac{1+\cosh\left[\frac{2\pi}{\beta}(q+l)\right]}{1+\cosh\left[\frac{2\pi}{\beta}(q-l)\right]}\right).
\end{aligned}
\end{equation} 
To clearly see the effect of the temperature on the complexity change, we consider the following two limits,
\begin{equation}
    C_2(q\rightarrow\infty)=\frac{2c}{3}\left(\frac{2l}{\epsilon}+\frac{\calT}{\sqrt{1-\calT^2}}\frac{2\pi}{\beta}l\right),
\end{equation}
and
\begin{equation}
	C_2(q=l\gg\beta)=\frac{2c}{3}\left[\frac{2l}{\epsilon}+\frac{\calT}{\sqrt{1-\calT^2}}\left(\frac{2\pi}{\beta}l-\frac{1}{2}\log2\right)\right].
\end{equation}
The extra complexity is proportional to the temperature. 
As we increase temperature, the entanglement between the left and right increases, so does the extra complexity. 
Actually, since our system is infinitely long, only the ratio $\beta:l:q$ matters. Increasing the temperature is equivalent to increasing $l$ and $q$ with $\beta$ fixed.

The entanglement entropy from these two candidate RT surfaces is given by 
\begin{equation}
	\begin{aligned}
		S_1-S_2&=\frac{c}{3}\left[\log\frac{1}{2}\left(\sqrt{\frac{a_2}{a_1}}-\sqrt{\frac{a_1}{a_2}}\right)-\log\sqrt{\frac{1+\calT}{1-\calT}}\right].
	\end{aligned}
\end{equation}
This quantity monotonically decreases with temperature. 
For high enough temperature, $S_1-S_2$ can always achieve a positive value, so the $C_2$ is dominant. 
Hence, there is a transition of complexity with the temperature. 

\subsection{Finite size system, measure the entire left}
 
Consider a thermofield double state with temperature $\beta$ of a finite system with spacial a periodicity $W$. 
The bulk geometry is either thermal AdS$_3$ or BTZ black hole, separated by a Hawking-Page transition~\cite{Hawking:1982dh} at $\frac{W}{\beta}=1$.

Below the critical temperature, the bulk geometry is thermal AdS$_3$. 
The complexity is $C_{AdS}=\frac{2c}{3}\left(\frac{W}{\epsilon}-2\pi\right)$. 
Above the critical temperature, the bulk is the BTZ black hole. 
The complexity of the entire right system is given by the volume between the horizon and the cut-off surface, whose Euler characteristic is zero. 
The complexity is $C_{BTZ}=\frac{2c}{3}\frac{W}{\epsilon}$.

\begin{figure}
    \centering
    \includegraphics[width=\textwidth]{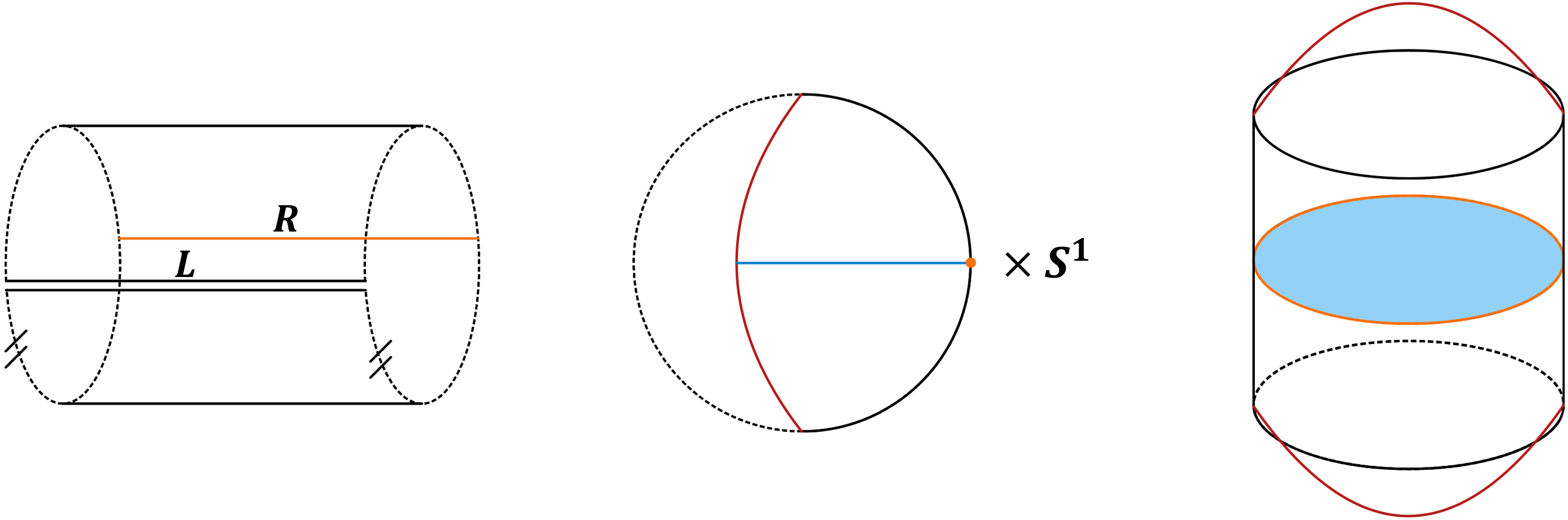}
    \caption{(a) Path integral manifold. The left system is projected onto a Cardy state. 
    (b) BTZ phase. 
    The left side is cut off by the end-of-the-world brane. 
    The $S^1$ denotes the contractible time direction.
    (c) AdS phase.
    Two end-of-the-world branes end on the left side.}
    \label{fig:TFD_finite_bulk}
\end{figure}

We project the entire left system onto a Cardy state $\ket{B}$. 
Now the boundary manifold is an annulus with a modular parameter $\frac{W}{\beta}$. 
This is different from the boundary state $\ket{\psi_B}$ defined in the last section, where the manifold has a single boundary. 
Here we have two boundaries with the same boundary condition.
Let $w=x+i\tau$ denote the boundary coordinate. 
This is a standard AdS/BCFT setup studied in \cite{Takayanagi:2011zk,Fujita:2011fp}. 
There is a connected (BTZ) phase, where $\tau$ is contractible, and a brane smoothly connects the two boundaries. 
There is also a disconnected (AdS) phase where $x$ is contractible. 
Two branes with the same tension are attached to the two boundaries.
These are illustrated in figure \ref{fig:TFD_finite_bulk}.

In the BTZ phase, the metric is given by
\begin{equation}
	ds^2=L^2\big(\cosh^2\rho d\tilde{x}^2+d\rho^2+\sinh^2\rho d\tilde{\tau}^2\big),\quad (\tilde{x},\tilde{\tau})=\frac{\pi}{\beta}(x,\tau),\quad\tilde{\tau}\sim\tilde{\tau}+2\pi.
\end{equation}
The brane profile is~\cite{Takayanagi:2011zk,Fujita:2011fp}
\begin{equation}
	\tilde{\tau}(r)=\tilde{\tau}_0\pm\arctan\frac{\calT}{\sqrt{\sinh^2\rho-\calT^2\cosh^2\rho}},
\end{equation}
where the minimum of $\rho$ is
\begin{equation}
	\rho_{\min}=\arctanh|\calT|.
\end{equation}
Let's focus on the case with a positive tension. 
The entanglement wedge lands on the brane at $\rho_{\min}$. 
So, the complexity of $A$ is the volume of the blue region in the figure. 
As the cut-off surface is at $\sinh\rho_\epsilon=\frac{\beta}{\pi\epsilon}$, we arrive at
\begin{equation}
	C_{BTZ}=\frac{2c}{3}\left(\int_0^{\rho_{\min}}d\rho+\int_0^{\frac{1}{\epsilon}}d\rho\right)\cosh\rho\cdot \frac{\pi W}{\beta}=\frac{2c}{3}\left(\frac{W}{\epsilon}+\frac{\pi W}{\beta}\frac{\calT}{\sqrt{1-\calT^2}}\right).
\end{equation}
The complexity increases with temperature.

This geometry is the same as the ``boundary state black hole'' studied in \cite{Hartman:2013qma,Almheiri:2018ijj}. 
But now the periodicity of $\tau$ is promoted to $2\beta$ due to the measurement---the Hawking temperature is decreased by half. 
This temperature change is not weird, as measurements can do very drastic things. 
For example, if we project on energy eigenstates on the left system, then we can get a whole range of different temperatures on the right system depending on the energy of that eigenstate.\footnote{We thank Zhenbin Yang for pointing this out.} 
The post-measurement state on the right system is 
\begin{equation}
	\ket{\psi_R}=\frac{e^{-\frac{\beta}{2}H} | B \rangle }{\sqrt{\bra{B}e^{-\beta H}\ket{B}}}.
\end{equation}
It is often viewed as a black hole microstate, whose real-time evolution can be obtained by gluing the Euclidean part $\tau<0$ with the analytically continued Lorentzian part~\cite{Almheiri:2018ijj}.

\begin{figure}
	\centering
	\includegraphics[width=5cm]{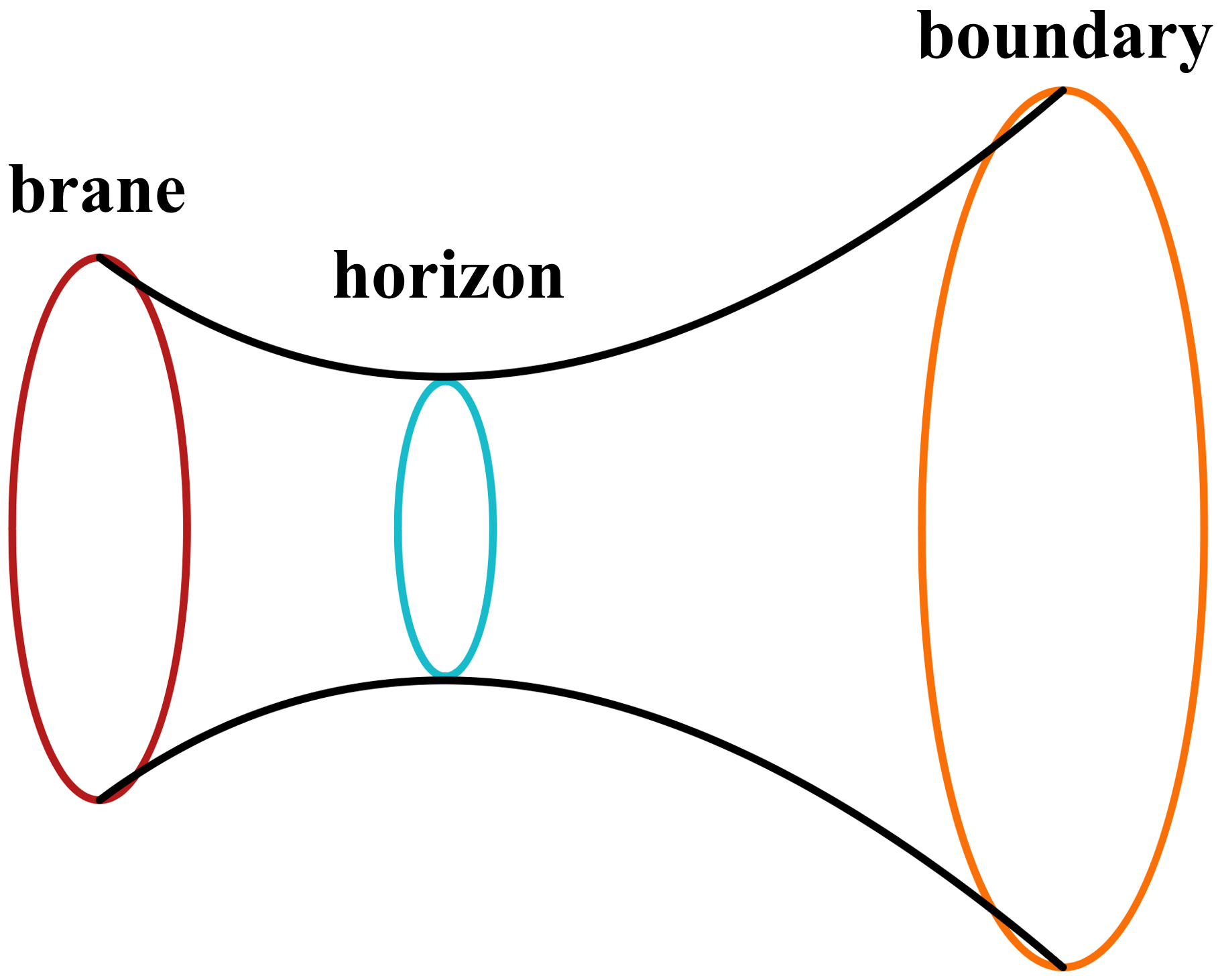}
	\caption{Spatial geometry in the BTZ phase.}
    \label{fig:TFD_finite_spacial_geometry}
\end{figure}

We emphasize that the boundary state black holes can be viewed as the result of measurement. The classical information of the measurement result can then be used to reconstruct the information that is teleported~\cite{Almheiri:2018ijj,Antonini:2022lmg}. 
Notice that there is also a minimal surface at the horizon if the tension is positive. 
The region between the surface on the brane and the surface at the horizon is the so called ``one-sided Python's lunch''~\cite{Brown:2019rox}, but here the lunch is cut off by the brane.
See figure~\ref{fig:TFD_finite_spacial_geometry} for an illustration.

In the AdS phase, the bulk metric is 
\begin{equation}
	ds^2=L^2\big(\cosh^2\rho d\tilde{\tau}^2+d\rho^2+\sinh^2\rho d\tilde{x}^2\big),\quad (\tilde{x},\tilde{\tau})=\frac{2\pi}{W}(x,\tau),\quad\tilde{x}\sim\tilde{x}+2\pi.
\end{equation}
The brane profile is given by
\begin{equation}
	\tilde{x}=\tilde{x}_0\pm\operatorname{arcsinh}\frac{1}{\cosh\rho}.
\end{equation}
Because the RT surfaces vanishes by shrinking to $\rho=0$, the complexity is given by the entire volume of the $(\tilde{x},\rho)$ disk inside the cut-off surface at $\sinh\rho_\epsilon=\frac{W}{2\pi\epsilon}$:
\begin{equation}
	C_{AdS}=\frac{2c}{3}\int_0^{2\pi}d\tilde{x}\int_0^{\rho_\epsilon}d\rho\sinh\rho=\frac{2c}{3}\left(\frac{W}{\epsilon}-2\pi\right). 
\end{equation}

The phase transition is at~\cite{Takayanagi:2011zk,Fujita:2011fp} \footnote{The role of space and time is opposite as in \cite{Takayanagi:2011zk,Fujita:2011fp}. 
The AdS (BTZ) phase in their papers correspond to the BTZ (AdS) phase here.}
\begin{equation}
	\frac{\beta}{W}=-\frac{1}{\pi}\arctanh\calT+\sqrt{\frac{1}{4}+\frac{1}{\pi^2}\arctanh^2\calT}.
\end{equation}
At higher temperatures and lower tension, the BTZ phase is favored, resulting in a significantly larger complexity. 
Note that when the temperature is very low, the black hole structure can be destroyed by the measurement above the Hawking-Page temperature.

\begin{figure}
    \centering
    \includegraphics[width=8cm]{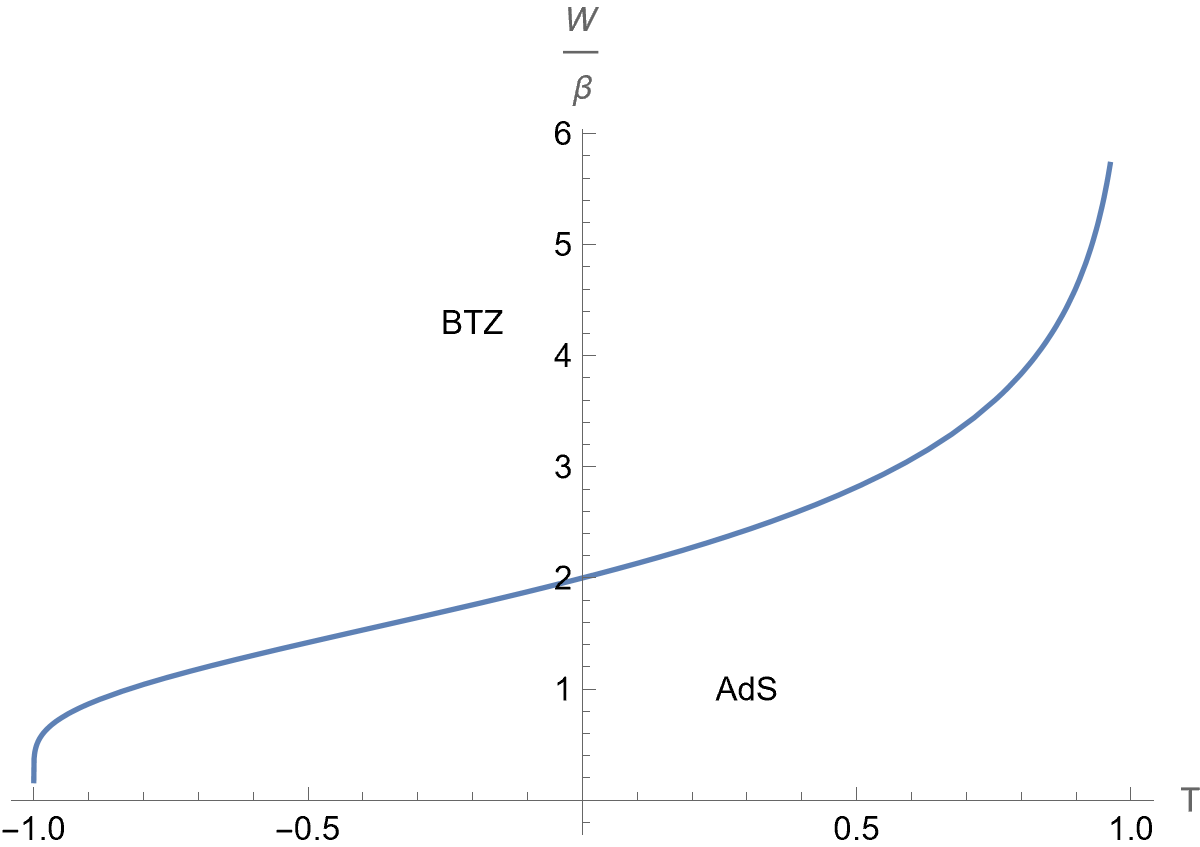}
    \caption{Phase diagram for the geometry after measurement.}
    \label{fig:TFD_finite_phase_diagram}
\end{figure}

Here we make a comparison with similar setups in 2d gravity~\cite{Kourkoulou:2017zaj,Goel:2018ubv}. 
In the BTZ phase of our 3d situation, a boundary observer in the right system will feel that the state looks thermal with temperature $T/2$. 
In the AdS phase, an observer will only see a zero temperature state. 
But in both cases, the bulk brane can be detected by non-local observables in the spacetime. 
In 2d gravity, it was claimed that KM pure states in the form of 
$e^{-\frac{\beta}{2}H}\ket{s}$ look thermal with temperature $T$~\cite{Kourkoulou:2017zaj}. 
This seems to be an important distinction between the 2d dual of KM states and the 3d dual of boundary states. 

~\\
\noindent\textbf{Comparison with python's lunch}

The python's lunch conjecture states that when there is another locally minimal surface apart from the globally minimal one, the complexity to reconstruct operators between these two surfaces (a region called the python's lunch) is exponentially large~\cite{Brown:2019rox}. 
Let $\gamma_l$ and $\gamma_g$ denote the length of the locally minimal surface and the globally minimal one, respectively.
In the reconstruction procedure, there are effectively $\frac{\gamma_l-\gamma_g}{4G\hbar}$ qubits to be post-selected, and the complexity to achieve the post-selection with unitary gates is exponential in this number, namely, the reconstruction complexity scales as $C\sim\exp\left(\frac{1}{2}\frac{\gamma_l-\gamma_g}{4G\hbar}\right)$. 
The complexity considered in our work is a little different, because the volume that we compute represents the minimal tensor network that describes the state. 
In other words, we are counting the number of tensors, and in doing that we allow single-qubit post-selection without extra complexity. 
This notion of complexity has gained more significance in recent times, as measurements and post-selection are becoming important tools for the state preparation~\cite{Ho:2021dmh,Choi:2021npc,Ippoliti:2022bsj,Cotler:2021pbc,Claeys:2022hts,piroli2021quantum,Tantivasadakarn:2021vel,Verresen:2021wdv,Bravyi:2022zcw,Lu:2022jax,Tantivasadakarn:2022ceu,Tantivasadakarn:2022hgp}.

\section{Black hole coupled to a zero temperature bath}\label{sec:Black hole coupled to zero temperature bath}

\begin{figure}
    \centering
    \includegraphics[width=14cm]{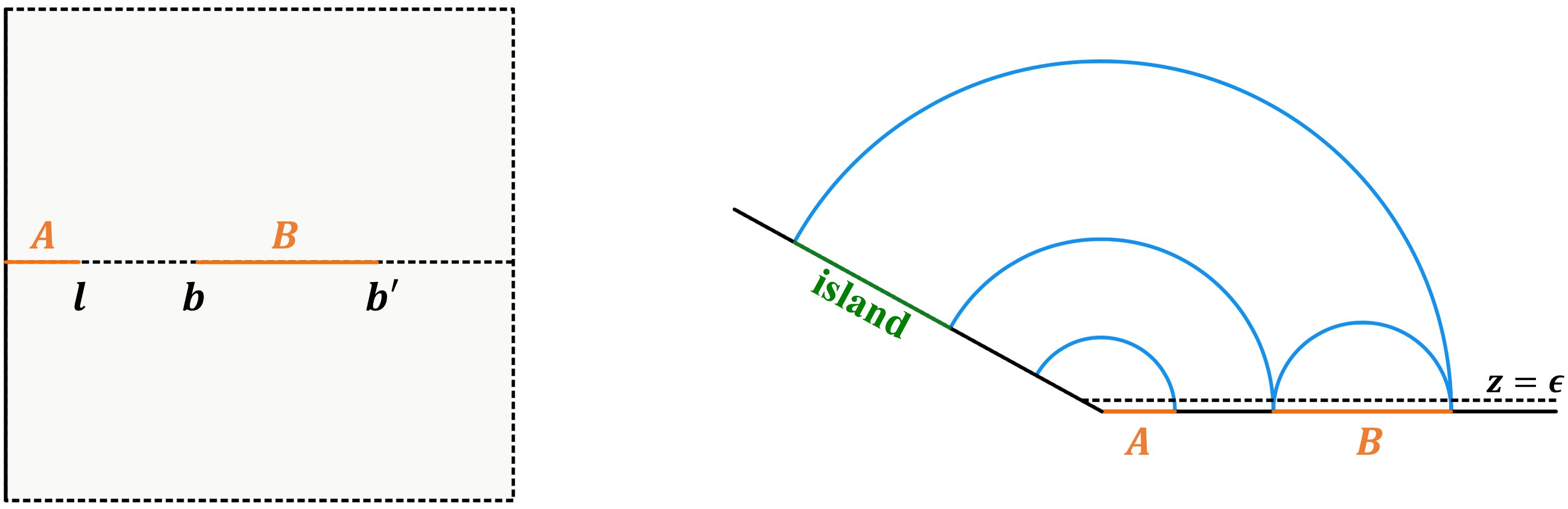}
    \caption{(left panel) Boundary picture of the black hole coupled to a zero temperature bath. 
    (right panel) Brane picture involving an AdS$_3$ bulk and an end-of-the-world brane ending at the boundary. }
    \label{fig:black_hole_bath}
\end{figure}

Consider a quantum dot coupled to a semi-infinite wire that hosts a CFT, and the ground state of such a joint system. 
Its partition function is given by a Euclidean path integral on the right half plane (left panel in figure \ref{fig:black_hole_bath}), terminated at the location of the quantum dot. 
Assuming a conformal boundary condition on the imaginary time axis, we have a BCFT on the right half plane, which is dual to a three-dimensional bulk via the AdS/BCFT correspondence with an end-of-the-world brane landing on the boundary.
This brane model and its extension have been extensively explored in the literature~\cite{Karch:2000ct,Karch:2000gx,Almheiri:2019hni,Chen:2019uhq,Rozali:2019day,Chen:2020jvn,Chen:2020uac,Chen:2020hmv,Hernandez:2020nem,Deng:2020ent,Geng:2020qvw,Geng:2020fxl,Geng:2021iyq,Geng:2021mic,Geng:2021hlu,Grimaldi:2022suv,Suzuki:2022xwv}. 
In particular, subsystem complexity has been studied in the brane model~\cite{Bhattacharya:2021jrn}. 
For a single brane coupled to a non-gravitating bath, there are three equivalent descriptions, which were referred to as the double holography.
The {\it boundary picture} is modeled by a $d$ dimensional CFT coupled to a $d-1$ dimensional boundary. 
The joint system of the quantum dot and the semi-infinite wire we considered corresponds to $d=2$.
The {\it bulk picture} describes a $d$ dimensional CFT coupled to the gravity in an asymptotic AdS$_d$ spacetime and another $d$ dimensional CFT living in a half-Minkowski space, where they are coupled by a transparent boundary condition. 
The {\it brane picture} is given by an Einstein gravity in an asymptotic AdS$_{d+1}$ spacetime with a $d$ dimensional end-of-the-world brane. 
The three-dimensional bulk with an end-of-the-world brane landing on the boundary in our description falls in the brane picture. 
In the bulk picture, our setup mimics a black hole coupled in equilibrium with a zero temperature bath, as different pictures can be related via the AdS/CFT duality:
\begin{itemize}

	\item Using the AdS$_2$/CFT$_1$ correspondence, the quantum dot can be dual to a 2d bulk that describes a zero temperature black hole. 
    See~\cite{Almheiri:2019yqk} for a detailed setup.
    
	\item Starting from the AdS$_3$ bulk with an end-of-the-world brane, one can integrate out the bulk degrees of freedom to get dynamical gravity on the brane coupled to CFT~\cite{Chen:2020uac,Chen:2020hmv}. 
    
    \item In reverse, one can start from AdS$_2$ with gravity coupled to holographic matter, and then apply holography to the holographic matter to get a three-dimensional bulk~\cite{Almheiri:2019hni}.
\end{itemize}
``Islands'' can appear in this equilibrium setup, without the need for real-time dynamics (right panel in figure \ref{fig:black_hole_bath}). 
In this work, we do not rely on these detailed interpretations, although we will assume that the brane has positive tension to be consistent with previous studies. 
We focus on the AdS$_3$/BCFT$_2$ correspondence, following~\cite{Rozali:2019day}.

We consider the region $A:0\le x<l$ that contains the quantum dot, and the region $B:b<x<b'$ that is a part of the bath.
Below we give a brief review of the calculation of the entanglement entropy of the region $B$ using the RT formula~\cite{Sully:2020pza,Ageev:2021ipd}. 
There are two candidate RT surfaces for the region $B$ (left panel in figure \ref{fig:black_hole_bath}): a connected surface and a disconnected surface
\begin{equation}
	S_1(B)=\frac{c}{3}\log\frac{b'-b}{\epsilon},\qquad S_2(B)=\frac{c}{6}\log\frac{2b'}{\epsilon}+\frac{c}{6}\log\frac{2b}{\epsilon}+\frac{c}{3}\log\sqrt{\frac{1+\calT}{1-\calT}},
\end{equation}
\begin{equation}
	S_2(B)-S_1(B)=\frac{c}{3}\left[\log\sqrt{\frac{1+\calT}{1-\calT}}-\log\frac{1}{2}\left(\sqrt{\frac{b'}{b}}-\sqrt{\frac{b}{b'}}\right)\right].
\end{equation}
When $b'/b$ is large enough such that
\begin{equation}
	\frac{1}{2}\left(\sqrt{\frac{b'}{b}}-\sqrt{\frac{b}{b'}}\right)>\sqrt{\frac{1+\calT}{1-\calT}},
\end{equation}
the connected RT surface is dominant, and the entanglement wedge of $B$ contains a portion of the brane---an island.

The complexity of the region $A$ is given by the volume enclosed by the EOW brane, the RT surface and the cut-off surface,
\begin{equation}
\begin{aligned}
    C_0&=\frac{1}{G_NL}\int_{-l\sin\theta}^{l}dx\int_{\min\{\epsilon,-x\cot\theta\}}^{\sqrt{l^2-z^2}}dz\frac{L^2}{z^2}
    =\frac{2c}{3}\left[\frac{l}{\epsilon}+\tan\theta+\tan\theta\cdot\log\frac{l\cos\theta}{\epsilon}-\theta-\frac{\pi}{2}\right].
\end{aligned}
\end{equation}

Since the projection measurement will also create an end-of-the-world brane, in the following we will refer to the brane dual to the quantum dot by the black hole brane or the system brane, and refer to the brane induced by measurements by the measurement brane. 

\subsection{Semi-infinite measurement}

\begin{figure}
    \centering
    \includegraphics[width=12cm]{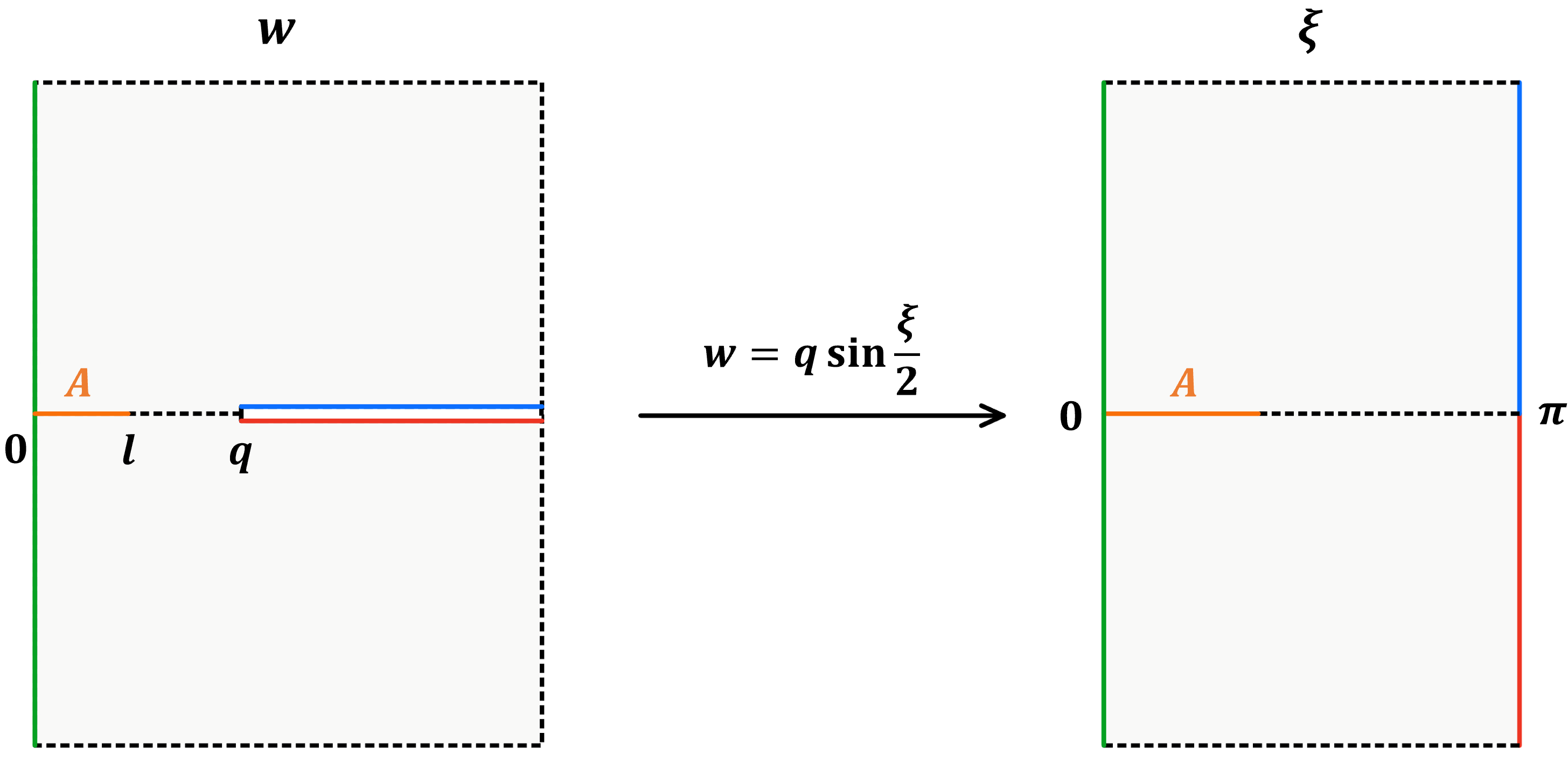}
    \caption{Conformal mapping to the infinite strip. Green (blue/red) line denotes the quantum dot (measurement).}
    \label{fig:bh-bath_infinite_map}
\end{figure}

We measure the region $B:q<x<\infty$ and study the entanglement and complexity of the region $A:0\le x<l$. 
In this case, only the ratio $q/l$ has the physical significance. 
The manifold (the left panel of figure \ref{fig:bh-bath_infinite_map}) can be mapped to the infinite strip (the right panel of figure \ref{fig:bh-bath_infinite_map}) by 
\begin{equation}
    w(\xi)=q\sin\frac{\xi}{2}.
\end{equation}
The black hole is mapped to the left boundary; the slit is mapped to the right boundary (the conformal map simply ``opens'' the slit); the region $A$ is mapped to $0\le\xi_1\le 2\arcsin\frac{l}{q}\equiv a$ and the left end of region $B$ is mapped to $\xi_1 = \pi$. 

Now we construct the bulk dual of the infinite strip. 
The boundary entropy corresponding to the black hole boundary stays the same, because it is an intrinsic property of the boundary and should not be affected by the presence of the other boundary. 
The bulk dual for general boundary conditions is not known, but the authors in \cite{Miyaji:2022dna,Biswas:2022xfw} constructed bulk duals containing two branes with different tensions, which may or may not intersect. 
In the non-intersecting case, the scaling dimension of the boundary-condition-changing (b.b.c.) operator can only be $\Delta_{\bcc}=\frac{c}{24}$; while in the intersecting case, we can achieve the range $\Delta_{\bcc}\in\left(0,\frac{c}{24}\right)$ by tuning the intersection angle (or the mass term on the intersection line). 
This boundary is related to the setup in~\cite{Geng:2021iyq,Liu:2022pan}, where two black holes couple to each other through a finite bath region in between. 
This resemblance stems from our slit description of the measurement.
In this work, we consider the bulk dual containing non-intersecting branes, or intersecting branes without additional structure.

\subsubsection{Non-intersecting configuration}\label{sec:bh-bath_infinite_non-intersecting}

When the bulk dual is Poincare AdS$_3$
\begin{equation}
	ds^2=\frac{L^2}{\eta^2}(d\eta^2+d\xi d\bar{\xi}),
\end{equation}
there are two disconnected branes ending on the black hole boundary and the measurement boundary, respectively.
Their (dimensionless) tensions are denoted by $\calT_{B}$ and $\calT_M$.
Let $z$ and $\eta$ denote the bulk direction in coordinate systems $w$ and $\xi$, such that $z$ and $\eta$ approach zero at the boundary. 
In Poincare coordinates $(\xi,\bar{\xi},\eta)$, the cut-off surface at $z=\epsilon$ is mapped to 
\begin{equation}
    \eta_\epsilon(\xi)=\left|\frac{d\xi}{dw}\right|\epsilon=\frac{2\epsilon}{q|\cos\frac{\xi}{2}|}.
\end{equation}

\begin{figure}
    \centering
    \includegraphics[width=0.7\textwidth]{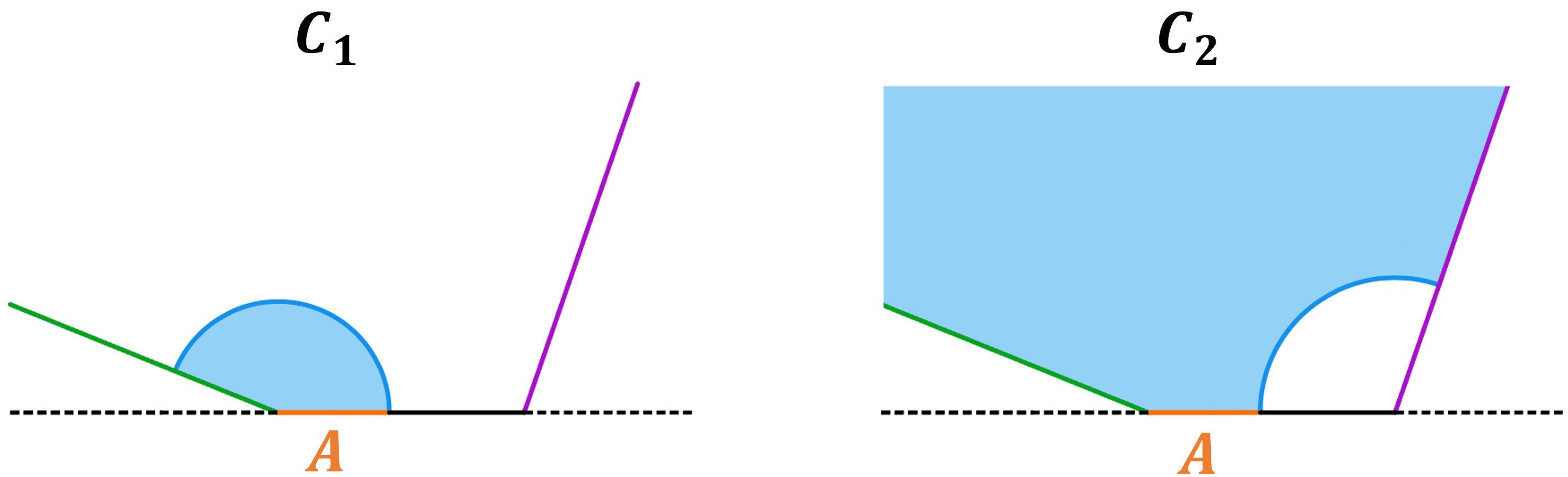}
    \caption{Two candidate RT surfaces and the corresponding volume.  
    (left panel) RT surface that lands on the system brane. (right panel) RT surface that lands on the measurement brane.}
    \label{fig:bh-bath_infinite_non-intersecting_bulk}
\end{figure}

The cut-off surface at the end point of region $A$ is $\eta_\epsilon(a)=\frac{2\epsilon}{\sqrt{q^2-l^2}}$. 
The RT surface of region $A$ can either land on the system brane or the measurement brane (figure \ref{fig:bh-bath_infinite_non-intersecting_bulk}). 
The corresponding entropy is
\begin{equation}
\begin{aligned}
	S_1(A)&=\frac{c}{6}\log\frac{2a}{\eta_\epsilon(a)}+\frac{c}{6}\log\sqrt{\frac{1+\calT_{B}}{1-\calT_B}}, \\
	S_2(A)&=\frac{c}{6}\log\frac{2(\pi-a)}{\eta_\epsilon(a)}+\frac{c}{6}\log\sqrt{\frac{1+\calT_M}{1-\calT_M}}. \\
\end{aligned}
\end{equation}

Since $\log\frac{a}{\pi-a}$ can take all values from $-\infty$ to $\infty$, there must be an exchange of dominance from the first surface to the second surface when we increase $a$ (enlarge the region $A$) or decrease $q$ (enlarge the measured region).
\footnote{If we fix $A$ and decrease $\calT_M$, there can be a transition from the first surface to the second surface. 
In the ``two black holes in equilibrium with each other'' picture, this corresponds to decreasing the boundary entropy (or the degree of freedom) of the right black hole. 
This can trigger a transition of the RT surface, making the left side lose entanglement.} 
The transition happens at
\begin{equation}
	l_*=q\sin\left[\frac{\pi}{2}\frac{\boldg_M}{\boldg_B+\boldg_M}\right],\qquad \boldg_B\equiv\sqrt{\frac{1+\calT_{B}}{1-\calT_B}},\qquad \boldg_M\equiv\sqrt{\frac{1+\calT_{M}}{1-\calT_M}} \label{lstar}
\end{equation}
When $\boldg_B$ is bigger, the transition happens at smaller $l$. This is analogous to the observation in section \ref{sec:Measurement induced complexity transition in random states}, where we find that the transition is easier to happen when $N_A$ gets bigger. There $N_A$ plays the role of $\boldg_B$.

\begin{figure}
    \centering
    \subfigure[]{\includegraphics[width=0.37\textwidth]{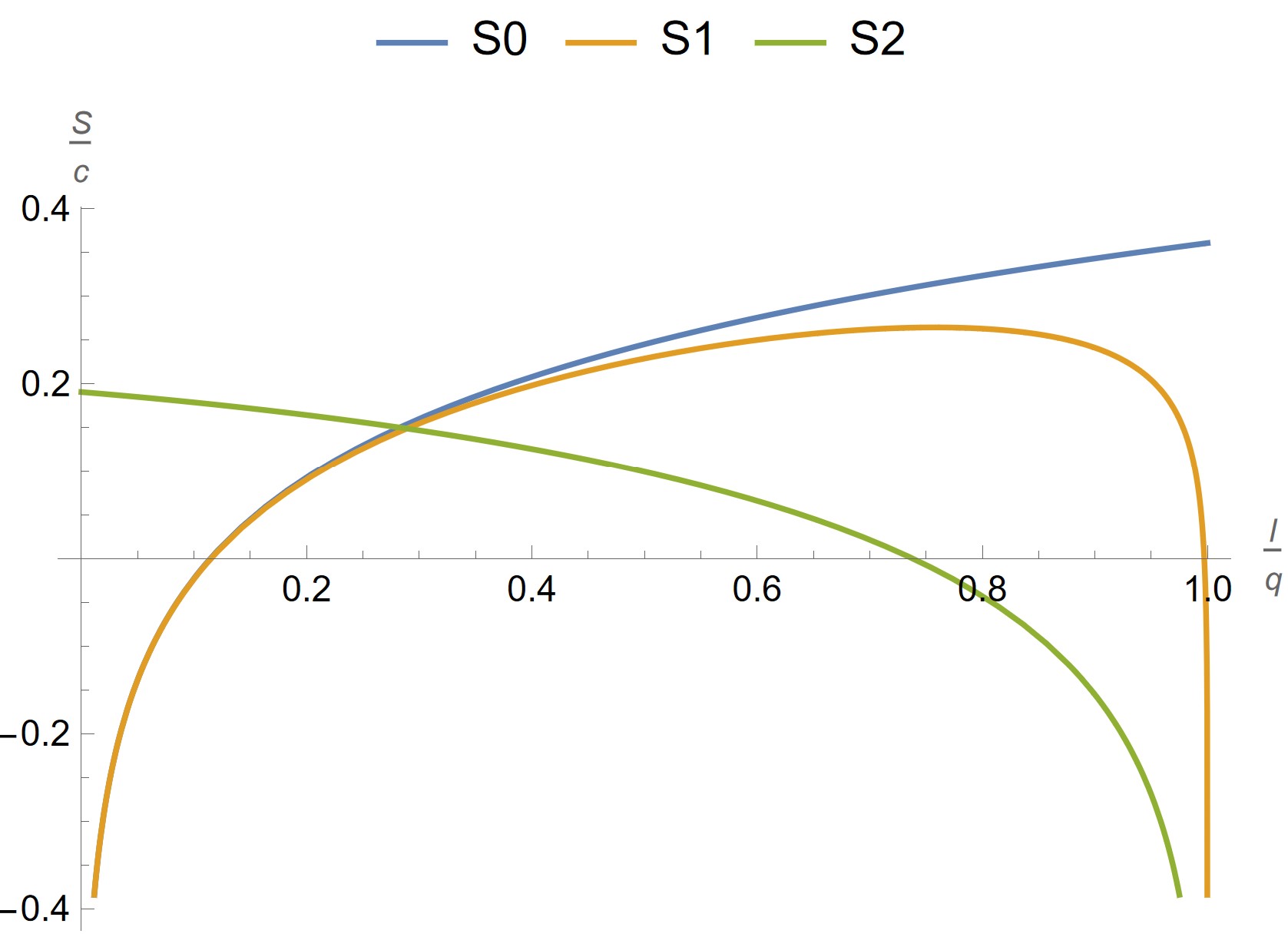}} 
    \quad
    \subfigure[]{\includegraphics[width=0.37\textwidth]{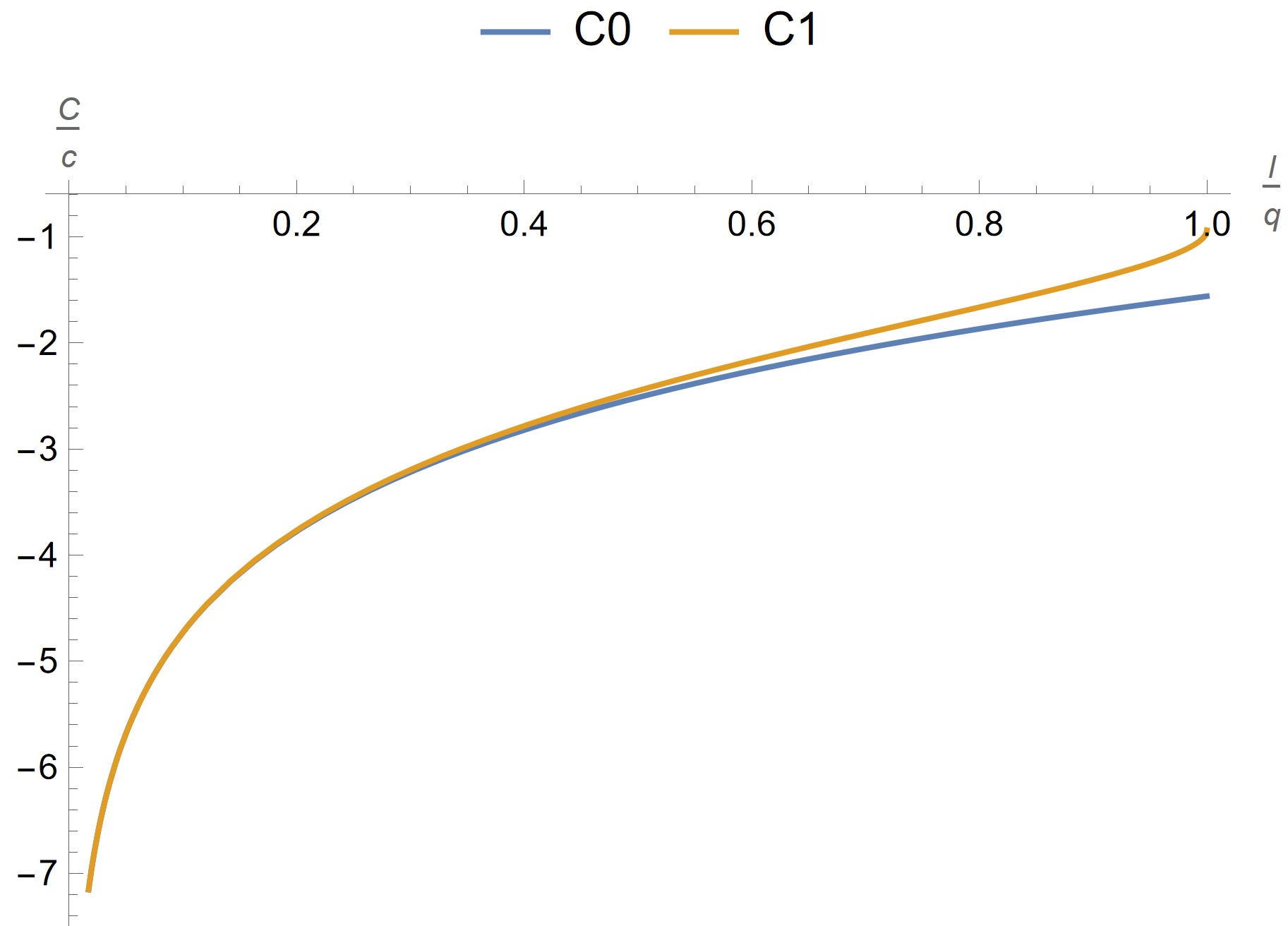}}
    \caption{Entanglement and complexity (regular part) with respect to $l$. We fix $\calT_B=0.9$, $\calT_M=0$.
    (left panel) $S_0 (S_{1,2})$ denotes the entanglement entropy of $A$ before (after) measurements. $S_1, S_2$ denotes the entanglement entropy corresponding to the RT surface ending on the system brane and the measurement brane, respectively. We neglect an infinite contribution from the UV cutoff $\log\epsilon^{-1}$. 
    (right panel) $C_0$ ($C$) denotes the complexity of $A$ before (after) measurements.}
\end{figure}

As in \eqref{eq:vacuum_infinite_line_brane_location}, we defined the brane angles
\begin{equation}
    \sin\theta_B=\calT_B,\qquad \sin\theta_M=\calT_M.
\end{equation}
The complexity corresponding to these two RT surfaces are given respectively by
\begin{equation}
\begin{aligned}
	C_1(A)&=\frac{2c}{3}\left[\frac{l}{\epsilon}+\tan\theta_B\cdot\log\frac{a\cos\theta_B}{\eta_\epsilon(0)}+\tan\theta_B-\theta_{B}-\frac{\pi}{2}\right] \\
	C_2(A)& \mbox{ is divergent}
\end{aligned}
\end{equation}
One difference in the behavior in complexity is that $C_1$ can increase before the transition. $C_1-C_0=\frac{2c}{3}\tan\theta_B\cdot\log\frac{\arcsin l}{l}>0$. It can happen because $A$ includes the quantum dot that represents the black hole. 

The complexity given by the second surface diverges. 
Later in section \ref{sec:Finite measurement}, we will see that the divergence disappears when we measure a finite region. 
So the divergence in the infinite measurement case might be due to the unphysical nature of measuring an infinite big region. 
Here is another possible explanation. 
The region that we measure is infinitely long and contains infinite amount of information. 
Therefore, we may need infinite amount of information to specify the particular state that we wish to prepare. 

\subsubsection{Intersecting configuration}

\begin{figure}
    \centering
    \includegraphics[width=\textwidth]{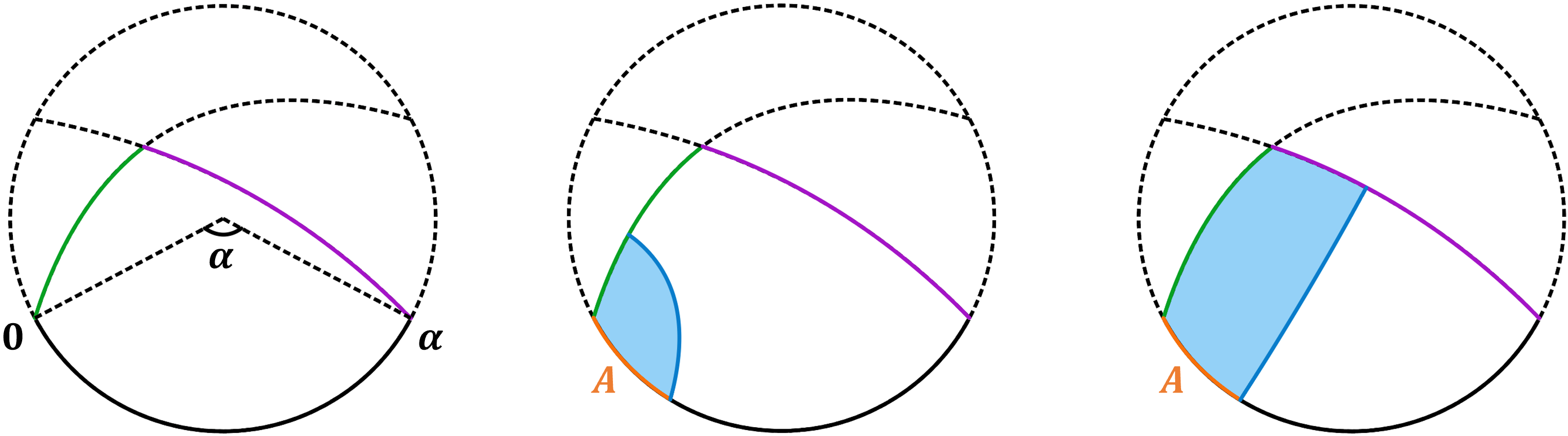}
    \caption{Bulk configuration in the global coordinate $(r,\phi,\tau)$ at a fixed $\tau$ slice.
    The green (purple) curve indicates the system (measurement) brane.
    (left panel) $\alpha$ is the angular extent of the boundary.
    (middle panel) RT surface that lands on the system brane. 
    (right panel) RT surface that lands on the measurement brane.}
    \label{fig:bh-bath_infinite_intersecting_bulk_global}
\end{figure}

When the bulk dual is global AdS$_3$
\begin{equation}
	ds^2=L^2\left[(r^2+1)d\tau^2+\frac{dr^2}{r^2+1}+r^2d\phi^2\right],\quad \phi\sim\phi+2\pi,
\end{equation}
the two branes with different tensions have to intersect at somewhere in the bulk. 
This is because of the presence of a boundary condition changing operator (b.c.c.) operator that connects the two distinct branes with different tensions. 
Let $\alpha<\pi$ be the angular extent of our boundary, it is determined by the scaling dimension of the b.c.c. operator as~\cite{Miyaji:2022dna}
\begin{equation}
    \Delta_{\bcc}=\frac{c}{24}\left(1-\frac{\alpha^2}{\pi^2}\right) .\label{bcc}
\end{equation}
For the same boundary state with the same tension, the b.c.c. operator is trivial and $\alpha = \pi$, which is consistent with the fact that the brane is anchored at antipodal points~\cite{Takayanagi:2011zk,Fujita:2011fp}.
We did a scale transformation to make the circumference of the boundary $2\pi$, $\phi=\frac{\alpha}{\pi}\xi_1$, so the end point of $A$ is at $\beta=\frac{2\alpha}{\pi}\arcsin\frac{l}{q}$. 
In the global coordinate, the brane profiles are
\begin{equation}
	\phi_B(r)=-\arctan\frac{\calT_B}{\sqrt{(1-\calT_B^2)r^2-\calT_B^2}},\qquad\phi_M(r)=\alpha+\arctan\frac{\calT_M}{\sqrt{(1-\calT_M^2)r^2-\calT_M^2}}.
\end{equation}
The Global coordinate $(r,\phi,\tau)$ is shown in figure \ref{fig:bh-bath_infinite_intersecting_bulk_global}.

Next, we switch to the Poincare coordinate $(\mu,\lambda,\bar{\lambda})$ with the black hole boundary at $\lambda_1=0$. 
One reason for doing this is that the brane configurations greatly simplified in the Poincare coordinate. 
Another reason is that we would like to use the coordinate transformation (\ref{bulktrans}) which starts from the Poincare coordinate to fix the position of the cut-off surface. 
The global coordinate $(r,\phi,\tau)$ is related to the Poincare coordinate $(\mu,\lambda,\bar{\lambda})$ by
\begin{equation}
\begin{aligned} \label{eq:transformation_poincare}
	\sqrt{r^2+1}\cosh\tau&=\frac{1+\mu^2+\lambda_1^2+\lambda_2^2}{2\mu}\\
	\sqrt{r^2+1}\sinh\tau&=\frac{\lambda_2}{\mu} \\
	r\sin\phi&=\frac{\lambda_1}{\mu} \\
	r\cos\phi&=\frac{1-\mu^2-\lambda_1^2-\lambda_2^2}{2\mu}
\end{aligned}
\end{equation}

\begin{figure}
    \centering
    \includegraphics[width=\textwidth]{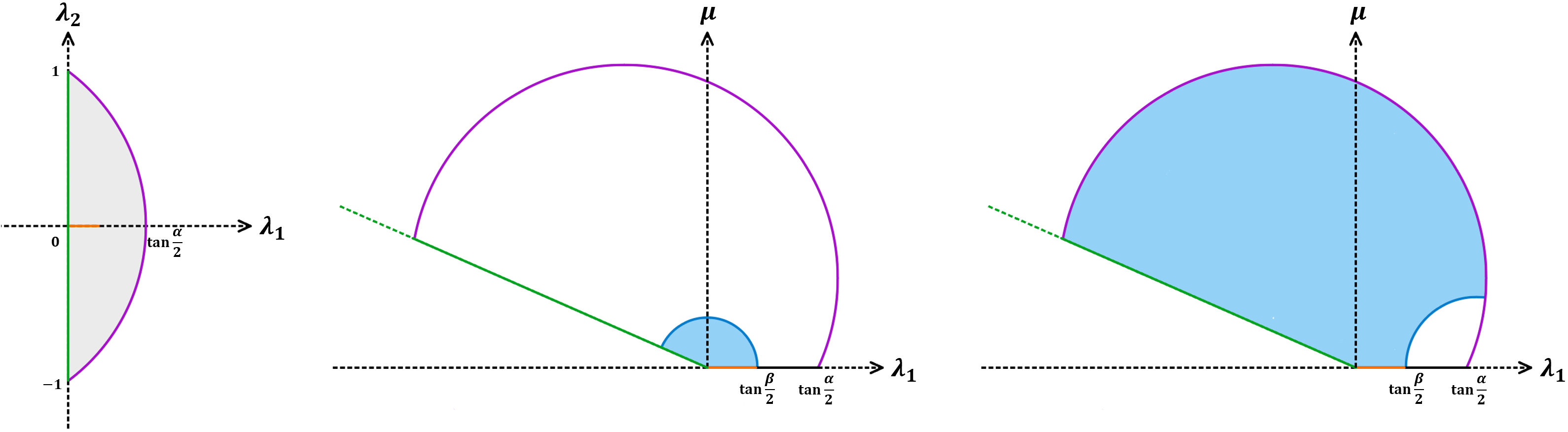}
    \caption{Boundary and bulk configuration in the Poincare coordinate.
    (left panel) The boundary in the Poincare patch. 
    The green line and the purple curve represent the system and the measurement, respectively.
    The orange interval is the subregion $A$.
    (middle/right panel) The bulk in the Poincare patch. 
    The green line and the purple curve represent the system brane and the measurement brane, respectively.
    The orange interval is the subregion $A$.
    The blue curve shows the candidate RT surface landing on the system brane (middle panel) and on the measurement brane (right panel).}
    \label{fig:bh-bath_infinite_intersecting_bulk_poincare}
\end{figure}

Set $\chi=\phi+i\tau$, at the boundary $r\rightarrow\infty$ and $\mu\rightarrow0$, the transformation is
\begin{equation}
	\lambda=\tan\frac{\chi}{2}.
\end{equation}
At the boundary, the system is mapped to $\lambda_1=0,\lambda_2\in(-1,1)$, and the measurement is an arc connecting $\lambda_1=0,\lambda_2=1$ and $\lambda_1=0,\lambda_2=-1$. 
The end point of $A$ is at $\lambda_1=\tan\frac{\beta}{2}=\tan\left(\frac{\alpha}{\pi}\arcsin\frac{l}{q}\right)$. 
See the left panel of figure \ref{fig:bh-bath_infinite_intersecting_bulk_poincare} for an illustration.

In the bulk, we have the system brane at
\begin{equation}
	\mu_B(\lambda)=-\lambda_1\cot\theta_B,
\end{equation}
and the measurement brane is given by a part of the following sphere~\footnote{We obatin this brane configuration by going to another Poincare coordinate. See appendix~\ref{app:bh-bath_details} for more detail.}
\begin{equation}
	(\lambda_1+\cot\alpha)^2+\lambda_2^2+\left(\mu-\frac{\tan\theta_M}{\sin\alpha}\right)^2=\frac{1}{\sin^2\alpha\cos^2\theta_M}
\end{equation}
See the last two panels in figure \ref{fig:bh-bath_infinite_intersecting_bulk_poincare} for an illustration of the branes. 
When we take the limit $\alpha\rightarrow 0$, the boundary becomes an infinite strip and the measurement brane becomes ``straight'': we recover the non-intersecting configuration discussed in section \ref{sec:bh-bath_infinite_non-intersecting}.

Consider the candidate RT surface of the region $A$. 
The candidate RT surface that lands on the system brane is a part of the semicircle as usual.\footnote{In some parameter region, the first candidate surface might intersect with the measurement brane when $A$ gets large enough. In this case, the first candidate ceases to exist. However, one should not worry about this because it can only happen when the second surface is already dominant.} 
The candidate surface that lands on the measurement brane can be determined by requiring that it intersects the brane orthogonally. 
To see the reason for this, we can switch to the other Poincare coordinate where the measurement brane is straight and find that this candidate RT surface is part of a semicircle that intersects the measurement brane orthogonally. 
These two candidate RT surfaces are shown in figure~\ref{fig:bh-bath_infinite_intersecting_bulk_poincare}. 

\begin{figure}
    \centering
    \subfigure[]{\includegraphics[width=0.47\textwidth]{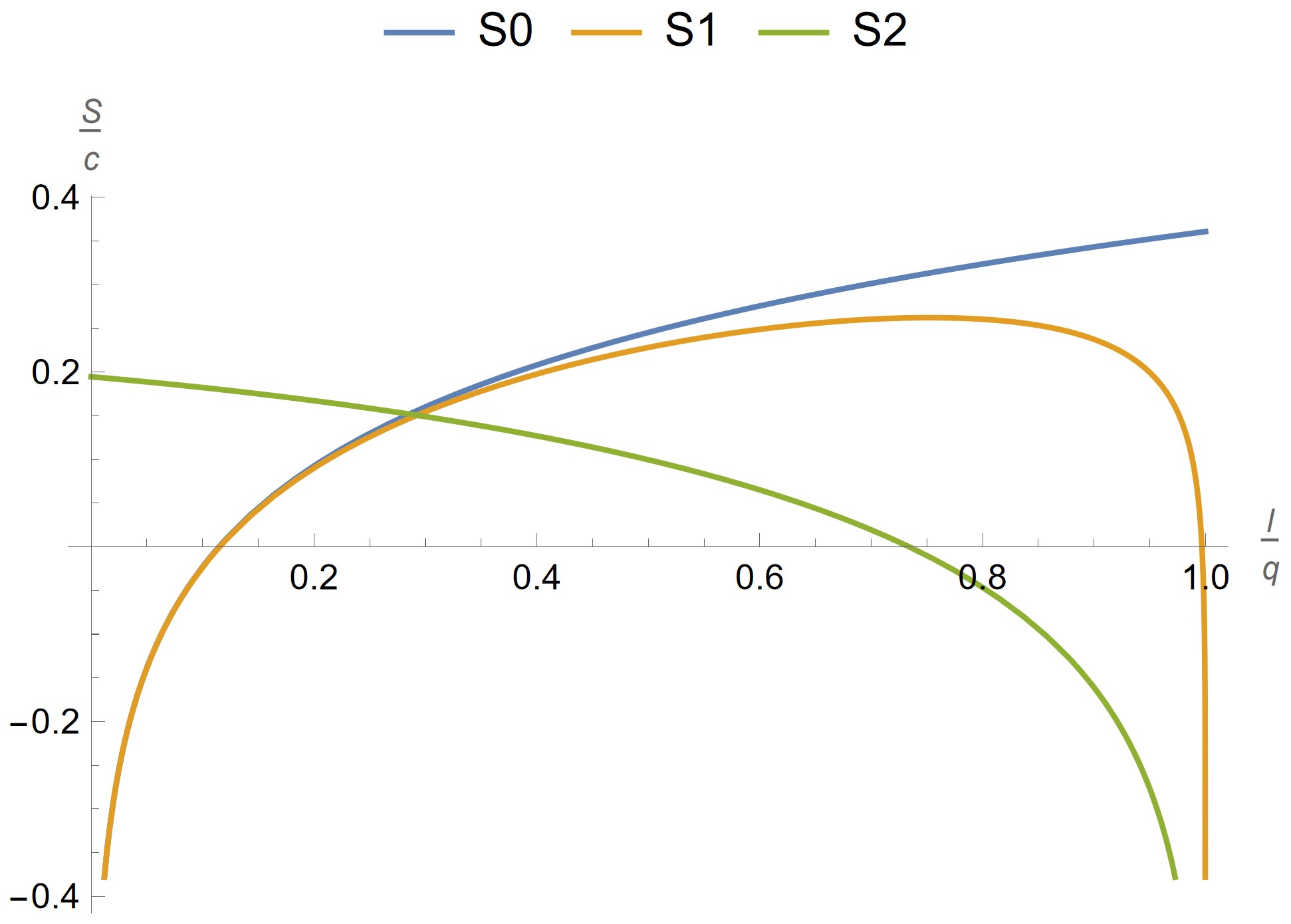}} 
    \subfigure[]{\includegraphics[width=0.47\textwidth]{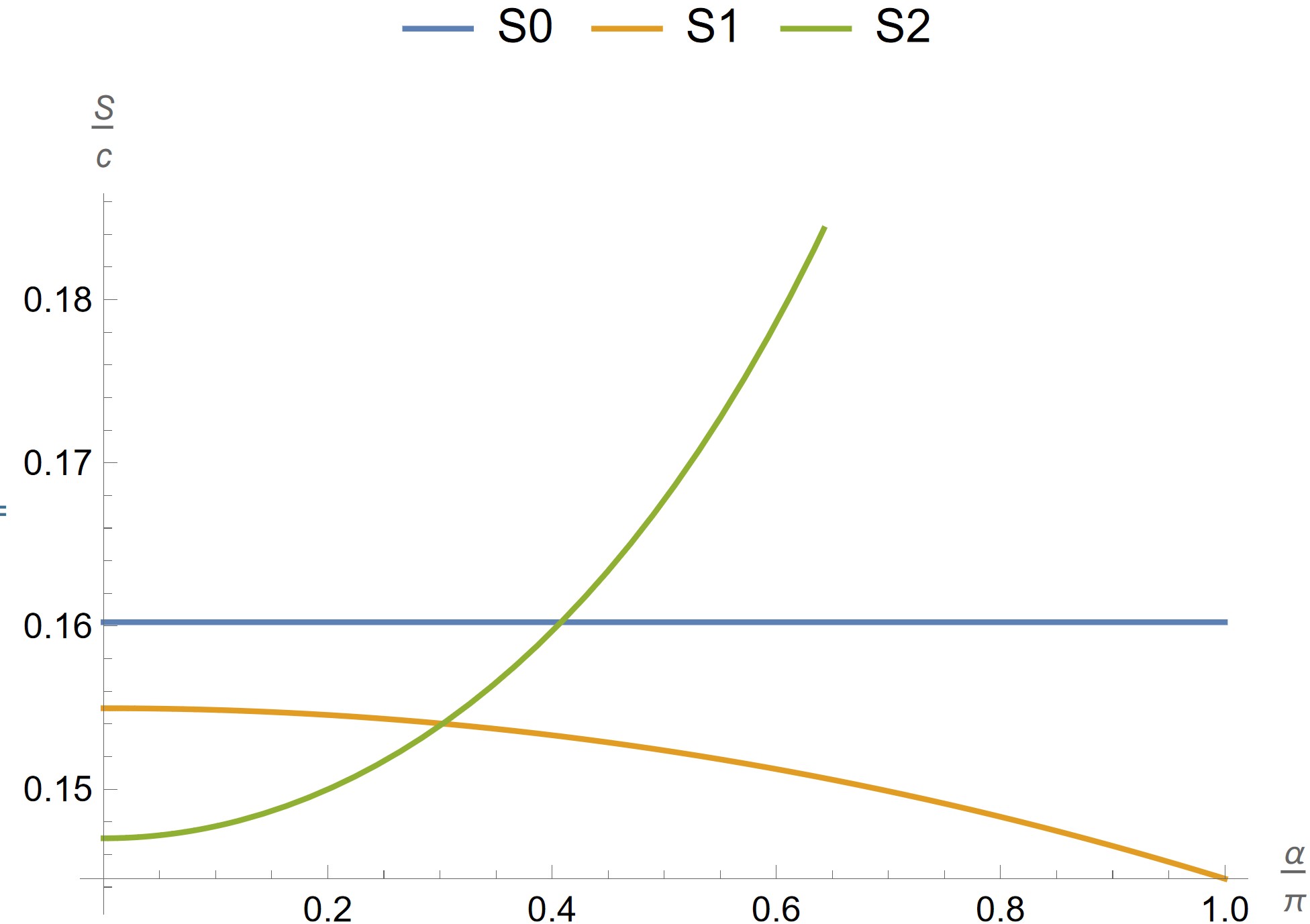}}
    \caption{Entanglement entropy at $\calT_B=0.9$, $\calT_M=0$. 
    $S_0$ ($S_{1,2}$) denotes the entanglement entropy of $A$ before (after) measurements.
    $S_1$, $S_2$ denotes the entanglement entropy corresponding to the RT surface ending on the system brane and the measurement brane, respectively.
    We neglect an infinite contribution from the UV cutoff $\log \epsilon^{-1}$.
    (a) Vary $l$, fix $\alpha=\frac{\pi}{2}$. (b) Vary $\alpha$, fix $l=0.3q$.}
    \label{fig:bh-bath_infinite_intersecting_entanglement}
\end{figure}

We are ready to discuss the transition between these two candidate RT surfaces.
We focus on $\calT_M>0$ here. 
Let's compute entanglement and complexity associated with the two candidate RT surfaces. 
Recall $\beta=\frac{2\alpha}{\pi}\arcsin\frac{l}{q}\in(0,\alpha)$ parametrizes the end point of $A$. 
The cut-off in the Poincare coordinate is at
\begin{equation}
	\mu_\epsilon(\chi)=\left|\frac{d\lambda}{dw}\right|\epsilon=\left|\frac{d\lambda}{d\chi}\right|\left|\frac{d\chi}{dw}\right|\epsilon=\frac{\alpha\epsilon}{\pi q\left|\cos\frac{\pi\chi}{2\alpha}\cos^2\frac{\chi}{2}\right|},
\end{equation}
where we have used $w=q\sin\frac{\pi\chi}{2\alpha}$. 
The entanglement entropy corresponding to the RT surface that ends on the system brane and the measurement brane is given respectively by
\begin{equation}
\begin{aligned}
    &S_1=\frac{c}{6}\log\frac{2\tan\frac{\beta}{2}}{\mu_\epsilon(\beta)}+\frac{c}{6}\log\sqrt{\frac{1+\calT_B}{1-\calT_B}},\\
    &S_2=\frac{c}{6}\log\frac{2\tan\frac{\alpha-\beta}{2}}{\mu_\epsilon(\beta)}+\frac{c}{6}\log\sqrt{\frac{1+\calT_M}{1-\calT_M}} . \label{eq:bh-bath_infinite_intersecting_entropy}
\end{aligned}
\end{equation}
For sufficiently small $l$ the first surface dominates. 
The second surface dominates when $l$ is greater than $l_*(\alpha)\in(0,q)$ determined by
\begin{equation}
	\frac{\boldg_M}{\boldg_B} = \frac{\tan\frac{\beta(l_*(\alpha))}{2}}{\tan\frac{\alpha-\beta(l_*(\alpha))}{2}}.
\end{equation}
In figure~\ref{fig:bh-bath_infinite_intersecting_entanglement}, we show the entanglement entropy before measurements, and the entanglement entropy of the two candidate RT surfaces after measurements.

\begin{figure}
    \centering
    \includegraphics[width=8cm]{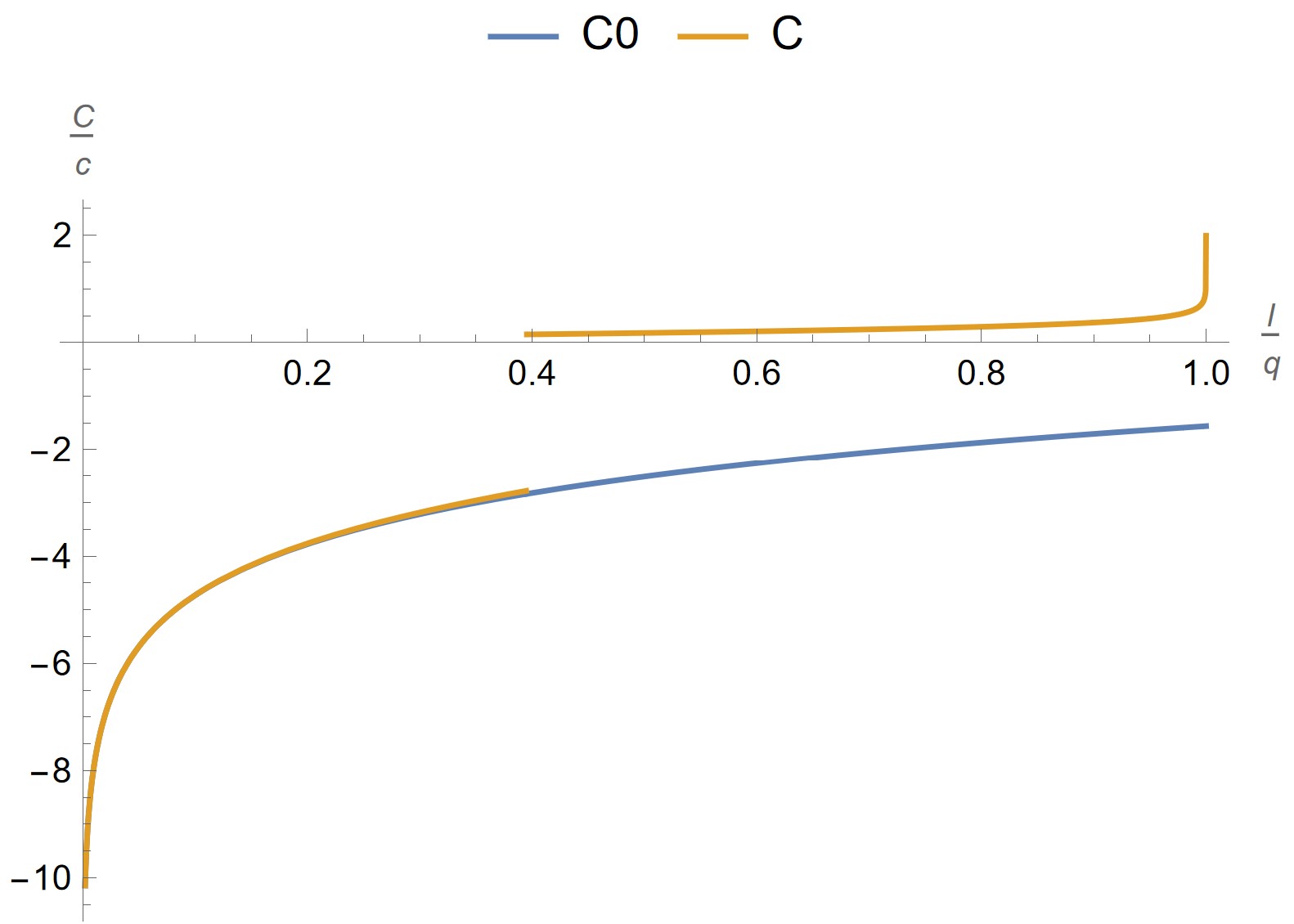}
    \caption{Complexity with respect to $l$. $C_0$ ($C$) denotes the complexity of $A$ before (after) measurements. 
    The jump in $C$ is due to the transition between two RT surfaces.
    $\calT_B=0.9$, $\calT_M=0.3$.}
    \label{fig:bh-bath_infinite_intersecting_connected_complexity}
\end{figure}

Next, we compute complexity using the Gauss-Bonnet theorem. 
Details can be found in appendix~\ref{app:bh-bath_details}.
The complexity corresponding to the first RT surface is
\begin{equation}
    C_1=\frac{2c}{3}\left[\frac{l}{\epsilon}+\tan\theta_B\cdot\log\frac{\tan\frac{\beta}{2}\cos\theta_B}{\mu_\epsilon(0)}+\tan\theta_B-\theta_B-\frac{\pi}{2}\right],\qquad\mu_\epsilon(0)=\frac{\alpha\epsilon}{\pi q} ,\label{eq:bh-bath_infinite_intersecting_complexity1}
\end{equation}
and the complexity corresponding to the second RT surface is
\begin{equation}
	C_2=\frac{2c}{3}\left[\frac{l}{\epsilon}+\tan\theta_B\cdot\log\frac{\mu_I}{\mu_\epsilon(0)}+\tan\theta_M\cdot\log\frac{\mu_I'}{\tan\frac{\alpha-\beta}{2}\cos\theta_M}+\tan\theta_B-\theta_B-\frac{\pi}{2}+\theta_I\right] ,\label{eq:bh-bath_infinite_intersecting_complexity2}
\end{equation}
where $\mu_I$ and $\mu_I'$ are related to the intersecting point of the two branes,
\begin{equation}
    \begin{split}
    \quad\mu_I=\frac{\cos\theta_B}{\sin\alpha}\left[(\cos\alpha\tan\theta_B+\tan\theta_M)+\sqrt{\frac{1}{\cos^2\theta_M}+(\cos\alpha\tan\theta_B+\tan\theta_M)^2}\right], \\
    \mu_I'=\frac{\cos\theta_M}{\sin\alpha}\left[(\cos\alpha\tan\theta_M+\tan\theta_B)+\sqrt{\frac{1}{\cos^2\theta_B}+(\cos\alpha\tan\theta_M+\tan\theta_B)^2}\right].
    \end{split}
\end{equation}
The behavior of complexity with respect to $l$ is plotted in figure \ref{fig:bh-bath_infinite_intersecting_connected_complexity}, where a jump from $C_1$ to $C_2$ appears as $l/q$ is increased.
The complexity without measurement $C_0$ is plotted for comparison. 

When the measurement region approaches the boundary of the region $A$, the complexity after measurement shows a logarithmic divergence.
In this case, we can take $q=l+\epsilon$, which gives $\beta=\alpha\left(1-\frac{2}{\pi}\sqrt{\frac{2\epsilon}{l}}\right)$ and
\begin{equation}
	C=\frac{2c}{3}\left[\frac{l}{\epsilon}+\tan\theta_B\cdot\log\frac{\pi q\mu_I}{\epsilon\alpha}+\tan\theta_M\cdot\log\left(\sqrt{\frac{l}{2\epsilon}}\frac{\pi\mu_I'}{\cos\theta_M}\right)+\tan\theta_B-\theta_B-\frac{\pi}{2}+\theta_I\right]. 
\end{equation}
The logarithm is clear in complexity.

\subsection{Finite measurement}\label{sec:Finite measurement}

\begin{figure}
    \centering
    \includegraphics[width=13cm]{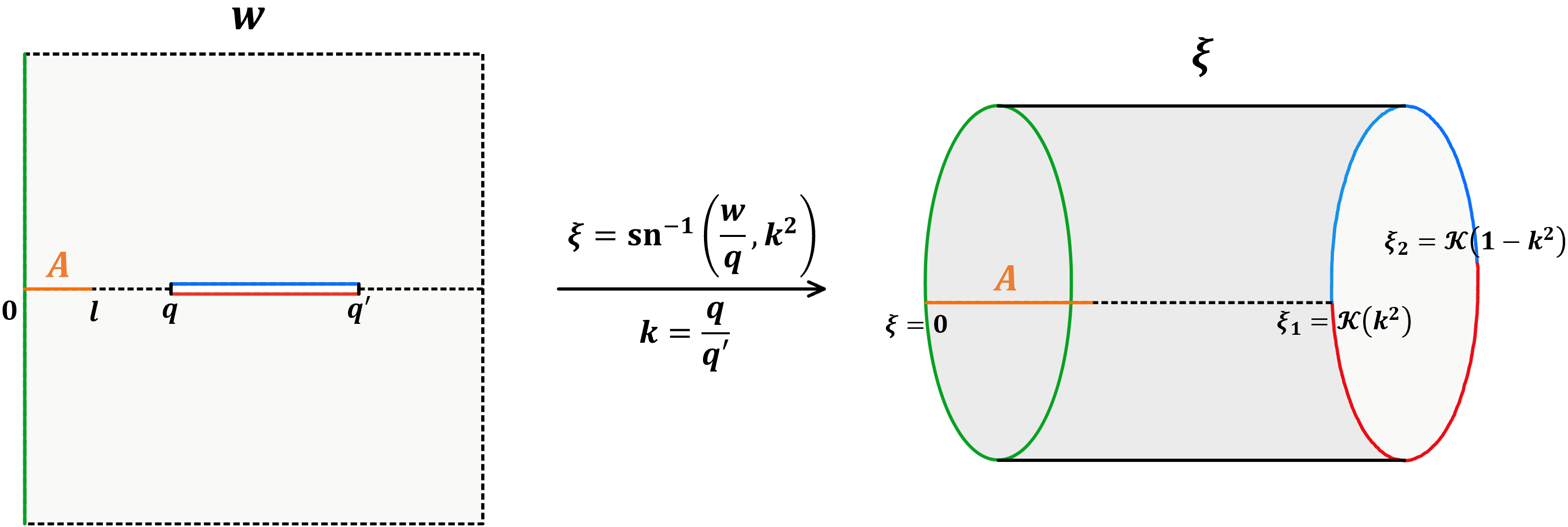}
    \caption{Conformal mapping from the half plane with a slit (left panel) to the annulus (right panel) for the finite measurement.
    $q$ and $q'$ are two endpoints of the measurement region. 
    The green (red/blue) line denotes the system (measurement) region.}
    \label{fig:bh-bath_finite_map}
\end{figure}

We consider the measurement of a finite region of the bath $B:q<w_1<q'$ as shown in the left panel of figure~\ref{fig:bh-bath_finite_map}. 
In this case, as we will see, the divergence in section \ref{sec:bh-bath_infinite_non-intersecting} disappears if the measured region is finite.
The manifold (the left panel of figure~\ref{fig:bh-bath_finite_map}) can be mapped to an annulus (the right panel of figure~\ref{fig:bh-bath_finite_map}) by
\begin{equation}
	\xi=\sn^{-1}\left(\frac{w}{q},k^2\right),\qquad k=\frac{q}{q'}.
\end{equation}
The width of the annulus is $\calK(k^2)$ and the periodicity is $2\calK(1-k^2)$. 
Here, $\calK$ is the elliptic integral of the first kind.

\begin{figure}
    \centering
    \includegraphics[width=\textwidth]{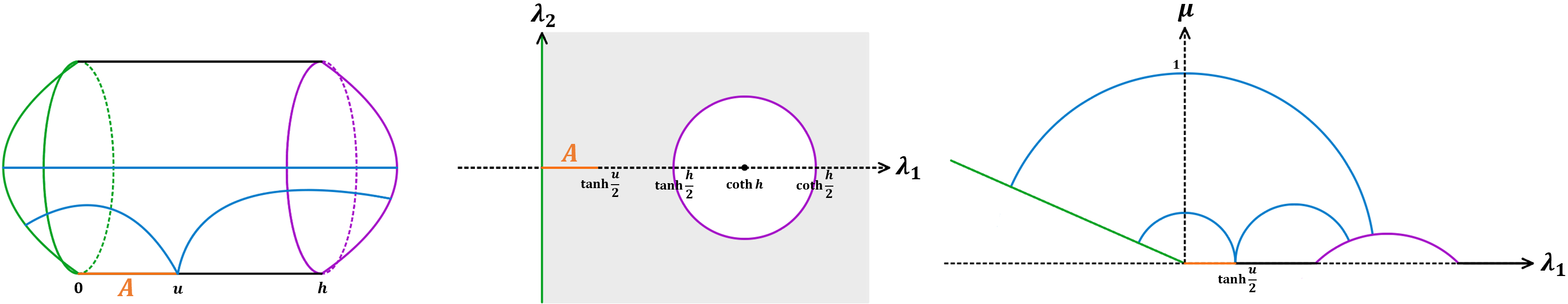}
    \caption{Gravity dual for the finite measurement. 
    Color green and purple denote the system and the measurement region (or brane).
    Blue curve in the left and right panel denotes the candidate RT surface.
    (left penal) The global coordinate $(x,\phi,r)$, $\chi=x+i\phi$. 
    (middle penal) The boundary coordinate $\lambda$. 
    (right penal) The bulk in the Poincare coordinate $(\mu, \lambda, \bar \lambda)$.}
    \label{fig:bh-bath_finite_non-intersecting_bulk}
\end{figure}

The gravity dual is a BTZ black hole ``capped off'' by end-of-the-world branes anchored at the two boundaries. 
The global metric reads\footnote{While we work in the global coordinate, it is easy to see that this is a BTZ black hole by a coordinate transformation~$z= \sqrt{r^2+1}$.}
\begin{equation}
\begin{aligned}
	&ds^2=L^2\left[(r^2+1)dx^2+\frac{dr^2}{r^2+1}+r^2d\phi^2\right], \\
	&\chi=\frac{\pi}{\calK(1-k^2)}\xi= x+i\phi,\quad\phi\sim\phi+2\pi,\quad	x\in[0,h],\quad h\equiv\pi\frac{\calK(k^2)}{\calK(1-k^2)}.
\end{aligned}
\end{equation}
The brane profile of the system brane and the measurement brane are, respectively,
\begin{equation}
	x_B=-\arcsinh\frac{\tan\theta_B}{\sqrt{r^2+1}},\qquad x_M=h+\arcsinh\frac{\tan\theta_M}{\sqrt{r^2+1}}.
\end{equation}
See the left panel in figure \ref{fig:bh-bath_finite_non-intersecting_bulk}.  
In the global coordinate, the end point of the $A$ region becomes $u=\frac{\pi}{\calK(1-k^2)}\sn^{-1}\left(\frac{l}{q},k^2\right) \in(0,h)$.

Similar to the previous discussion, we switch to the Poincare coordinate to compute entanglement and complexity using the same coordinate transformation~\eqref{eq:transformation_poincare}.
The calculation is similar, so we leave the detail in appendix~\ref{app:bh-bath_details} and only present key results here.
The boundary and bulk configuration in the Poincare coordinate is illustrated in figure \ref{fig:bh-bath_finite_non-intersecting_bulk}. 
At the boundary, the system boundary is at $\lambda_1=0$ and as usual, the system brane is given by
\begin{equation}
	\lambda_1=-\mu\tan\theta_B.
\end{equation}
On the other hand, the measurement brane is given by the following sphere:
\begin{equation}
	(\lambda_1-\coth h)^2+\lambda_2^2+\left(\mu+\frac{\tan\theta_M}{\sinh h}\right)^2=\frac{1}{\sinh^2h\cos^2\theta_M}. 
\end{equation}
For later convenience, we need the cut-off surface in the global coordinate at
\begin{equation}
\begin{aligned}
	&\mu_\epsilon(u)=\frac{\pi}{2\cosh^2\frac{\chi}{2}\calK(1-k^2)q\sn'\left(\frac{\calK(1-k^2)u}{\pi}\right)}\epsilon, \qquad \mu_\epsilon(0)=\frac{\pi}{2\calK(1-k^2)q}\epsilon.
\end{aligned}
\end{equation}

\begin{figure}
    \centering
    \subfigure[]{\includegraphics[width=0.47\textwidth]{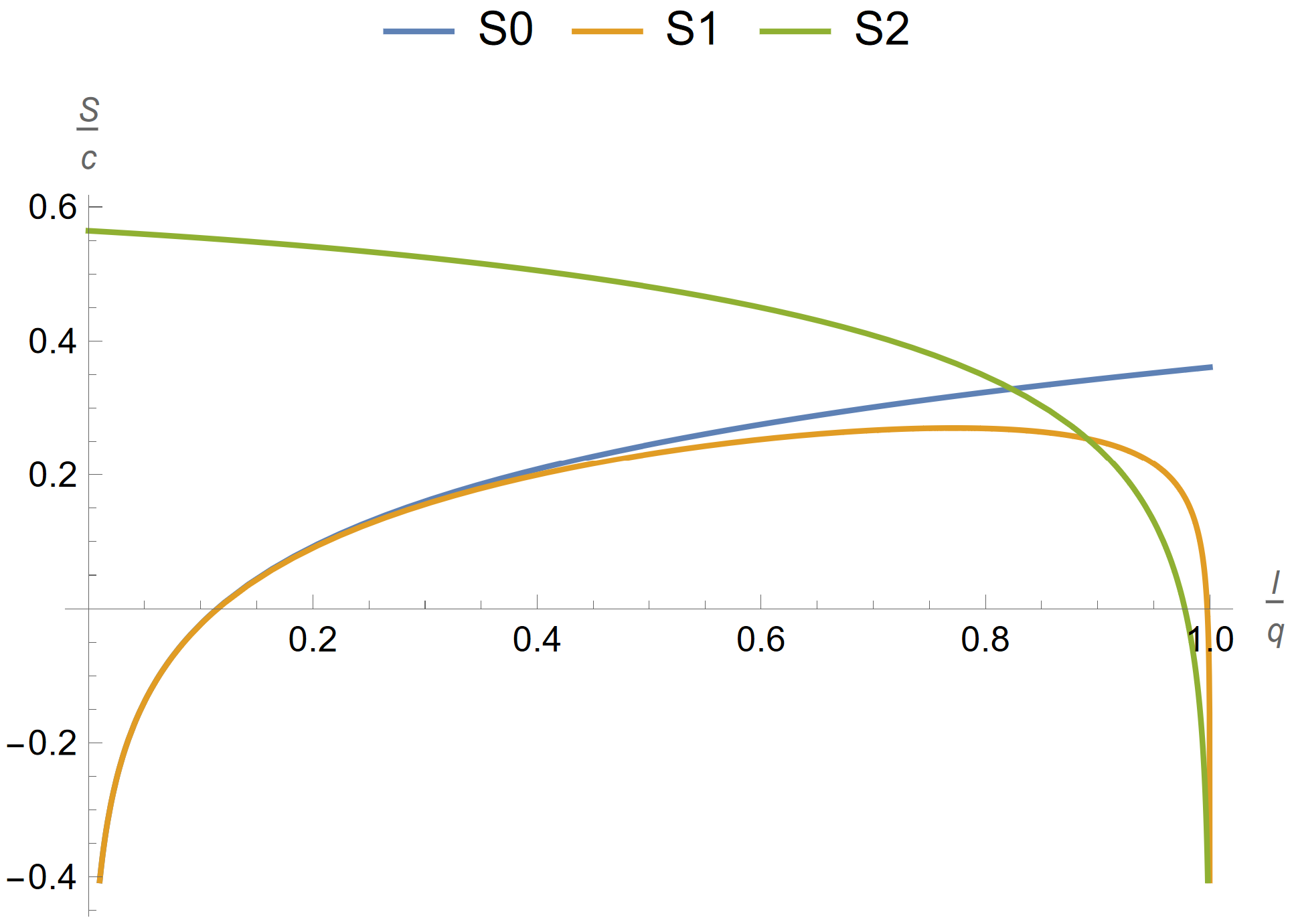}} 
    \subfigure[]{\includegraphics[width=0.47\textwidth]{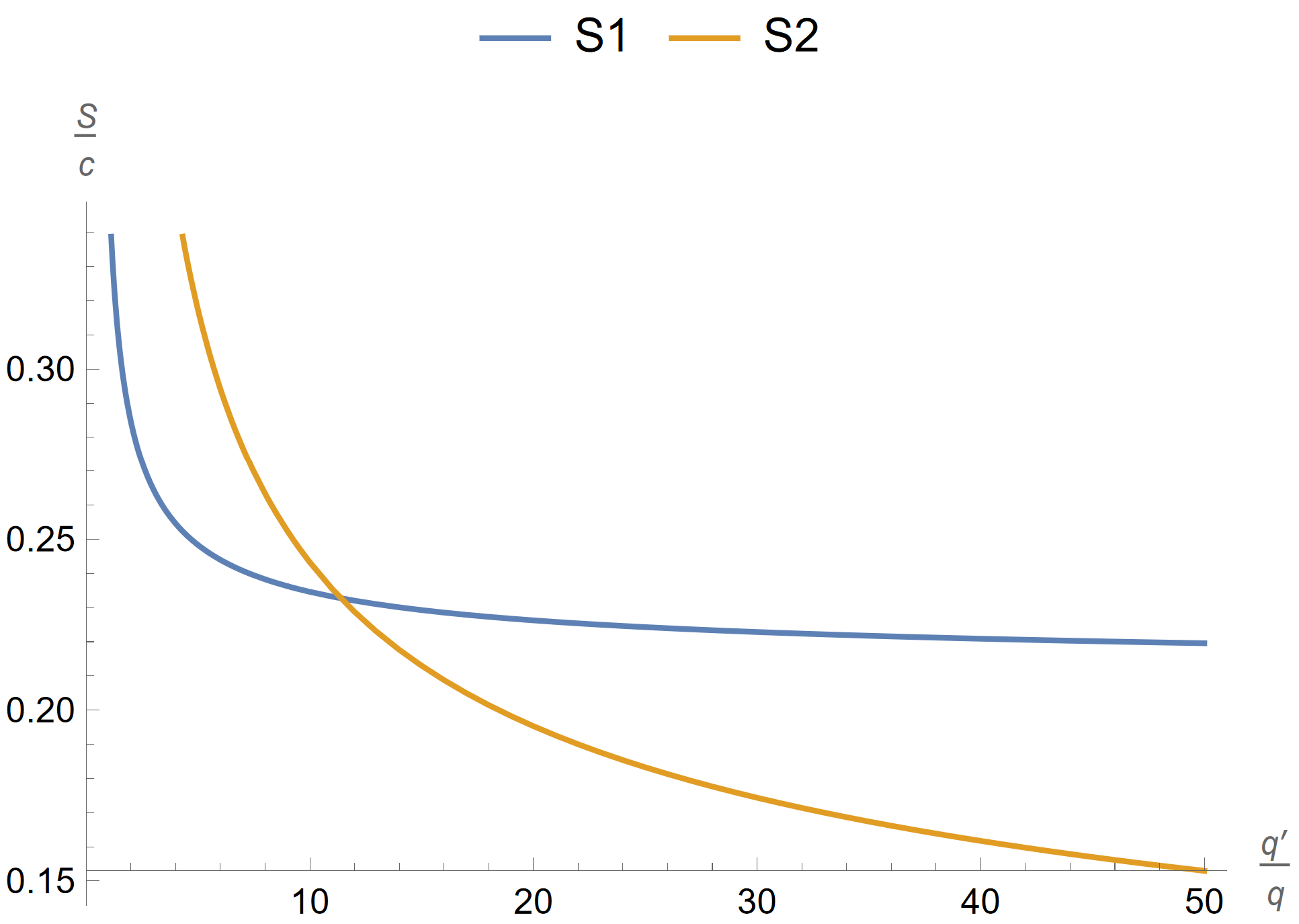}}
    \caption{Regular part of entanglement for the finite measurement and non-intersecting case at $\calT_B=0,9$, $\calT_M=0$. 
    $S_0$ ($S_{1,2}$) denotes the entanglement entropy of $A$ before (after) measurements.
    $S_1$, $S_2$ denotes the entanglement entropy corresponding to the RT surface ending on the system brane and the measurement brane, respectively.
    We neglect an infinite contribution from the UV cutoff $\log \epsilon^{-1}$.
    (a) $\frac{q'}{q}=100$, change $l$. 
    (b) $l=0.95q$, change $\frac{q'}{q}$.}
    \label{fig:bh-bath_finite_non-intersecting_entanglement}
\end{figure}

There are two candidate surfaces for the region $A$. 
The first candidate lands on the system brane. 
The second candidate consists of two pieces: one piece starts from the end point of $A$ and lands on the measurement brane; the second piece connects the two branes at $r=0$ (which is the horizon) in global coordinates. 
See the left panel in figure~\ref{fig:bh-bath_finite_non-intersecting_bulk}.
In the Poincare coordinate, the second piece is located at $\lambda_1^2+\mu^2=1$. 
See the right panel in figure~\ref{fig:bh-bath_finite_non-intersecting_bulk}.
This piece is the key difference from the infinite measurement case, because it ensures that the bulk volumes are finite.
The entanglement entropy corresponding to the two candidate RT surfaces is given by
\begin{equation}
	\begin{aligned}
		&S_1=\frac{c}{6}\log\frac{2\tanh\frac{u}{2}}{\mu_\epsilon(u)}+\frac{c}{6}\log\sqrt{\frac{1+\calT_B}{1-\calT_B}}, \\
		&S_2=\frac{c}{6}\log\frac{2\tanh\frac{h-u}{2}}{\mu_\epsilon(u)}+\frac{c}{6}\log\sqrt{\frac{1+\calT_M}{1-\calT_M}}+\frac{c}{6}(h+\arcsinh\tan\theta_B+\arcsinh\tan\theta_M).
	\end{aligned}
\end{equation}
Since $\frac{\tanh\frac{u}{2}}{\tanh\frac{h-u}{2}}$ can take all positive values as we vary $u\in(0,h)$, there has to be a transition to the second surface when $u$ gets close enough to $h$. 
The length of the horizon is long, so generally the transition happens when $u$ is close to $h$. 
See the left panel in figure \ref{fig:bh-bath_finite_non-intersecting_entanglement}.

On the other hand, there is a transition as $q'/q$ is increased, if $S_1-S_2$ is positive at $q'\rightarrow\infty$. 
One can check that $S_1-S_2$ is a monotonically increasing function of $q'$. In the limit of $q'/q\rightarrow\infty$ , we have $k=q/q'\rightarrow 0$, $\calK(k^2)\rightarrow\frac{\pi}{2}$, $\calK(1-k^2)=\log\frac{q'}{q}$, $\sn(\xi,k^2)\rightarrow\sin\xi$, $h\rightarrow \frac{\pi^2}{2\log(q'/q)}$, $u\rightarrow\frac{\pi}{\log(q'/q)}\arcsin\frac{l}{q}$, and
\begin{equation}
\begin{aligned}
   S_1-S_2\rightarrow&\frac{c}{6}\log\frac{2\arcsin(l/q)}{\pi-2\arcsin(l/q)}+\frac{c}{6}\log\sqrt{\frac{1+\calT_B}{1-\calT_B}}-\frac{c}{6}\log\sqrt{\frac{1+\calT_M}{1-\calT_M}} \\
   &-\frac{c}{6}(\arcsinh\tan\theta_B+\arcsinh\tan\theta_M).
\end{aligned}
\end{equation}
The first line is just the answer for the infinite measurement and non-intersecting case. 
The second line is the extra contribution from the horizon. 
Whether this quantity is positive or negative depends on $l/q$, $\calT_B$ and $\calT_M$. 
If this quantity is positive, then a transition from $S_1$ to $S_2$ happens at a finite $q'/q$. 

\begin{figure}
    \centering
    \subfigure[]{\includegraphics[width=0.5\textwidth]{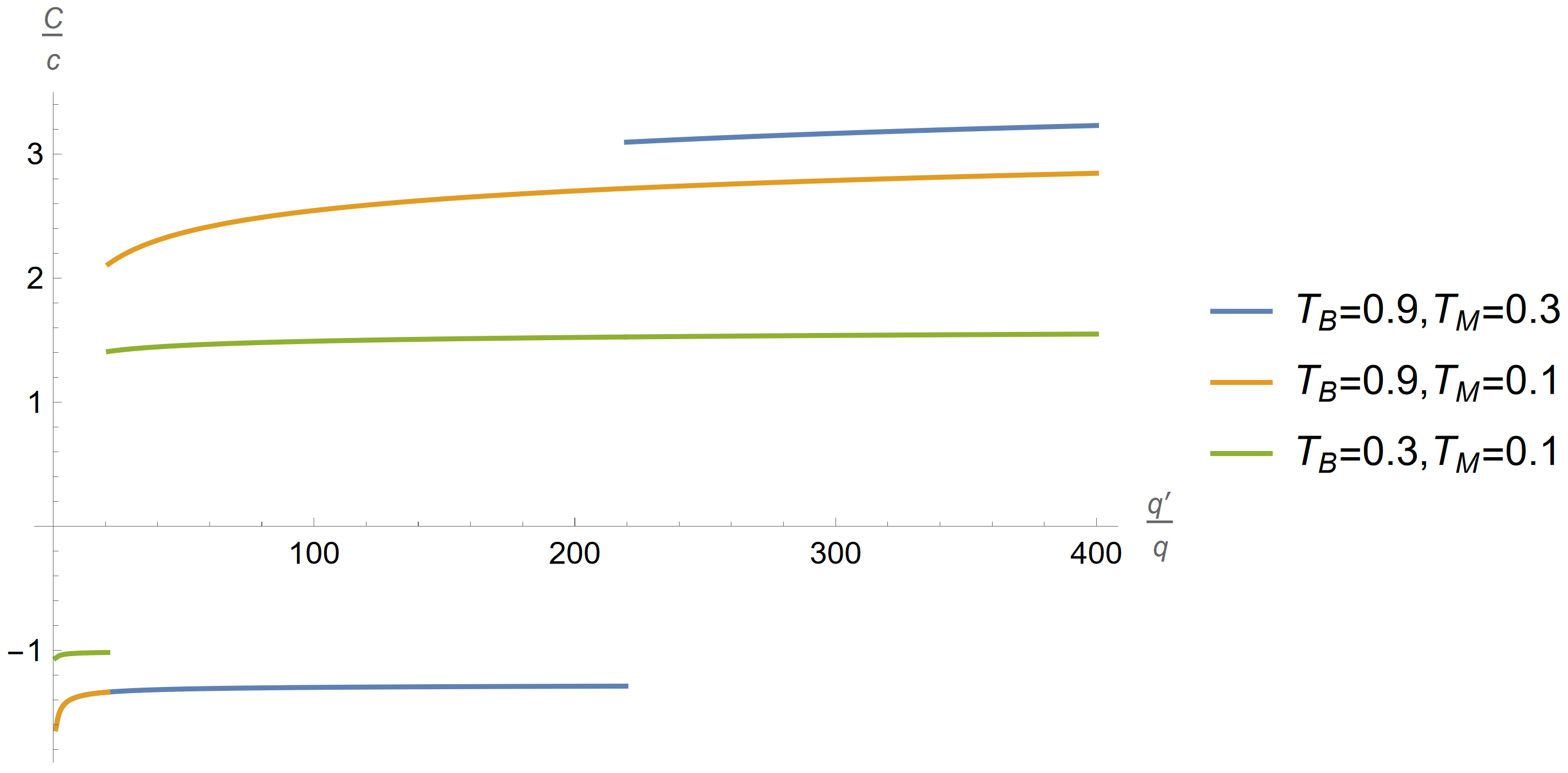}} 
    \quad
    \subfigure[]{\includegraphics[width=0.37\textwidth]{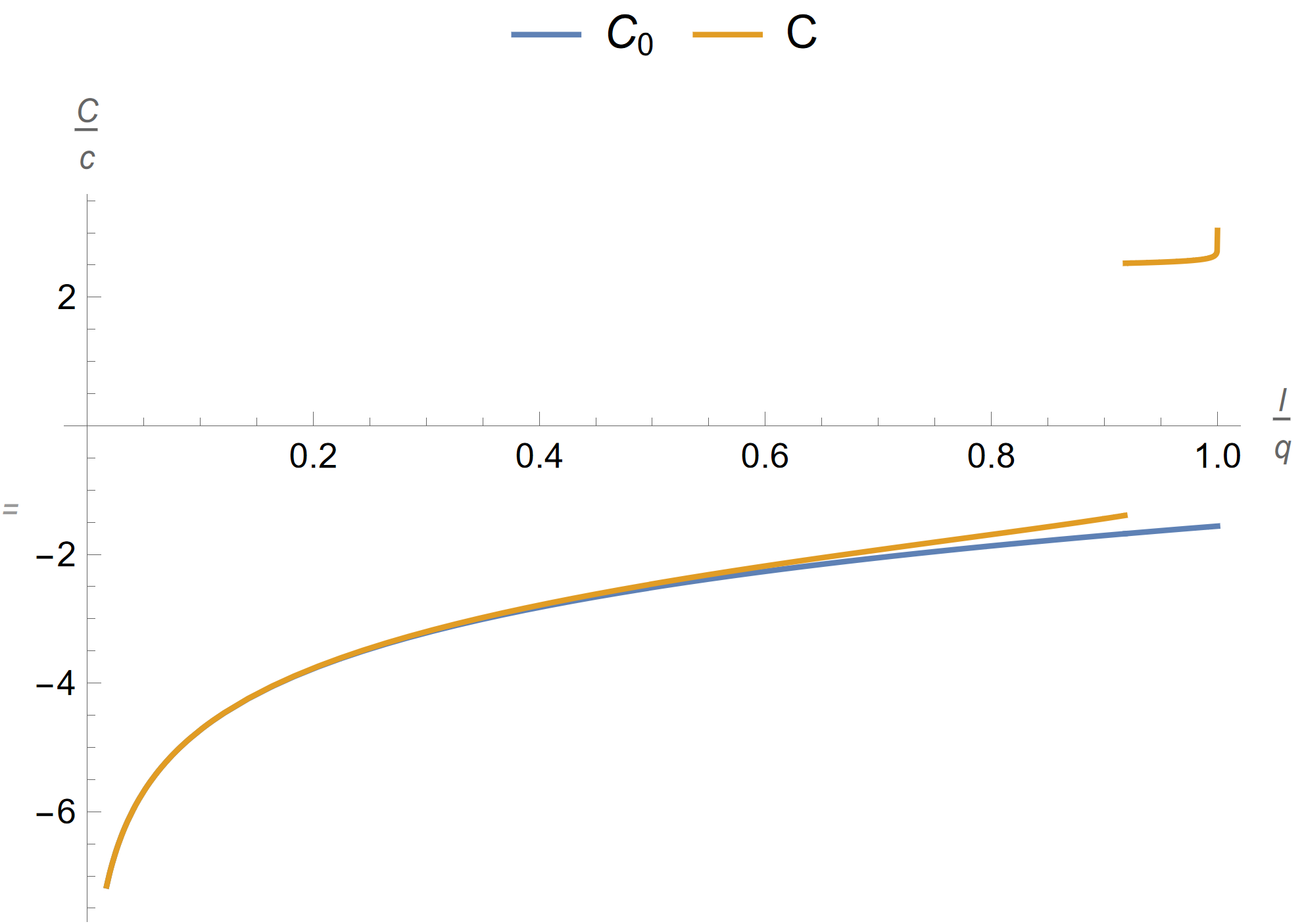}}
    \caption{Complexity for the finite measurement, non-intersecting case. 
    (a) Subsystem complexity after measurement as a function of $\frac{q'}{q}$. $l=0.95q$ is fixed.
    (b) Subsystem complexity as a function of $l/q$. $C_0$ ($C$) denotes the complexity before (after) measurements. 
    We fix $\frac{q'}{q}=100$, $\calT_B=0,9$, $\calT_M=0.1$.}
    \label{fig:bh-bath_finite_non-intersecting_complexity}
\end{figure}

The complexity corresponding to the two RT surfaces are
\begin{equation}
\begin{aligned}
	C_1&=\frac{2c}{3}\left[\frac{l}{\epsilon}+\tan\theta_B\cdot\log\frac{\tanh\frac{u}{2}\cos\theta_B}{\mu_\epsilon(0)}+\tan\theta_B-\theta_B-\frac{\pi}{2}\right], \\
	C_2&=\frac{2c}{3}\left[\frac{l}{\epsilon}+\tan\theta_B\cdot\log\frac{\cos\theta_B}{\mu_\epsilon(0)}+\tan\theta_B-\theta_B+\frac{\pi}{2}+\tan\theta_M\cdot\log\frac{1}{\tanh\frac{h-u}{2}}\right].
\end{aligned}
\end{equation}
One can check that for $\calT_B,\calT_M>0$, $C_2$ is a monotonically increasing function of $q'$. In this case, if there is a complexity jump when we increase $q'$, then complexity continues to increase after the jump. In the $q'/q\rightarrow\infty$ limit, $C_2$ diverges as $\log\log q'$. 
\begin{equation}
\begin{aligned}
    C_2\rightarrow \frac{2c}{3}&\left[\frac{l}{\epsilon}+\tan\theta_B\cdot\log\frac{2\cos\theta_B\log(q'/q)q}{\pi\epsilon}+\tan\theta_M\cdot\log\frac{4\log(q'/q)}{\pi(\pi-2\arcsin(1/q))}\right. \\
    & \left.+\tan\theta_B-\theta_B+\frac{\pi}{2}\right] 
\end{aligned}
\end{equation}
The behavior of complexity is plotted in figure \ref{fig:bh-bath_finite_non-intersecting_complexity}.

\begin{figure}
    \centering
    \includegraphics[width=0.4\textwidth]{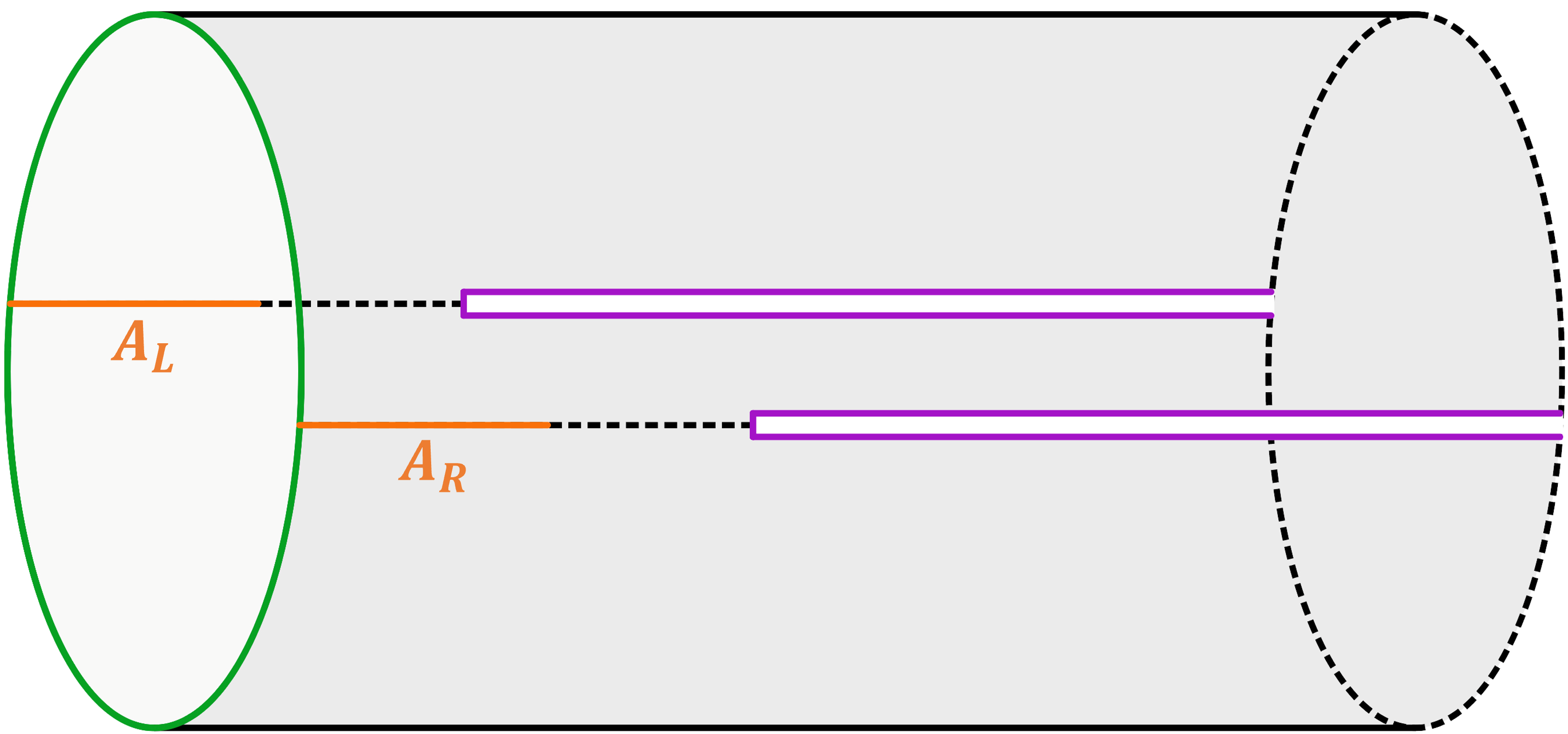}
    \caption{Boundary picture of the black hole and bath system at finite temperature. 
    The measurement introduces the slits.}
    \label{fig:bh-bath_finite_temperature}
\end{figure}

Finally, we comment on the joint system of a black hole coupled to a bath at finite temperature. 
In the boundary picture, we have a thermofield double state between the left and right system. 
Consider region $A=A_L\cup A_R$, where $A_L$ is $[0,a]_L$ in the left system and $A_R$ is $[0,a]_R$ in the right system. We measure the region $B=B_L\cup B_R$, where $B_{L,R}=[b,+\infty)_{L,R}$. 
The boundary manifold, as shown in figure \ref{fig:bh-bath_finite_temperature}, can be conformally mapped to the annulus, just like the finite measurement case at zero temperature. 
The resemblance comes from the compactness of the system boundary. 
The bulk dual can be constructed in the same way. 
Again, depending on the scaling dimension of the b.c.c. operator, we have non-intersecting or intersecting branes in the bulk.


\section{Conclusion and outlook}
\label{sec:conclusion}

We studied subsystem complexity within holographic systems subjected to projection measurements. 
The projection measurement is effectively modeled by a slit within the Euclidean path integral, incorporating an end-of-the-world brane through the AdS/BCFT correspondence. 
Various holographic setups have been explored, where complexity jumps to a higher value upon measurements have been found. 
We conclude with some remarks and possible future directions.

An intriguing avenue for future exploration pertains to the real-time evolution of complexity in the presence of such projection measurements.
While the temporal evolution of complexity has been extensively studied, its interplay with projection measurements remains relatively less comprehended within the framework of holographic models. 
While previous research has investigated the real-time evolution of a post-measurement state~\cite{Goto:2022fec}, there is still an interesting need to investigate the evolution of subsystem complexity in these scenarios.

In a broader context, the phenomenon of complexity transition has attracted recent attention within the realm of random quantum circuits featuring projection measurements~\cite{Suzuki:2023wxw,Niroula:2023meg,Bejan:2023zqm,Fux:2023brx}. 
The state complexity within these circuits experiences a transition from an exponential to a polynomial at late times, influenced by the introduction of measurements. 
A noteworthy topic for further exploration involves extending this study to the domain of subsystem complexity. 
As demonstrated in~\cite{Ho:2021dmh,Choi:2021npc,Ippoliti:2022bsj,Cotler:2021pbc,Claeys:2022hts} and also in our paper, while measurements lead to a decrease in the complexity of the full pure state, intriguingly, the complexity of the subsystem can be enhanced compared to its unmeasured counterpart. 
Consequently, delving into the dynamics of subsystem complexity in a broader context will unveil more intriguing insights.

A geometric approach, exemplified by accessible dimension~\cite{Haferkamp:2021uxo,Suzuki:2023wxw}, has emerged as a crucial methodology for quantification of complexity. 
A potential avenue for refinement lies in revisiting the counting argument presented in the section~\ref{sec:Counting argument for a complexity jump} to achieve a more accurate quantification of mixed state complexity. 
Furthermore, the quest for a versatile measure of quantum complexity, applicable to both pure and mixed states, stands as a significant objective. 

Finally, the relationship between the definition of complexity in quantum information or many-body systems and the complexity notion in holography remains an outstanding question. 
Solvable models with holographic duality, like the SYK model, may lead to useful insights. 
For example, the frame potential has been explored in brownian SYK models~\cite{Jian:2022pvj,Tiutiakina:2023ilu}. 
Our findings present an additional test for the complexity-volume duality, specifically within the context of subsystems and measurements.

\acknowledgments
We are grateful to Bartłomiej Czech for many inspiring conversations and comments on the manuscript. 
We would also like to thank Yingfei Gu, Xiaoliang Qi, Yiming Chen, Zhenbin Yang, Raghu Mahajan, Yifei Wang, Haimeng Zhao for helpful discussions.
SKJ would like to thank Stefano Antonini, Brianna Grado-White, and Brian Swingle for many useful discussions on related topics in previous collaborations.
The work of SKJ is supported by a start-up fund at Tulane university.

\appendix
\section{Computation of volume using Gauss-Bonnet}\label{app:Computation of volume using Gauss-Bonnet}
\begin{figure}
    \centering
    \subfigure[]{\includegraphics[width=0.24\textwidth]{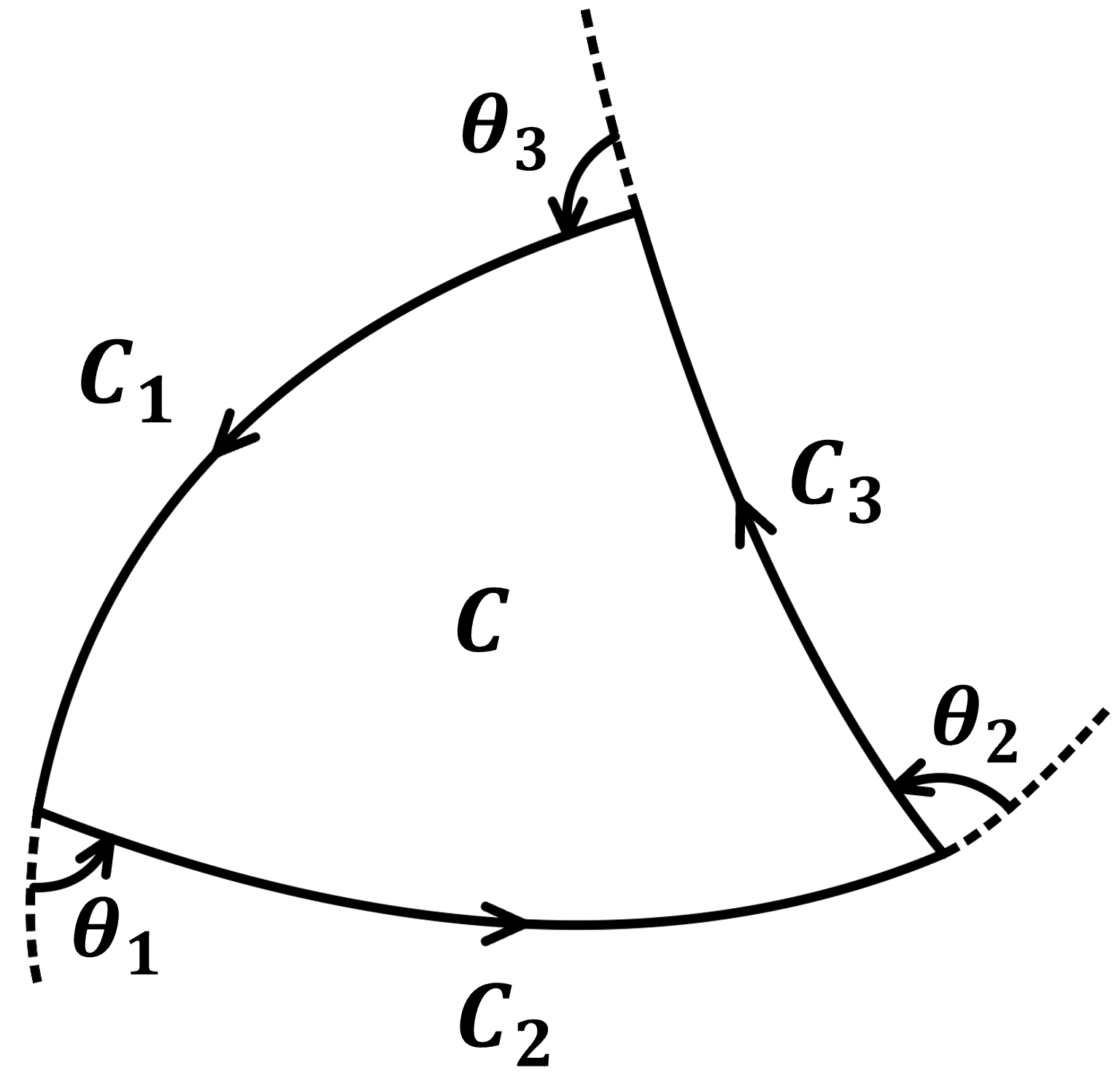}\label{fig:Gauss-bonnet}}
    \quad
    \subfigure[]{\includegraphics[width=0.31\textwidth]{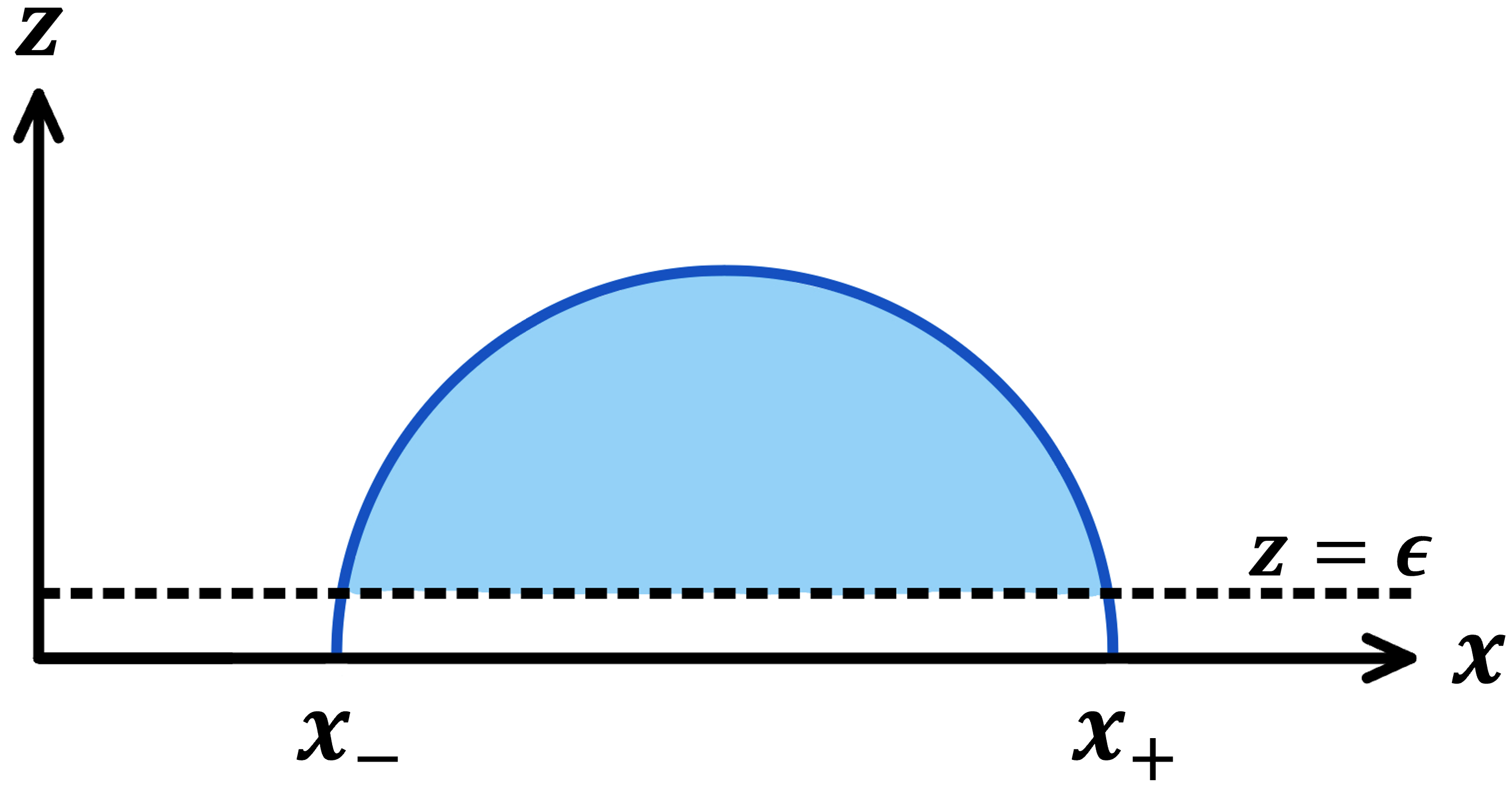}\label{fig:Gauss-bonnet_vacuum}} 
    \quad
    \subfigure[]{\includegraphics[width=0.37\textwidth]{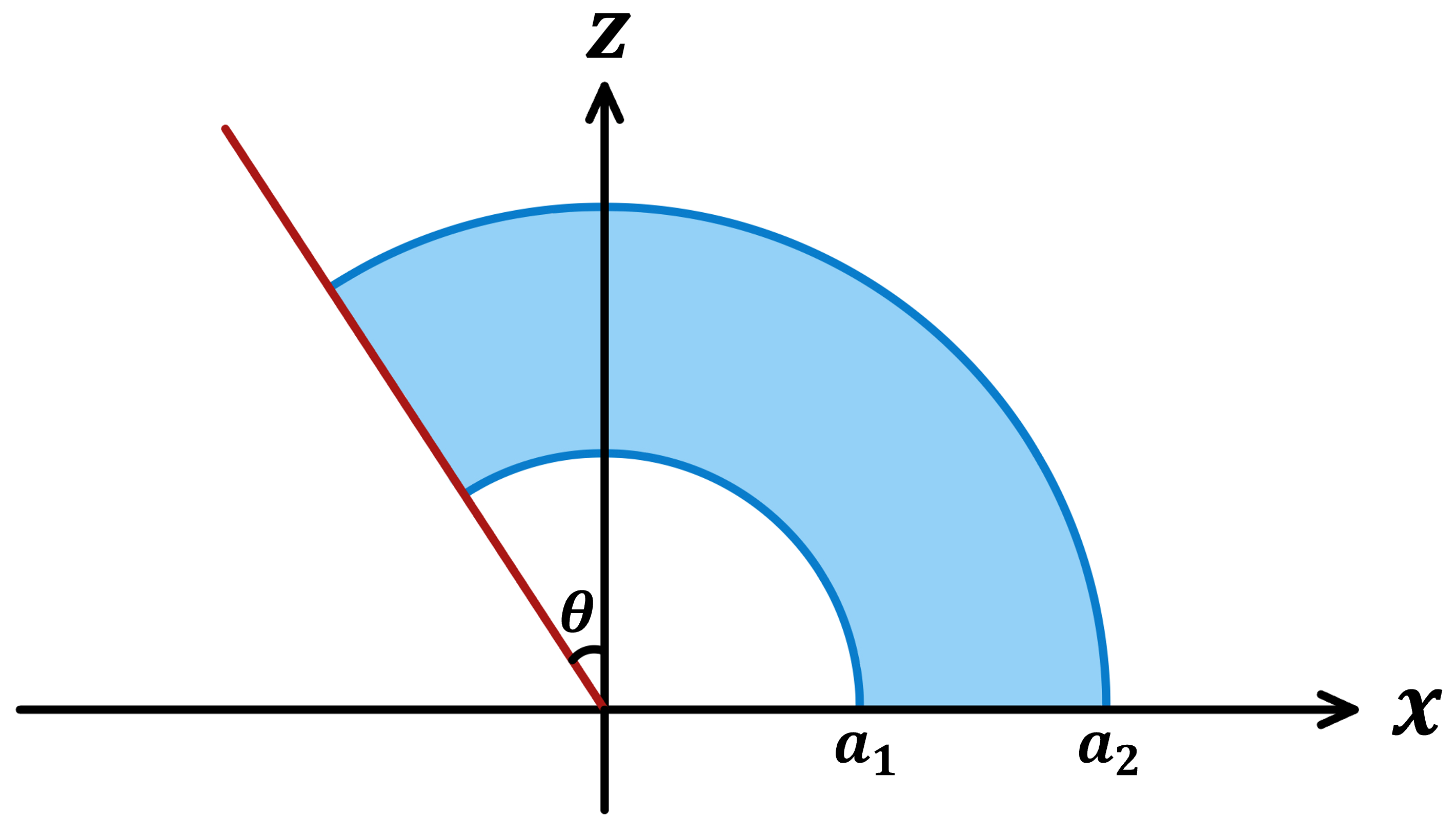} \label{fig:Gauss-Bonet_brane}}
    \caption{Computation of volume using the Gauss-Bonnet theorem. (a) Demonstration of the Gauss-Bonnet theorem. (b) Subsystem in the vacuum. (b) Subsystem complexity corresponding to the RT surface landing on an end-of-the-world brane (brane contribution).}
\end{figure}

The observation of \cite{Abt:2017pmf} is that since the constant $\tau$ slice has constant Gaussian curvature $R_{\operatorname{Gau}}=-\frac{2}{L^2}$, the volume of $\Sigma$ can be cast into topological quantities of $\partial\Sigma$ using the Gauss-Bonnet theorem:
\begin{equation}
	V=\int_\Sigma d\sigma=L^2\left[-\frac{1}{2}\int_{\partial\Sigma}R_{\operatorname{Gau}}d\sigma\right]=L^2\left[\int_{\partial\Sigma}k_gds-2\pi\chi(\Sigma)\right]
\end{equation} 
$k_g$ is the geodesic curvature defined as $k_g\equiv\left|\frac{Du}{ds}\right|$ where $u$ is the unit tangent vector. $\chi$ is the Euler characteristic and will mostly be $1$ in this work. When $\partial\Sigma$ have corners where $k_g$ is singular, then $\int_{\partial\Sigma}k_gd\sigma$ is the integral over the smooth pieces plus the sum of deficit corner angles. For example, in the figure below we have open segments $C_1$, $C_2$, and $C_3$. The deficit corner angles are $\theta_1$, $\theta_2$ and $\theta_3$. 
See figure \ref{fig:Gauss-bonnet} for a plot.

\begin{equation}
    \int_{C}k_gds=\sum_i\int_{C_i}k_gds+\sum_i\theta_i
\end{equation}

In all cases, $\Sigma$ will be enclosed by geodesics, branes, and cut-off surfaces. Geodesics has zero geodesic curvature, hence do not contribute.

~\\
\noindent\textbf{Subregion of vacuum on the infinite line}

For a subregion with length $l$, the volume we should compute is shown in figure \ref{fig:Gauss-bonnet_vacuum}. In the Poincare coordinates $(z,x,\tau)$, the relevant Christoffel symbols are
\begin{equation}
	\Gamma^z_{zz}=-\frac{1}{z},\qquad\Gamma^z_{xx}=\frac{1}{z},\qquad\Gamma^x_{xz}=\Gamma^x_{zx}=-\frac{1}{z}
\end{equation}
The unit tangent vector on the cut-off line is $u=\frac{\epsilon}{L}(0,1)$. 
\begin{equation}
    \frac{Du}{dx}=(\Gamma^z_{xx}u^x,0)=\left(\frac{1}{L},0\right)
\end{equation}
The geodesic curvature is
\begin{equation}
    k_g\equiv\left|\frac{Du}{dx}\right|=\frac{1}{\epsilon}
\end{equation}
There are two corners, each contributing a $\frac{\pi}{2}$. Hence the volume is 
\begin{equation}
	V=L^2\left(\frac{l}{\epsilon}-\pi\right)
\end{equation}

~\\
\noindent\textbf{Brane contribution}

We compute the brane contribution When part of the integration is on the brane (figure \ref{fig:Gauss-Bonet_brane}).
\begin{equation}
	u=\frac{z}{L}(-\cos\theta,\sin\theta)
\end{equation}
\begin{equation}
\begin{aligned}
    \nabla_zu&=0 \\
    \nabla_xu&=(\Gamma^z_{xx}u^x,\Gamma^x_{zx}u^z)=\frac{1}{L}(\sin\theta,\cos\theta) \\
    \frac{Du}{dx}&=\frac{dz}{dx}\nabla_zu+\nabla_xu=\frac{1}{L}(\sin\theta,\cos\theta)
\end{aligned}
\end{equation}
\begin{equation}
	k_g\equiv\left|\frac{Du}{dx}\right|=\frac{1}{z}
\end{equation}
\begin{equation}
	\int_{\mbox{brane}}k_gdx=\int_{-a_2\sin\theta}^{-a_1\sin\theta}\frac{\tan\theta}{-x}=\tan\theta\cdot\log\frac{a_2}{a_1}
\end{equation}
The surface intersects the brane orthogonally, contributing a $\pi$. So the overall contribution from the brane is 
\begin{equation}
	V_{\mbox{brane}}=L^2\left(\tan\theta\cdot\log\frac{a_2}{a_1}+\pi\right)
\end{equation}

\section{Details for the black hole coupled to bath setup}\label{app:bh-bath_details}

\subsection{Infinite measurement, intersecting configuration}
In the global coordinates, the brane profiles are
\begin{equation}
	\phi_B(r)=-\arctan\frac{\calT_B}{\sqrt{(1-\calT_B^2)r^2-\calT_B^2}},\qquad\phi_M(r)=\alpha+\arctan\frac{\calT_M}{\sqrt{(1-\calT_M^2)r^2-\calT_M^2}}
\end{equation}
See figure \ref{fig:bh-bath_infinite_intersecting_bulk_global} for an illustration. 

\begin{figure}
    \centering
    \includegraphics[width=\textwidth]{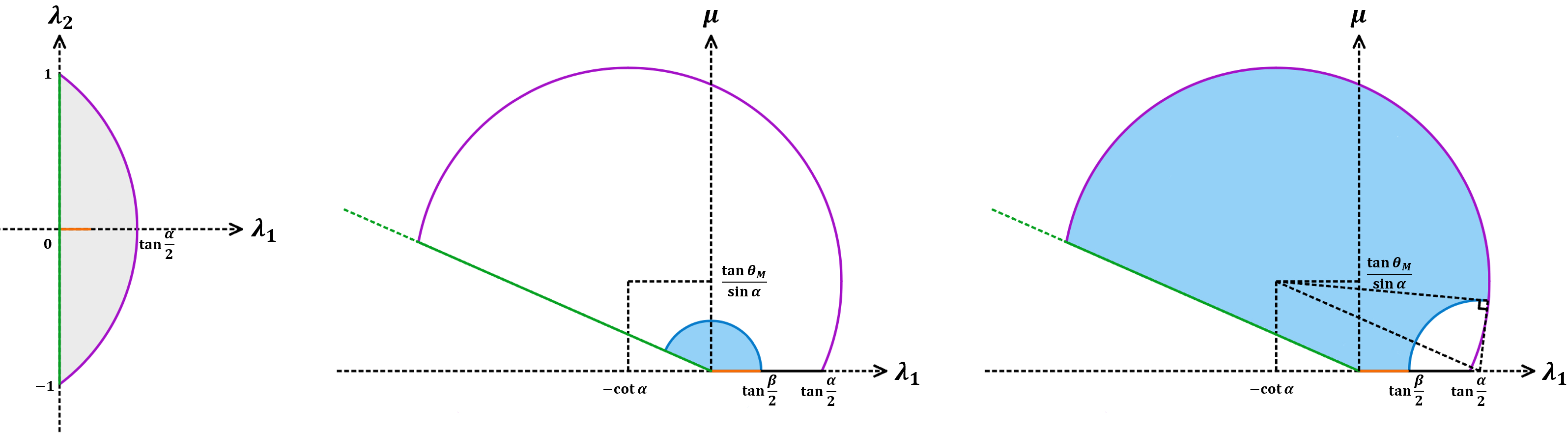}
    \caption{Boundary and bulk configuration in the Poincare coordinate.
    (left panel) The boundary in the Poincare patch. 
    The green line and the purple curve represent the system and the measurement, respectively.
    The orange interval is the subregion $A$.
    (middle/right panel) The bulk in the Poincare patch. 
    The green line and the purple curve represent the system brane and the measurement brane, respectively.
    The orange interval is the subregion $A$.
    The blue curve shows the candidate RT surface landing on the system brane (middle panel) and on the measurement brane (right panel).}
    \label{fig:app_bh-bath_infinite_intersecting_bulk_poincare}
\end{figure}

Next, we switch to the Poincare coordinate $(\mu,\lambda,\bar{\lambda})$ with the black hole boundary at $\lambda_1=0$. 
One reason for doing this is that the brane configurations greatly simplified in the Poincare coordinate. 
Another reason is that we would like to use the coordinate transformation (\ref{bulktrans}) which starts from Poincare coordinates to fix the position of the cut-off surface. 
The global coordinate $(r,\phi,\tau)$ is related to the Poincare coordinate $(\mu,\lambda,\bar{\lambda})$ by
\begin{equation}
\begin{aligned}
	\sqrt{r^2+1}\cosh\tau&=\frac{1+\mu^2+\lambda_1^2+\lambda_2^2}{2\mu}\\
	\sqrt{r^2+1}\sinh\tau&=\frac{\lambda_2}{\mu} \\
	r\sin\phi&=\frac{\lambda_1}{\mu} \\
	r\cos\phi&=\frac{1-\mu^2-\lambda_1^2-\lambda_2^2}{2\mu}
\end{aligned}
\end{equation}
Set $\chi=\phi+i\tau$. 
At the boundary $r\rightarrow\infty$ and $\mu\rightarrow0$, the dual transformation is
\begin{equation}
	\lambda=\tan\frac{\chi}{2}
\end{equation}
The black hole boundary is mapped to $\lambda_1=0,\lambda_2\in(-1,1)$ and the measurement boundary is an arc connecting $\lambda_1=0,\lambda_2=1$ and $\lambda_1=0,\lambda_2=-1$. The end point of $A$ is at $\lambda_1=\tan\frac{\beta}{2}=\tan\left(\frac{\alpha}{\pi}\arcsin\frac{l}{q}\right)$. In the bulk, we have the system brane at
\begin{equation}
	\mu_B(\lambda)=-\lambda_1\cot\theta_B
\end{equation}
To obtain the position of the measurement brane, we notice that in another Poincare coordinate $(\mu',\lambda',\bar{\lambda}')$ where the measurement boundary is at $\lambda_1'=0$, the measurement brane takes the simple form
\begin{equation}
	\mu_M'(\lambda')=\lambda_1'\cot\theta_M \label{minpoincare}
\end{equation}
The new Poincare coordinate is related to the global coordinates with an additional angle $\phi\rightarrow\phi-\alpha$. Therefore, the $(\mu',\lambda',\bar{\lambda}')$ coordinates are related to the $(\mu,\lambda,\bar{\lambda})$ coordinates in the following way:
\begin{equation}
	\frac{\lambda_1'}{\mu'}=r\sin(\phi-\alpha)=r\sin\phi\cos\alpha-r\cos\phi\sin\alpha=\frac{\lambda_1}{\mu}\cos\alpha-\frac{1-\mu^2-\lambda_1^2-\lambda_2^2}{2\mu}\sin\alpha.
\end{equation}
Plugging in \eqref{minpoincare}, we get a sphere:
\begin{equation}
	(\lambda_1+\cot\alpha)^2+\lambda_2^2+\left(\mu-\frac{\tan\theta_M}{\sin\alpha}\right)^2=\frac{1}{\sin^2\alpha\cos^2\theta_M}.
\end{equation}
See figure \ref{fig:app_bh-bath_infinite_intersecting_bulk_poincare} for an illustration. When we take the $\alpha\rightarrow 0$ limit, the boundary becomes an infinite strip and the measurement brane becomes ``straight'': we recover the non-intersecting configuration discussed in Sec \ref{sec:bh-bath_infinite_non-intersecting}.

The surface that lands on the system brane is part of the semicircle as usual. 
The surface that lands on the measurement brane can be determined by requiring that it intersects the brane orthogonally. 
To see the reason for this, we can switch to the Poincare coordinate $(\mu',\lambda',\bar{\lambda}')$ and find that the surface is part of a semicircle that intersects the measurement brane orthogonally. In some parameter region, the first candidate surface might intersect with the measurement brane when $A$ gets large enough. In this case the first candidate ceases to exist. However, one shouldn't worry about this because it can only happen when the second surface is already dominant. 

To compute complexity, we also need the location of the intersection point of the two branes. Let $(\lambda_I,\mu_I)$ denote its location in Poincare coordinates. It is fixed by
\begin{equation}
\begin{aligned}
	&(\lambda_{I}+\cot\alpha)^2+\left(\mu_I-\frac{\tan\theta_M}{\sin\alpha}\right)^2=\frac{1}{\sin^2\alpha\cos^2\theta_M} \\
	&\lambda_I = -\mu_I\tan\theta_B \\
	\Rightarrow&\quad\mu_I=\frac{\cos\theta_B}{\sin\alpha}\left[(\cos\alpha\tan\theta_B+\tan\theta_M)+\sqrt{\frac{1}{\cos^2\theta_M}+(\cos\alpha\tan\theta_B+\tan\theta_M)^2}\right],
\end{aligned}
\end{equation}
where the first line is the profile of the measurement brane, and the second line is the profile of the system brane. Similarly, the intersection point in the Poincare coordinate  $(\mu',\lambda',\bar{\lambda}')$ is at
\begin{equation}
	\quad\mu_I'=\frac{\cos\theta_M}{\sin\alpha}\left[(\cos\alpha\tan\theta_M+\tan\theta_B)+\sqrt{\frac{1}{\cos^2\theta_B}+(\cos\alpha\tan\theta_M+\tan\theta_B)^2}\right].
\end{equation}

\subsection{Finite measurement, non-intersecting configuration}
\begin{figure}
    \centering
    \includegraphics[width=\textwidth]{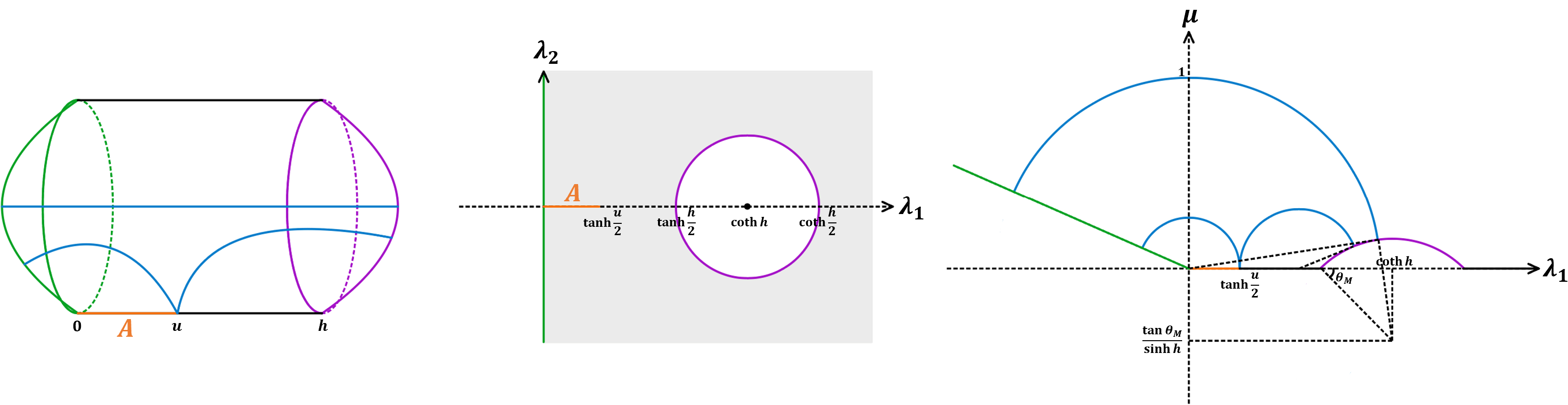}
    \caption{Gravity dual for the finite measurement, non-intersecting case. 
    Color green and purple denote the system and the measurement region (or brane).
    Blue curve in the left and right panel denotes the candidate RT surface.
    (left penal) The global coordinate $(x,\phi,r)$, $\chi=x+i\phi$. 
    (middle penal) The boundary coordinate $\lambda$. 
    (right penal) The bulk in the Poincare coordinate $(\mu, \lambda, \bar \lambda)$.}
    \label{fig:app_bh-bath_finite_non-intersecting_bulk}
\end{figure}

In this configuration the bulk is global AdS$_3$ ``capped off'' by end-of-the-world branes anchored at the two boundaries. 
An important consequence is that the divergence in Sec \ref{sec:bh-bath_infinite_non-intersecting} disappears. 
The global metric reads
\begin{equation}
\begin{aligned}
	&ds^2=L^2\left[(r^2+1)dx^2+\frac{dr^2}{r^2+1}+r^2d\phi^2\right] \\
	&\chi=\frac{\pi}{\calK(1-k^2)}\xi= x+i\phi,\quad\phi\sim\phi+2\pi,\quad	x\in[0,h],\quad h\equiv\pi\frac{\calK(k^2)}{\calK(1-k^2)}	.
\end{aligned}
\end{equation}
Let $u=\frac{\pi}{\calK(1-k^2)}\sn^{-1}\left(\frac{l}{q},k^2\right) \in(0,h)$ denote the end point of $A$. 
The brane profile of the system brane and the measurement brane is respectively,
\begin{equation}
	x_B=-\arcsinh\frac{\tan\theta_B}{\sqrt{r^2+1}},\qquad x_M=h+\arcsinh\frac{\tan\theta_M}{\sqrt{r^2+1}}.
\end{equation}
See the left figure in figure \ref{fig:app_bh-bath_finite_non-intersecting_bulk}. 
To compute entanglement and complexity, we switch to the Poincare coordinate defined by
\begin{equation}
    \begin{aligned}
	\sqrt{r^2+1}\cosh x&=\frac{1+\mu^2+\lambda_1^2+\lambda_2^2}{2\mu}\\
	\sqrt{r^2+1}\sinh x&=\frac{\lambda_1}{\mu} \\
	r\sin\phi&=\frac{\lambda_2}{\mu} \\
	r\cos\phi&=\frac{1-\mu^2-\lambda_1^2-\lambda_2^2}{2\mu}
    \end{aligned}
\end{equation}
The boundary and bulk configuration in the Poincare coordinate is demonstrated in figure \ref{fig:app_bh-bath_finite_non-intersecting_bulk}. 
At the boundary, this transformation becomes
\begin{equation}
	\lambda=\tanh\frac{\chi}{2}.
\end{equation}
The black hole boundary is at $\lambda_1=0$. 
As usual, the system brane is at
\begin{equation}
	\lambda_1=-\mu\tan\theta_B.
\end{equation}
To find the position of the measurement boundary, we notice that in another Poincare coordinate $(\lambda',\bar{\lambda}',\mu')$ where the measurement boundary is at $\lambda_1'=0$, the measurement brane shoots out radially:
\begin{equation}
	\lambda_1'=\mu'\tan\theta_M \label{measurmentbrane}.
\end{equation}
The new Poincare coordinate is related to the global coordinates with a shift $x\rightarrow x-h$. We have
\begin{equation}
\begin{aligned}
    \frac{\lambda_1'}{\mu'}&=\sqrt{r^2+1}\sinh(x-h)=\sqrt{r^2+1}(\sinh x\cosh h-\cosh x\sinh h)\\
    &=\frac{\lambda_1}{\mu}\cosh h-\frac{1+\mu^2+\lambda_1^2+\lambda_2^2}{2\mu}\sinh h.
\end{aligned}
\end{equation}
Plugging in \eqref{measurmentbrane}, we get a sphere for the measurement brane:
\begin{equation}
	(\lambda_1-\coth h)^2+\lambda_2^2+\left(\mu+\frac{\tan\theta_M}{\sinh h}\right)^2=\frac{1}{\sinh^2h\cos^2\theta_M}.
\end{equation}

The cut-off surface is at
\begin{equation}
\begin{aligned}
	&\mu_\epsilon=\left|\frac{d\lambda}{dw}\right|\epsilon=\left|\frac{d\lambda}{d\chi}\right|\left|\frac{d\chi}{d\xi}\right|\left|\frac{d\xi}{dw}\right|\epsilon
	=\frac{\pi}{2\cosh^2\frac{\chi}{2}\calK(1-k^2)q\sn'(\xi)}\epsilon, \\
	&\mu_\epsilon(0)=\frac{\pi}{2\calK(1-k^2)q}\epsilon,\qquad\mu_\epsilon(u)=\frac{\pi}{2\cosh^2\frac{\chi}{2}\calK(1-k^2)q\sn'\left(\frac{\calK(1-k^2)u}{\pi}\right)}\epsilon.
\end{aligned}
\end{equation}

\subsection{Facts about the $\sn$ function}
\begin{equation}
	\sn^2\xi+\cn^2\xi=1
\end{equation}
\begin{equation}
	\dn^2\xi+k^2\sn^2\xi=1
\end{equation}
\begin{equation}
	\dn^2\xi-k^2\cn^2\xi=1-k^2
\end{equation}
\begin{equation}
	\sn'\xi=\cn\xi\cdot\dn\xi=\sqrt{(1-\sn^2\xi)(1-k^2\sn^2\xi)}
\end{equation}

\bibliographystyle{JHEP}
\bibliography{reference.bib}

\end{document}